\PassOptionsToPackage{square,numbers,sort&compress}{natbib}
\documentclass[pdftex,twocolumn,epjc3]{svjour3}

\usepackage{graphicx}
\usepackage{bm}
\usepackage{epsf}
\usepackage{epsfig}
\usepackage{amsmath}
\usepackage{amssymb}
\usepackage{amsfonts}
\usepackage{float}

\usepackage{dsfont}
\usepackage{slashed}
\usepackage{booktabs}

\usepackage{multirow}
\usepackage{siunitx}
\usepackage{listings}
\usepackage{etoolbox}
\usepackage{cuted}
\usepackage[dvipsnames]{xcolor}

\usepackage[mathscr]{euscript}

\usepackage[pdftex,colorlinks=true,
    linkcolor=blue,
    filecolor=blue,
    urlcolor=blue,
    citecolor=blue,
    plainpages=false,
    hyperfootnotes=false,
    bookmarksopen=true]{hyperref} % always put hyperref last

\journalname{Eur. Phys. J. C}

\newcommand{\MulRow}{\multirow}
\newcommand{\MulCol}{\multicolumn}

\newcommand{\diag}{\mathop{\rm diag}}

\newcommand{\tr}{\mathop{\rm tr}}

\newcommand{\vslash}{\ensuremath{v\kern-0.55em/\kern0.05em}}
\newcommand{\vhslash}{\ensuremath{\hat{v}\kern-0.45em/\kern0.05em}}
\newcommand{\eslash}{\ensuremath{\kern0.1em\epsilon\kern-0.4em/\kern-0.1em}}
\newcommand{\ebslash}{\ensuremath{\kern0.1em\bar\epsilon\kern-0.4em/\kern-0.1em}}
\renewcommand{\case}[2]{{\displaystyle\frac{1}{2}}}
\newcommand{\half}{\ensuremath{\case{1}{2}}}

\newcommand{\bOneCol}{\begin{strip}\rule[-1ex]{\columnwidth}{1pt}\rule[-1ex]{1pt}{1.5ex}}
\newcommand{\eOneCol}{\hfill\rule[1ex]{1pt}{1.5ex}\rule[2.3ex]{\columnwidth}{1pt}\end{strip}}
 
\definecolor{MyDarkGreen}{rgb}{0.0, 0.545098, 0.0}
\definecolor{FNALblue}{HTML}{004C97}
\definecolor{Teal}{HTML}{008080}

\newcommand{\str}[1]{}                            % text to be deleted and not shown, to keep the PDF readable

\newcommand\Tstrut{\rule{0pt}{2.4ex}}       % top strut
\newcommand\Bstrut{\rule[-1.3ex]{0pt}{0pt}} % bottom strut
\newcommand{\Hs}[1]{\hspace{#1}}            % Hspace for siunitx labels
\newcolumntype{R}{S[table-text-alignment=right]}

\newcommand{\epm}[3]{\Hs{-#1cm}$^{+#2}_{-#3}$}
\newcommand{\cnt}[1]{$\times 10^{#1}$}

\newbool{PRD}
\newcommand{\basePlotDir}{plots}
\boolfalse{PRD}

\RequirePackage{natbib}[square,numbers,sort,compress]
\setcitestyle{square,numbers,sort&compress}

\begin{document}
%\preprint{FERMILAB-PUB-21/261-T}

\setcitestyle{square,numbers,comma,sort&compress}

\title{\boldmath Semileptonic form factors for $B\to D^\ast\ell\nu$ at nonzero recoil from \\ $2+1$-flavor lattice QCD}

\date{\today}
\author{A.~Bazavov\thanksref{msu}\and
        C.E.~DeTar\thanksref{utah}\and
        D.~Du\thanksref{syr}\and
        A.X.~El-Khadra\thanksref{uiuc,icasu}\and
        E.~G\'amiz\thanksref{ugr}\and
        Z.~Gelzer\thanksref{uiuc}\and
        S. Gottlieb\thanksref{IU}\and
        U.M.~Heller\thanksref{aps}\and
        A.S.~Kronfeld\thanksref{fnal}\and
        J.~Laiho\thanksref{syr}\and
        P.B.~Mackenzie\thanksref{fnal}\and
        J.N.~Simone\thanksref{fnal}\and
        R.~Sugar\thanksref{ucsb}\and
        D.~Toussaint\thanksref{ua}\and
        R.S.~Van~de~Water\thanksref{fnal}\and
        A.~Vaquero\thanksref{em,utah}\\
        (Fermilab Lattice and MILC collaborations)}

\institute{Department of Computational Mathematics, Science and Engineering, and Department of Physics and Astronomy,
           Michigan State University, East Lansing, Michigan 48824, USA\label{msu}\and
           Department of Physics and Astronomy, University of Utah, Salt Lake City, Utah 84112, USA\label{utah}\and
           Department of Physics, Syracuse University, Syracuse, New York 13244, USA\label{syr}\and
           Department of Physics, University of Illinois, Urbana, Illinois 61801, USA\label{uiuc}\and
           Illinois Center for Advanced Studies of the Universe, University of Illinois, Urbana, Illinois 61801, USA\label{icasu}\and
           Departamento de Física Teórica y del Cosmos, Universidad de Granada, E-18071 Granada, Spain\label{ugr}\and
           Department of Physics, Indiana University, Bloomington, Indiana 47405, USA\label{IU}\and
           American Physical Society, Ridge, New York 11961, USA\label{aps}\and
           Fermi National Accelerator Laboratory, Batavia, Illinois 60510, USA\label{fnal}\and
           Department of Physics, University of California, Santa Barbara, California 93106, USA\label{ucsb}\and
           Department of Physics, University of Arizona, Tucson, Arizona 85721, USA\label{ua}}

\thankstext{em}{e-mail: alexv@unizar.es}
%\author{A.~Bazavov}
%\affiliation{\msu}
%\author{C.E.~DeTar}
%\affiliation{\utah}
%\author{Daping Du}
%\affiliation{\syr}
%\author{A.X.~El-Khadra}
%\affiliation{\uiuc}
%\affiliation{\icasu}
%\author{E.~G\'amiz}
%\affiliation{\ugr}
%\author{Z.~Gelzer}
%\affiliation{\uiuc}
%\author{Steven Gottlieb}
%\affiliation{\IU}
%\author{U.M. Heller}
%\affiliation{\aps}
%\author{A.S.~Kronfeld}
%\affiliation{\fermilab}
%\author{J.~Laiho}
%\affiliation{\syr}
%\author{P.B.~Mackenzie}
%\affiliation{\fermilab}
%\author{J.N. Simone}
%\affiliation{\fermilab}
%\author{R. Sugar}
%\affiliation{\ucsb}
%\author{D. Toussaint}
%\affiliation{\ua}
%\author{R.S.~Van de Water}
%\affiliation{\fermilab}
%\author{A.~Vaquero}
%\email{alexvaq@physics.utah.edu}
%\affiliation{\utah}

%\collaboration{Fermilab Lattice and MILC Collaborations}
%\noaffiliation

\maketitle

\begin{abstract}
We present the first unquenched lattice-QCD calculation of the form factors for the decay $B\rightarrow D^\ast\ell\nu$ at nonzero recoil.
Our analysis includes 15 MILC ensembles with $N_f=2+1$ flavors of asqtad sea quarks, with a strange quark mass close to its physical mass.
The lattice spacings range from $a\approx 0.15$ fm down to $0.045$ fm, while the ratio between the light- and the strange-quark masses ranges from 0.05 to 0.4.
The valence $b$ and $c$ quarks are treated using the Wilson-clover action with the Fermilab interpretation, whereas the light sector employs asqtad staggered fermions.
We extrapolate our results to the physical point in the continuum limit using rooted staggered heavy-light meson chiral perturbation theory.
Then we apply a model-independent parametrization to extend the form factors to the full kinematic range.
With this parametrization we perform a joint lattice-QCD/experiment fit using several experimental datasets to determine the CKM matrix element $|V_{cb}|$.
We obtain $\left|V_{cb}\right| = (38.40 \pm 0.68_{\textrm{th}} \pm 0.34_{\textrm{exp}} \pm 0.18_{\textrm{EM}})\times 10^{-3}$. 
The first error is theoretical, the second comes from experiment and the last one includes electromagnetic and electroweak uncertainties, with an overall
$\chi^2\text{/dof} = 126/84$,  which illustrates the tensions between the experimental data sets, and between theory and experiment.
This result is in agreement with previous exclusive determinations, but the tension with the inclusive determination remains. 
Finally, we integrate the differential decay rate obtained solely from lattice data to predict $R(D^\ast) = 0.265 \pm 0.013$, which
confirms the current tension between theory and experiment.
 % \hfill
\end{abstract}

\sisetup{uncertainty-mode = compact-marker, table-alignment-mode = format, group-digits = none, group-minimum-digits = 6}

\section{Introduction}

High precision tests of the standard model (SM) offer exciting possibilities for discovering new physics.  In particular, the flavor
sector of the SM is very rich in phenomena that can be used to explore physics beyond the standard model (BSM).  Most flavor physics
revolves around the Cabibbo-Kobayashi-Maskawa (CKM) matrix, which relates the mass and flavor eigenstates of the quarks.  Since it
is a basis transformation, the CKM matrix is constrained by unitarity, so violations of this rule could indicate the influence of
new physics.  Weak processes that are loop-suppressed in the SM may also expose new physics.  To determine CKM matrix elements to
high precision and to perform precision tests of the SM in measurements of rare decay processes, it is essential to know the
strong-interaction environment in which these processes occur.

Among the CKM matrix elements, $\left|V_{cb}\right|$ has arguably been one of the most perplexing.  There is a long-standing tension
between the determination of this element via exclusive and inclusive decays.  The operator product expansion (OPE) is used to
analyze inclusive decay experiments measuring semileptonic decays $B \to X_c\ell\nu$, where $X_c$ represents any charmed hadron or
combination of hadrons with a single $c$ quark.  On the other hand, exclusive decay experiments focus on decays with a specific
charmed hadron in the final state, for example, $B \to D\ell\nu$ or $B \to D^\ast\ell\nu$.  We expect both types of experiments to
yield consistent results for $|V_{cb}|$; however, there is a $\sim 3\sigma$ discrepancy between the inclusive and exclusive
determinations~\cite{Amhis:2019ckw,Zyla:2020zbs}.  Since contributions from new physics are unlikely to explain these differences
\cite{Crivellin:2014zpa,Jung:2018lfu}, the disagreement presents an obstacle to higher precision tests of the SM.

We turn now to the determination of $|V_{cb}|$ from the exclusive decay $B \to D^\ast\ell\nu$, which has an interesting history.  As
is detailed below in Eq.~\eqref{DiffDecayRate}, to determine $|V_{cb}|$, a measurement of the differential decay rate $d\Gamma/dw$
is needed and the form factors must be computed by theory.  Using lattice QCD, it has been possible to determine the key form factor
at zero recoil.  However, the differential decay rate vanishes at that point because of kinematic factors, so it is necessary to use
the differential decay rate at nonzero recoil to extrapolate the form factors to zero recoil.  Since the 1990s, there have been two
parametrizations of the form factors, one by Boyd, Grinstein, and Lebed (BGL)~\cite{Boyd:1995sq,Boyd:1995cf,Boyd:1997kz}, and the
other by Caprini, Lellouch, and Neubert (CLN)~\cite{Caprini:1997mu}. The CLEO and BaBar experiments~\cite{Aubert:2007qs,Adam:2002uw,Aubert:2008yv},
for example, relied on the CLN parametrization to analyze the dependence of the recoil parameter of their data.
This was also true for earlier reviews from the Heavy Flavor Averaging Group (HFLAV)~\cite{Amhis:2012bh}.

The situation changed in 2017 with the publication of unfolded data from a $B \to D^\ast\ell\nu$ experiment by the Belle
collaboration~\cite{Abdesselam:2017kjf}.  This data release was quickly followed by several theoretical
analyses~\cite{Bigi:2017njr,Bigi:2017jbd,Grinstein:2017nlq,Jaiswal:2017rve} comparing the effect of the choice of
parametrization on $|V_{cb}|$.  They found that the CLN parametrization~\cite{Caprini:1997mu}, at least as it is usually employed to
extrapolate the experimental data to the zero-recoil point (see for instance, Sec.V.A of Ref.~\cite{Waheed:2018djm}), does not provide
a good description of the experimental data, whereas the BGL parametrization~\cite{Boyd:1995sq,Boyd:1995cf,Boyd:1997kz} describes the
data properly and yielded exclusive determinations of $|V_{cb}|$ that were compatible with the inclusive
ones~\cite{Grinstein:2017nlq,Bigi:2017njr,Bigi:2017jbd,Jaiswal:2017rve}.
However, more recent analyses by the Belle collaboration using the much larger untagged dataset~\cite{Waheed:2018djm} and by the
BaBar collaboration performing a new analysis of their old data~\cite{Dey:2019bgc} contradicted this picture and reinforced the
long-standing tension between the inclusive and the exclusive determinations.  Newer theoretical analyses using Belle's
untagged dataset also found agreement between CLN and BGL results~\cite{Gambino:2019sif,Jaiswal:2020wer}. These analyses
also conclude that CLN is still a useful parametrization given the current errors in the experimental measurements.  Unfortunately,
previous unquenched lattice-QCD calculations of the $B\to D^*$ form factor~\cite{Bailey:2014tva,McLean:2019sds} cannot provide
constraints on the shape, because they are limited to zero recoil.  Hence, a precise calculation from first principles of the form
factors involved in the exclusive process performed for a range of nonzero-recoil momenta, could be extremely helpful.

Another motivation to study this process is the existing tension between experimental measurements and SM predictions of several
lep\-ton-\-fla\-vor-\-u\-ni\-ver\-sal\-i\-ty-\-vi\-o\-la\-ting (LFUV) observables in $B$-meson semileptonic decays~\cite{Bifani:2018zmi}.
The ratios of branching fractions of the semitauonic and other semileptonic $B\to D^{(*)}$ transitions
\begin{equation}
    R(D^{(*)})\equiv\frac{{\cal B}(B\to D^{(*)}\tau\nu_\tau)}{{\cal B}(B\to D^{(*)}\ell\nu_\ell)},\quad\quad \ell=e,\mu
    \label{RDstDef}
\end{equation}  
disagree with the SM at the $\sim 3\sigma$ level when $R(D)$ and $R(D^\ast)$ are taken together~\cite{Amhis:2019ckw}.  Although the
last HFLAV average~\cite{Amhis:2019ckw} shows a large discrepancy between theory and experiment, the most recent measurements from
the BaBar, Belle, and LHCb collaborations find $R(D^\ast)$ to be closer to SM
expectations~\cite{Hirose:2016wfn,Hirose:2017dxl,Aaij:2017uff,Aaij:2017deq,Belle:2019rba}.  However, a complete lattice-QCD
calculation of $R(D^\ast)$ is still lacking.  In view of current tensions in these observables, the limitations of the available
theoretical predictions, and the future improvements on the experimental side expected from LHCb and Belle II forthcoming data, an
independent theoretical calculation with a tight control of systematic errors that could help to either confirm or reduce the
tension is urgently needed.

In this work, we use lattice QCD to address these two points, i.e., tensions between exclusive and inclusive determinations of
$|V_{cb}|$, and between the SM theoretical prediction and experiment for $R(D^*)$.  Although lattice QCD has previously been used to
extract $|V_{cb}|$ from experimental data for $B\to D^\ast\ell\nu$, the relevant decay amplitude has always been computed at zero
recoil~\cite{Hashimoto:2001nb,Bernard:2008dn,Bailey:2014tva,Harrison:2017fmw}, except for an early study in the quenched
approximation~\cite{deDivitiis:2008df}.  Here we compute the form factors that contribute to the $B\to D^\ast\ell\nu$ decay for
nonzero values of the recoil parameter in full QCD with $2+1$ flavors of dynamical sea quarks and extrapolate their behavior to the
large recoil region.  Instead of using the standard procedure of extrapolating experimental results to zero recoil and then
extracting $\left|V_{cb}\right|$ using the calculated value of the form factor at zero recoil, we do a joint fit of lattice and
experimental data where $\left|V_{cb}\right|$ is one of the free parameters.  Once the decay amplitude is determined, we integrate
over the whole kinematic space to find the branching ratios with and without a $\tau$ in the decay.  We calculate $R(D^\ast)$ and
compare our results with existing experimental determinations.  Preliminary reports of this analysis were presented in
Refs.~\cite{Qiu:2013ofa,Aviles-Casco:2017nge,Aviles-Casco:2019vin,Vaquero:2019ary,Aviles-Casco:2019zop}.  In keeping with previous
work on the same ensembles~\cite{Lattice:2015tia,Bailey:2015nbd,Bailey:2015dka,Bazavov:2019aom}, this analysis was blinded until a
systematic error budget was finalized. The final results were then frozen, apart from unblinding.

This article is organized as follows.
In Sec.~\ref{FormFactors}, we introduce the formalism and present the form factors that take part in our calculation.
Section~\ref{Analysis} gives details on the ensembles available and also describes the analysis of the lattice data up to the
chiral-continuum extrapolation.
In Sec.~\ref{ErrorBudget}, we discuss the systematic errors, and Sec.~\ref{Results} shows the $z$ expansion and the joint fit with
experimental data that leads to our final results for $\left|V_{cb}\right|$ and $R(D^\ast)$.
Section~\ref{Conclusions} presents our conclusions.
\ref{ApChiral} includes details on the fit function, employed in the chiral-continuum extrapolation.
\ref{ApMatch} outlines the calculation of the matching factors and its errors.
\ref{ApHQCorr} explains in detail how the $\kappa$~tuning correction for the heavy quarks was calculated.
Finally, \ref{ApFullRes} provides a guide to ancillary files containing complete results for the form factors, with a full correlation matrix.

\ifbool{PRD}{\begin{figure}[b]
  \includegraphics[width=0cm]{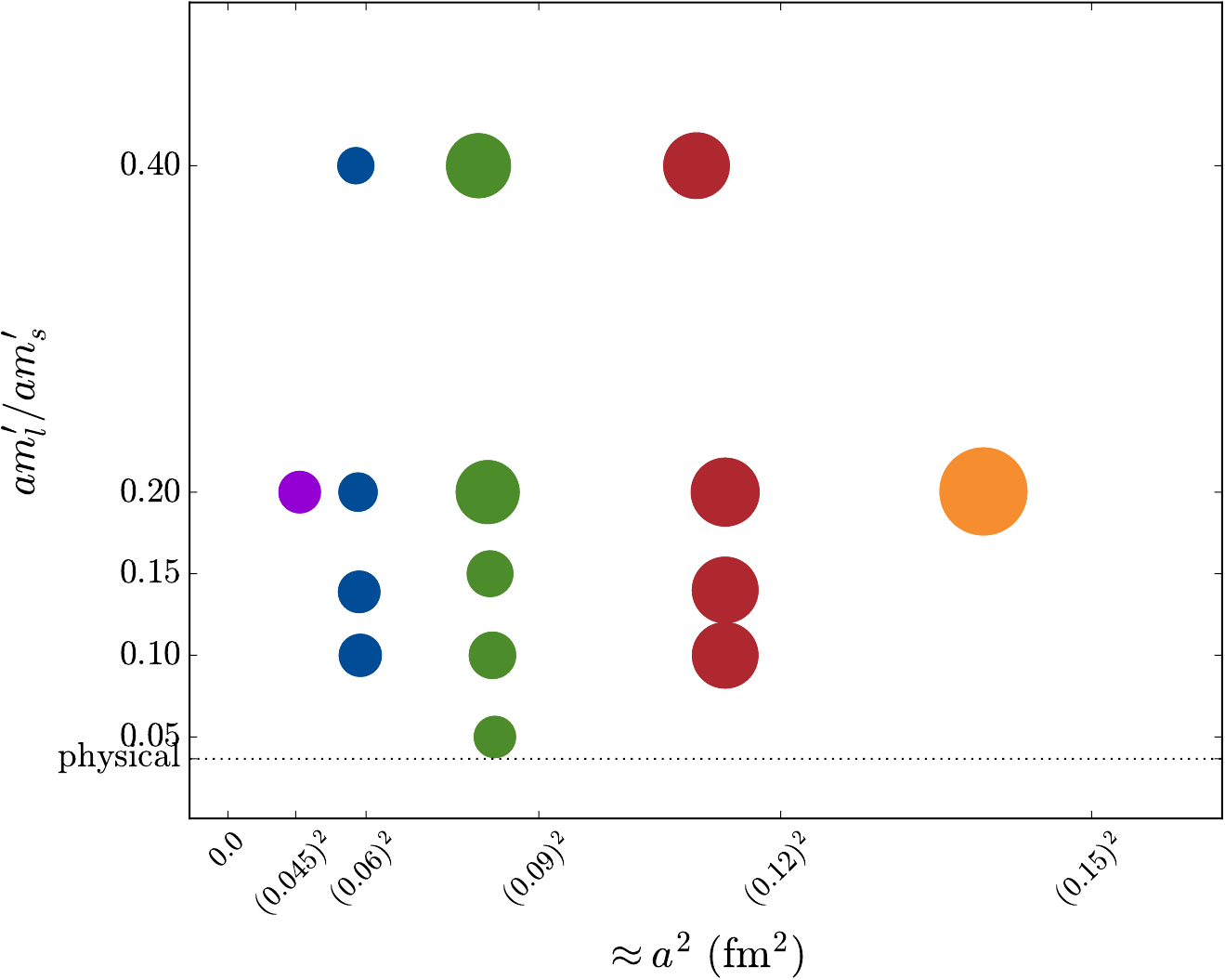}
\end{figure}}{}

\section{Form Factor Definitions}
\label{FormFactors}

In this section, we set the definitions and notation used in the following sections for the form factors, ratio of correlators,
currents, and renormalization factors, among others.

\subsection{Form factors in the continuum}

The $B\to D^\ast\ell\nu$ process is mediated by the axial $\mathcal{A}^\mu = \bar{c}\gamma_\mu\gamma_5b$ and the vector
$\mathcal{V}^\mu = \bar{c}\gamma_\mu b$ electroweak currents.
The transition matrix for this process is usually decomposed into form factors inspired by the heavy quark effective theory 
(HQET)~\cite{Falk:1992wt}:
\begin{align} % irrelevant signs and factors of i may disagree with A.S.K. note
    \frac{\langle D^\ast(p_{D^\ast},\varepsilon) |\mathcal{A}^\mu| B(p_B)\rangle}{\sqrt{M_{D^\ast}M_B}} &=
        i\varepsilon^*_\nu \Big[g^{\mu\nu}(w+1)\,h_{A_1}(w) \nonumber \\
        & \hspace{-1cm} - v_B^\nu(v_B^\mu\,h_{A_2}(w) + v_{D^\ast}^\mu\,h_{A_3}(w))\Big],
    \label{axCur} \\
    \frac{\langle D^\ast(p_{D^\ast},\varepsilon) |\mathcal{V}^\mu| B(p_B)\rangle}{\sqrt{M_{D^\ast}M_B}} &= 
        \epsilon^{\mu\nu}_{\phantom{\mu\nu}\rho\sigma}\varepsilon^*_\nu v_B^\rho v_{D^\ast}^\sigma\,h_V(w),
    \label{vcCur}
\end{align}
where $\varepsilon$ is the polarization vector of the $D^\ast$ meson, $M_Y$ is the mass of the $Y=B$, $D^\ast$ meson and $p_Y$ its
respective momentum.
From the four-velocities $v_Y=p_Y/M_Y$, one can define the recoil parameter $w=v_B\cdot v_{D^\ast}$.

To express the differential decay rate, it is convenient to introduce helicity amplitudes according to the polarization of 
the off-shell $W$~boson~\cite{Korner:1989qb}:
\begin{align} % irrelevant signs disagree with A.S.K. note
    H_\pm(w) &=   (w+1)\left[h_{A_1}(w) \mp \sqrt{\frac{w-1}{w+1}} h_V(w)\right],
    \label{eq:helicity:T} \\
    H_0(w)   &= y (w+1) \Big\{(w-r) h_{A_1}(w) \nonumber \\
             & \phantom{=} - (w-1)\left[r h_{A_2}(w) + h_{A_3}(w) \right] \Big\},
    \label{eq:helicity:L} \\
    H_S(w)   &= y \sqrt{w^2-1} \Big[(w+1) h_{A_1}(w) \nonumber \\
             &\phantom{=} - (1-wr) h_{A_2}(w) - (w-r) h_{A_3}(w)\Big],
    \label{eq:helicity:S}
\end{align}
where $r=M_{D^\ast}/M_B$ and $y^2(1 - 2wr + r^2)=1$.
The differential decay rate for $B^-\to D^{0\ast}\ell^-\bar\nu$ is then
\begin{align}
\frac{d\Gamma}{dw} &= |V_{cb}|^2 |\eta_\text{EW}|^2 \frac{G_F^2 M_B^5}{16\pi^3}
    \left(1 - \frac{m_\ell^2}{q^2}\right)^2 r^3(w^2-1)^{1/2}
    \nonumber \\ &\phantom{=}%\hspace*{7em}
    \Bigg\{ \frac{1}{3y^2} \left(1 + \frac{m_\ell^2}{2q^2} \right) \left[ |H_+|^2 + |H_-|^2 + |H_0|^2 \right] \nonumber \\
    & \phantom{=} + \frac{m_\ell^2}{2M_B^2} |H_S|^2 \Bigg\},
    \label{DiffDecayRate}
\end{align}
where $\eta_\text{EW}$ is a short-distance electroweak correction~\cite{Sirlin:1981ie}, $G_F$ is the Fermi constant determined
from muon decay, and $m_\ell$ is the charged lepton mass.
Note that the scalar helicity amplitude's contribution is suppressed by $(m_\ell/M_B)^2$.
In practice, it is neglected for semielectronic and semimuonic decays.
The process $\bar{B}^0\to D^{+ \ast}\ell^-\bar{\nu}$ needs an extra factor $(1+\alpha\pi)$ on the right-hand side
of Eq.~\eqref{DiffDecayRate} in order to account for the Coulomb attraction among the charged decay
products~\cite{Ginsberg:1968pz,Ginsberg:1969jh,Atwood:1989em}.
Other electromagnetic effects, including structure-dependent corrections, are smaller, of order $\alpha/\pi$ instead of 
$\alpha\pi$~\cite{Ginsberg:1968pz,Ginsberg:1969jh,Atwood:1989em}.
For the determination of $|V_{cb}|$, the full angular information of the decay chain $B\to D^*\ell\nu\;(D^*\to D\pi)$ is used, as 
discussed in Sec.~\ref{sec:Vcb}.

As noted in the introduction, experimental measurements of the ratio of branching fractions in
Eq.~\eqref{RDstDef} are in tension with the SM.
To date, the $B\to D^\ast$ form factors have been estimated with HQET, QCD sum rules, and input from experiment.
With the results presented below, however, we can compute this ratio directly from (lattice) QCD:
\begin{equation}
    \mathcal{B}(B\to D^\ast\ell\nu) = \tau_B \int_1^{w_{\text{Max},\ell}} dw\,\frac{d\Gamma}{dw},
    \label{eq:BR}
\end{equation}
where $w_{\text{Max},\ell} = (1+r^2 - m_\ell^2/M_B^2)/2r$.
In $R(D^\ast)$, the $B$-meson lifetime $\tau_B$ drops out, so we form it from the ratio of partial widths.

Many papers in the literature introduce a decay amplitude $\mathcal{F}(w)$, defined by
\begin{equation}
    |\mathcal{F}(w)|^2 = \frac{1-2wr+r^2}{w+1} \frac{|H_+|^2 + |H_-|^2 + |H_0|^2}{(5w+1)(1-r)^2 -8rw(w-1)},
    \label{probAmp}
\end{equation}
and refer to $\mathcal{F}$ as a ``form factor.''
For example, experimental results are often reported as $|V_{cb}|^2|\eta_\text{EW}|^2|\mathcal{F}(w)|^2$.
At zero recoil, $\mathcal{F}(1)=h_{A_1}(1)$.
Thus, previous work in lattice QCD on this decay has focused on this single number, rather than the four independent functions, 
$h_{A_1} (w)$, $h_{V} (w)$, $h_{A_2} (w)$, and $h_{A_3} (w)$, computed in this work. 

\subsection{Extracting the form factors from lattice matrix elements}
\label{eFfMe}

Our heavy quarks ($b,c$) are simulated using the Fermilab action~\cite{ElKhadra:1996mp}, as discussed in Sec.~\ref{Latsetup}.
In this framework, the lattice currents for the quark transition $y\to x$ are
\begin{align}
  V_{xy}^\mu &= \bar{\Psi}_x\gamma^\mu\Psi_y,\label{vClat} \\
  A_{xy}^\mu &= \bar{\Psi}_x\gamma^\mu\gamma_5\Psi_y,\label{aClat}
\end{align}
where $x,y$ indicates the flavor $c,b$ and $\Psi$ is the Fermilab-improved field,
\begin{equation}
  \Psi = (1+d_1\bm{\gamma}\cdot\bm{D}_\textrm{lat})\psi.
\end{equation}
In this expression, the original heavy-quark field $\psi$ is rotated in order to reduce the discretization errors.
In particular, the coefficient $d_1$ must be calculated for each value of the quark mass in order to
remove the $O(a)$ terms.

The lattice current $J_{xy}^\mu$ is related to its equivalent in the continuum $\mathcal{J}_{xy}^\mu$ through the renormalization
factors,
\begin{equation}
  \mathcal{J}_{xy}^\mu \dot{=} Z_{J_{xy}^\mu} J_{xy}^\mu,\label{cRen}
\end{equation}
where the $\dot{=}$ symbol means that both sides of the equation have the same matrix elements.
In practice, the renormalization factors are only calculated approximately up to some order in $a$ and $\alpha_s$.
In this work, we use a technique called \emph{mostly nonperturbative renormalization}~\cite{Harada:2001fj,ElKhadra:2001rv} that
eliminates most of the nonperturbative dependence of the renormalization factors by defining the factors,
\begin{equation}
    \rho^2_{J^\mu} = \frac{Z_{J_{cb}^\mu}Z_{J_{bc}^\mu}}{Z_{V^4_{cc}}Z_{V^4_{bb}}}.
    \label{cRat}
\end{equation}
When taking appropriate ratios of three-point correlators, the dominant, nonperburbative contribution to the renormalization of the
currents, collected in the flavor-diagonal renormalization factors $Z_{V^4_{xx}}$, cancels.
The remaining matching factors $\rho_{J^\mu}$ are amenable to a perturbative calculation~\cite{Harada:2001fj}.
We compute these matching factors to one-loop in perturbation theory, with the full $m_ca$ dependence at zero 
recoil, but $m_ca=0$ at nonzero recoil.
Of course, these simplifications introduce a variety of errors that must be kept under control.
The truncation in the perturbative expansion is expected to be small, because $\alpha_s$ ranges in our case from $0.20$ to $0.35$.
In addition, the coefficients of the expansion are small due to several cancelations~\cite{Harada:2001fj}.
The errors coming from the other two approximations are estimated in~\ref{ApMatch} and taken into account accordingly.

Not only does the use of ratios reduce the error in the calculation of the matching factors, but it also reduces the statistical
fluctuations from the correlators.
We set up the calculation in the rest frame of the $B$ meson while the $D^\ast$ meson carries a momentum $\bm{p}$, which 
determines the recoil~$w$.
The first ratio is the nonzero-recoil version of the double ratio~\cite{Hashimoto:2001nb},
\begin{equation}
    R^2_{A_1} = \frac{\langle D^\ast(\bm{p}_\bot)|A_j|B(\bm{0})\rangle \langle B(\bm{0})| A_j |D^\ast(\bm{p}_\bot)\rangle}
                 {\langle D^\ast(\bm{0})|V^4|D^\ast(\bm{0})\rangle \langle B(\bm{0})| V^4 |B(\bm{0})\rangle}
    \label{defRA1},
\end{equation}
where the $\bot$ symbol in the momentum $\bm{p}_\bot$ indicates that the polarization of the $D^\ast$ is aligned with the current
and perpendicular to the momentum (i.e., transverse polarization).
A parallel symbol $\parallel$ is used for longitudinal polarization.
This double ratio yields $h_{A_1}$, which is the only form factor that survives at zero recoil.

The following single ratios
\begin{align}
    X_V &= \frac{\langle D^\ast(\bm{p}_\bot     )| V_j |B(\bm{0})\rangle}{\langle D^\ast(\bm{p}_\bot)| A_j |B(\bm{0})\rangle}
    \label{defXv}, \\
    X_0 &= \frac{\langle D^\ast(\bm{p}_\parallel)| A^4 |B(\bm{0})\rangle}{\langle D^\ast(\bm{p}_\bot)| A_j |B(\bm{0})\rangle}
    \label{defR0}, \\
    X_1 &= \frac{\langle D^\ast(\bm{p}_\parallel)| A_j |B(\bm{0})\rangle}{\langle D^\ast(\bm{p}_\bot)| A_j |B(\bm{0})\rangle}
    \label{defR1},
\end{align}
yield the remaining form factors.
Last, the ratio~\cite{Bailey:2012rr,Lattice:2015rga}
\begin{equation}
%    \bm{x}_f = \frac{\langle D^\ast(\bm{p}_{(\alpha)})| \bm{V} |D^\ast(\bm{0})\rangle}
    x_f = \frac{\langle D^\ast(\bm{p}_{(\alpha)})| V_j |D^\ast(\bm{0})\rangle}
        {\langle D^\ast(\bm{p}_{(\alpha)})| V^4 |D^\ast(\bm{0})\rangle}
    \label{wParm}
\end{equation}
yields the recoil parameter $w$ and involves only the flavor-diagonal transition $D^\ast\to D^\ast$.
Here $(\alpha)=\bot,\parallel$ is the polarization, and must be the same in numerator and denominator to achieve the right
cancelation of form factors.

The ratio in Eq.~\eqref{wParm} yields the three-velocity,
\begin{equation}
%    \bm{x}_f = \frac{\bm{v}_{D^\ast}}{w+1},\label{xfTow}
    x_f = \frac{v_{D^\ast}}{w+1},\label{xfTow}
\end{equation}
from which it is straightforward to calculate $w$.
In the $B$-meson rest frame ($\bm{v}_B=\bm{0}$),
\begin{equation}
    w = \frac{1 + x_f^2}{1 - x_f^2}.
    \label{wDef0}
\end{equation}
The ratio $X_1$ defined in Eq.~\eqref{defR1} can be used to extract $h_{A_3}(w)$ as
\begin{equation}
    X_1(w) = w - \frac{(w^2-1)h_{A_3}(w)}{(w+1)h_{A_1}(w)}.
\end{equation}
The matching factors for these two ratios, $x_f$ and $X_1$, are $\rho = 1 + O(\alpha_s^2)$, and thus no renormalization is required
at LO.
In contrast, the remaining ratios require several nontrivial matching factors,
\begin{align}
    X_0    (w) &= \frac{\rho_{A_j}}  {\rho_{A^4}} \sqrt{w^2-1} \left(1 - \frac{h_{A_2}(w) + wh_{A_3}(w)}{(w+1)h_{A_1}(w)}\right), \\
    X_V    (w) &= \frac{\rho_{A_j}}{\rho_{V_j}} \frac{\sqrt{w-1}}{\sqrt{w+1}}\frac{h_V(w)}{h_{A_1}(w)}, \\
    R_{A_1}(w) &= \frac{w+1}{2}\frac{h_{A_1}(w)}{\rho_{A_j}}.
\end{align}
From these equations, it is quite easy to extract all form factors as a function of the ratios defined in
Eqs.~(\ref{defRA1})-(\ref{wParm}),
\begin{align}
    h_{A_1}(w) &= \rho_{A_j}\frac{2R_{A_1}}{w+1}, \label{eq:matched-hA1} \\
    h_{A_2}(w) &= \rho_{A_j}\frac{2R_{A_1}}{w^2-1}(wX_1 - \sqrt{w^2-1}\frac{\rho_{A^4}}{\rho_{A_j}}X_0 - 1),
    \label{eq:matched-hA2} \\
    h_{A_3}(w) &= \rho_{A_j}\frac{2R_{A_1}}{w^2-1}(w - X_1), \label{eq:matched-hA3} \\
    h_V    (w) &= \rho_{A_j}\frac{2R_{A_1}}{\sqrt{w^2-1}}\frac{\rho_{V_j}}{\rho_{A_j}}X_V. \label{eq:matched-hV}
\end{align}
These expressions determine the four form factors up to discretization and matching errors.

\section{Analysis}
\label{Analysis}

\subsection{Lattice setup}
\label{Latsetup}

\begin{figure}%[h]
  \centering
  \includegraphics[width=\linewidth]{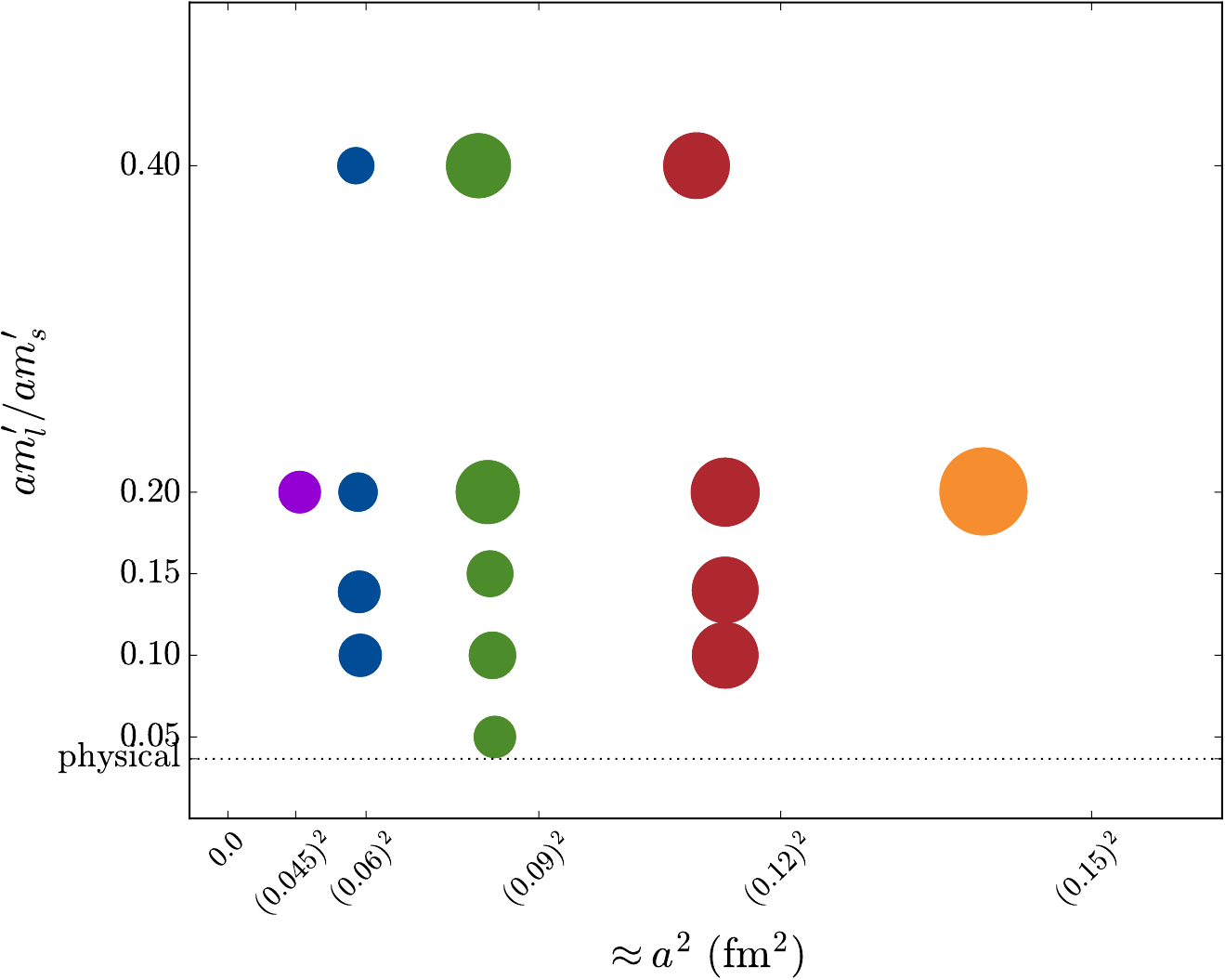}
  \caption{Ensembles used in this work.
      The vertical (horizontal) axis shows the ratio between the light- and strange-quark masses (the lattice spacing).
      The area of each circle is proportional to the total sample size available on each ensemble.
      The horizontal line marks the value of the $m_l/m_s$ ratio that results in pions with physical mass.}
   \label{PlotEnsembles}
\end{figure}

In this analysis, we use 15 ensembles of gauge-field configurations, generated by the MILC collaboration~\cite{Bazavov:2009bb,Aubin:2004wf,Bernard:2001av}.
These ensembles include three flavors of asqtad-improved staggered sea quarks at five different lattice spacings, ranging from
$0.15$~fm in the coarsest case to $0.045$~fm in the finest case.
The mass of the strange sea quark is tuned to be close to its physical value, while the two light sea-quark masses are set equal,
and cover a range of values that correspond to pion masses from $M_\pi\approx 560$~MeV to $M_\pi\approx 180$~MeV.
The simulation parameters of all ensembles employed in this analysis are given in Table~\ref{TableEnsembles}, while
Fig.~\ref{PlotEnsembles} provides a visual summary of the range of lattice spacings, sea-quark light-to-strange-mass ratios, and
number of statistical samples.

\begin{table*}
  \caption{List of ensembles used in this work.
      The columns, from left to right, list the approximate lattice spacing, the scale-setting parameter $r_1/a$ in lattice units,
      the ratio $am_l/am_s$ between the light- and the strange-quark masses, the spatial length of the lattice in fm, the mass of
      the lightest pseudoscalar meson $M_\pi^P$ in MeV, the dimensionless factor $M_\pi^P L$, the dimensions of the lattice in
      lattice units, the total sample size expressed as the number of sources $\times$ the number of configurations, and the
      tadpole-improvement factor $u_0$ obtained from the average plaquette.}
  \label{TableEnsembles}
  
  \begin{tabular*}{\textwidth}{@{\extracolsep{\fill}}
                               cS[table-format = 1.4(2)]
                                %S[table-format = 1.5]@{}c@{}
                                %S[table-format = 1.4]ccccrcl
                                cccccrcl
                                S[table-format=1.6, group-minimum-digits = 6]}
    \hline
    \hline
            $a$ (fm)\Tstrut   &     {$r_1/a$}     &   $am_l / am_s$   & $L$ (fm) & $M^P_\pi$ (MeV) & $M_\pi^P L$ &   Lattice size   &  Sources   &          &    Configs    & {$u_0$}  \\
    \hline
           $\approx 0.15$     &    2,2215(57)     &   0.0097/0.0484   &    2.4   &      340        &     3.9     & $16^3\times 48$  &         24 & $\times$ & 631           & 0.8604   \\
    \hline
\MulRow{4}{*}{$\approx 0.12$} &    2,8211(28)     &     0.02/0.05     &    2.4   &      560        &     6.2     & $20^3\times 64$  &          4 & $\times$ & 2052          & 0.8688   \\
                              &    2,7386(33)     &     0.01/0.05     &    2.4   &      390        &     4.5     & $20^3\times 64$  &          4 & $\times$ & 2259          & 0.8677   \\
                              &    2,7386(33)     &    0.007/0.05     &    2.4   &      320        &     3.7     & $20^3\times 64$  &          4 & $\times$ & 2110          & 0.8678   \\
                              &    2,7386(33)     &    0.005/0.05     &    2.9   &      270        &     3.8     & $24^3\times 64$  &          4 & $\times$ & 2099          & 0.8678   \\
    \hline
\MulRow{5}{*}{$\approx 0.09$} &    3,8577(32)     &   0.0124/0.031    &    2.4   &      500        &     5.8     & $28^3\times 96$  &          4 & $\times$ & 1996          & 0.8788   \\
                              &    3,7887(34)     &   0.0062/0.031    &    2.4   &      350        &     4.1     & $28^3\times 96$  &          4 & $\times$ & 1931          & 0.8782   \\
                              &    3,7716(34)     &  0.00465/0.031    &    2.7   &      310        &     4.1     & $32^3\times 96$  &          4 & $\times$ & 984           & 0.8781   \\
                              &    3,7546(34)     &   0.0031/0.031    &    3.4   &      250        &     4.2     & $40^3\times 96$  &          4 & $\times$ & 1015          & 0.8779   \\
                              &    3,7376(34)     &  0.00155/0.031    &    5.5   &      180        &     4.8     & $64^3\times 96$  &          4 & $\times$ & 791           & 0.877805 \\
    \hline
\MulRow{4}{*}{$\approx 0.06$} &    5,399(17)      &   0.0072/0.018    &    2.9   &      450        &     6.3     & $48^3\times 144$ &          4 & $\times$ & 593           & 0.8881   \\
                              &    5,353(17)      &   0.0036/0.018    &    2.9   &      320        &     4.5     & $48^3\times 144$ &          4 & $\times$ & 673           & 0.88788  \\
                              &    5,330(16)      &   0.0025/0.018    &    3.4   &      260        &     4.4     & $56^3\times 144$ &          4 & $\times$ & 801           & 0.88776  \\
                              &    5,307(16)      &   0.0018/0.018    &    3.8   &      220        &     4.3     & $64^3\times 144$ &          4 & $\times$ & 827           & 0.88764  \\
    \hline
       $\approx 0.045$ \Bstrut&    7,208(54)      &   0.0028/0.014    &    2.9   &      320        &     4.6     & $64^3\times 192$ &          4 & $\times$ & 801           & 0.89511  \\
    \hline
    \hline
  \end{tabular*}
\end{table*}

In the light sector, we use the same value for the masses of the valence and sea quarks.  The heavy quarks employ the clover action
with the Fermilab interpretation, and since the regularization used for the light quarks has a different Dirac structure, we promote
the staggered propagators to ``naive" ones, so we can apply the standard Dirac spin algebra and combine them with the Wilson-like
heavy quark propagators to construct heavy-light mesons~\cite{Wingate:2002fh}.  The heavy quark masses are tuned so that the kinetic
masses of the $D_s$ and the $B_s$ mesons are equal to their physical values (see Appendix~C of Ref.~\cite{Bailey:2014tva}).  In
Table~\ref{TableParms}, we gather the parameters we used to calculate the heavy quark propagators for each ensemble.  The simulation
values chosen for the heavy quark masses to generate the meson correlators are close to, but not exactly the same as, our best-tuned
values, which were determined \emph{a posteriori}.  Hence we apply a correction to the form factors to account for this slight
mistuning which is described in \ref{ApHQCorr}.

\begin{table*}
  \caption{Parameters used on each ensemble to generate the propagators for the valence, heavy quarks $c$ and $b$. The approximate 
      lattice spacing and the masses of the sea quarks (light and strange) in the first two columns identify the ensemble. The 
      remaining columns show the clover term coefficient $c_{\rm SW}$, the bare hopping parameter $\kappa$, the rotation parameter 
      $d_1$ of the Fermilab action, and the values of the available source/sink Euclidean-time separations $T$ in lattice units for 
      the computed three-point correlators. The primes on the $\kappa$ indicates that this is the value used in the simulation, as 
      opposed to the physical (tuned) value, see \ref{ApHQCorr}.}
  \label{TableParms}
  \begin{tabular*}{\textwidth}{@{\extracolsep{\fill}}ccS[table-format = 1.4]ccccc}
    \hline
    \hline
            $a$ (fm)          &    $am_l$/$am_s$   & $c_{\rm SW}$ & $\kappa'_b$ & $d_{1b}$ & $\kappa'_c$ & $d_{1c}$  & Sink times \\
    \hline 
         $\approx 0.15$       &    0.0097/0.0484   &    1.567     &    0.0781   &  0.08354 &    0.1218   &  0.08825  &   10,11    \\
    \hline
\MulRow{4}{*}{$\approx 0.12$} &      0.02/0.05     &    1.525     &    0.0918   &  0.09439 &    0.1259   &  0.07539  &   12,13    \\
                              &      0.01/0.05     &    1.531     &    0.0901   &  0.09334 &    0.1254   &  0.07724  &   12,13    \\
                              &     0.007/0.05     &    1.530     &    0.0901   &  0.09332 &    0.1254   &  0.07731  &   12,13    \\
                              &     0.005/0.05     &    1.530     &    0.0901   &  0.09332 &    0.1254   &  0.07733  &   12,13    \\
    \hline
\MulRow{5}{*}{$\approx 0.09$} &    0.0124/0.031    &    1.473     &    0.0982   &  0.09681 &    0.1277   &  0.06420  &   17,18    \\
                              &    0.0062/0.031    &    1.476     &    0.0979   &  0.09677 &    0.1276   &  0.06482  &   17,18    \\
                              &   0.00465/0.031    &    1.477     &    0.0977   &  0.09671 &    0.1275   &  0.06523  &   17,18    \\
                              &    0.0031/0.031    &    1.478     &    0.0976   &  0.09669 &    0.1275   &  0.06537  &   17,18    \\
                              &   0.00155/0.031    &    1.4784    &    0.0976   &  0.09669 &    0.1275   &  0.06543  &   17,18    \\
    \hline
\MulRow{4}{*}{$\approx 0.06$} &    0.0072/0.018    &    1.4276    &    0.1048   &  0.09636 &    0.1295   &  0.05078  &   24,25    \\
                              &    0.0036/0.018    &    1.4287    &    0.1052   &  0.09631 &    0.1296   &  0.05055  &   24,25    \\
                              &    0.0025/0.018    &    1.4293    &    0.1052   &  0.09633 &    0.1296   &  0.05070  &   24,25    \\
                              &    0.0018/0.018    &    1.4298    &    0.1052   &  0.09635 &    0.1296   &  0.05076  &   24,25    \\
    \hline
         $\approx 0.045$      &    0.0028/0.014    &    1.3943    &    0.1143   &  0.08864 &    0.1310   &  0.03842  &   32,33    \\
    \hline
    \hline
  \end{tabular*}
\end{table*}

\subsection{Correlation functions}
\label{sec:corfunc}

The two- and three-point correlation functions are calculated using four sources, equally-spaced in time, except for the case of the
coarsest ensemble that employs 24 sources.  The sources are randomly shifted in space and time from one configuration to another in
order to reduce correlations between successive gauge-field configurations within the same ensemble.  A standard blocking analysis
of the correlator data, ranging from block size 1 to block size 8, reveals that the autocorrelations in our ensembles are negligible
and that the errors in the correlator points stay approximately constant as we increase the block size, in line with our previous analyses
that employed the same gauge configurations, fermion formulations, and source
set-up~\cite{Bailey:2014tva,Lattice:2015rga,Lattice:2015tia,Bailey:2015nbd,Bailey:2015dka,Bazavov:2016nty,Bazavov:2017weg,Bazavov:2019aom}.
Therefore, we do not block the data in this work, and the correlators are processed through a single-elimination jackknife.

Two previous analyses with the asqtad ensembles~\cite{Bernard:2008dn,MILC:2010iid} found that blocking the configurations by 4 or 8 was
necessary in order to suppress autocorrelations. However, these analyses refer either to global observables (the topological
susceptibility), or did not use the randomization procedure for the sources, which greatly reduces the autocorrelations in our data.

Given that one of our ensembles has a very fine lattice spacing $a\approx 0.045$ fm, one might be worried about the topology freezing
and its effect in the final results of the form factors. We did not perform a topology freezing analysis in the asqtad ensembles, but we expect
the behavior to be similar to that of the HISQ ensembles. Based on Refs.~\cite{FermilabLattice:2019ycs,Bazavov:2017lyh}, we expect topology
freezing to introduce a negligible bias in the chiral-continuum limit of the form factors.

The correlation functions described in the following two subsections contain the desired ground-state matrix elements, energies and
form factors, but they also include contributions from excited states, which we must remove.  For this purpose, we use two different
kinds of interpolating operators per source: a local operator $d$ and a smeared operator based on the Richardson $1S$ wave function
\cite{Richardson:1978bt,Bazavov:2011aa}, and therefore we fix the configuration to Coulomb gauge.
For each meson, the radius of the smearing operator is the same in physical units for all ensembles.
We refer the reader to Ref.~\cite{Bazavov:2011aa} for further details.
The smeared operator increases the overlap with the
ground state, allowing for a more precise determination of the lowest energy level and its overlap factors.  The inclusion of a
local operator gives us a useful handle on the excited states.  Further, to quantify excited-state contributions and obtain robust
estimates of the associated uncertainties, we use Bayesian constraints with Gaussian priors and fit functions that include varying
numbers of excited states~\cite{Lepage:2001ym}.

To implement the Bayesian constraints, we follow the procedure of Appendix~B  of Ref.~\cite{Bazavov:2016nty}.
We minimize the augmented $\chi^2_{\rm aug}$, as defined in Eq. (B3) of Ref.~\cite{Bazavov:2016nty},
but for the goodness of fit use the data-only $\chi^2$ (evaluated at the minimum of $\chi^2_{\rm aug}$) and subtract the number of parameters from the number of data.
Below we refer to this $\chi^2$ and counting of degrees of freedom as the \emph{deaugmented} $\chi^2/{\rm dof}$.
In our experience, the $p$~value calculated with a $\chi^2$ and a number of dof as defined in Eq. (B5) of Ref.~\cite{Bazavov:2016nty}
is a good indicator of goodness of fit, but it has not been proven rigorously to follow a uniform distribution.
When calculating $p$~values, we further process the $\chi^2$/dof ratio to take into account finite sample size~\cite{Toussaint:2008ke}.

\subsection{Two-point functions}
\label{SubSec:2pt}
The $D^\ast$ and $B$-meson two-point functions are needed to extract the overlap factors and the energy states, for these are
required inputs for the ratio fits.
The two-point functions are constructed using interpolating operators $\mathcal{O}_{Y_a}(\bm{p}, t)$, where $Y=\{B,D^\ast\}$ is the
meson of interest, $a=\{d,1S\}$ represents the smearing (point and Richardson), $t$ is the time and $\bm{p}$ is the spatial
momentum.
These operators are constructed with the same quantum numbers as a pseudoscalar for $Y=B$ and a vector for $Y=D^\ast$.
In terms of the interpolating operators, the two-point correlators are then
\begin{equation}
C^{\textrm{2pt}}_{Y_a\to Y_b}(\bm{p}, t) = \left\langle\mathcal{O}_{Y_b}(\bm{p},t) \,\mathcal{O}_{Y_a}^\dagger(\bm{p}, 0)\right\rangle.
\end{equation}
Inserting a complete set of states between the interpolating operators, we obtain the spectral decomposition:
\begin{align}
C^{\textrm{2pt}}_{Y_a\to Y_b}(\bm{p}, t) &= \sum_n s_n(t)\frac{\sqrt{Z_{Y_a,n}(\bm{p})\, Z_{Y_b,n}(\bm{p})}}{2E_n(\bm{p})}\nonumber\\
 &\phantom{=}\left(e^{-E_n(\bm{p})t} + e^{-E_n(\bm{p})(L_t-t)}\right),\label{2ptDecomp}
\end{align}
with $\sqrt{Z_{Y_{a,b}}(\bm{p})}$ the overlap factors, $L_t$ the temporal extent of our lattice and $s_n(t)$ the extra sign that arises due to the presence of particles with the opposite parity in the staggered
regularization for the fermions,
\begin{equation} % always use equation numbers
    s_n(t) = \left\{\begin{matrix}
       1      & \textrm{Correct  parity} \\
    -(-1)^{t} & \textrm{Opposite parity}
    \end{matrix}\right. .
\end{equation}
Most correlators are available in four different configurations according to the smearing of the source and the sink: $d$-$d$, $d$-$1S$,
$1S$-$d$ and $1S$-$1S$.
In the case of the $D^\ast$ meson, eight different momenta are available, namely, $(0,0,0)$, $(1,0,0)$, $(1,1,0)$, $(1,1,1)$, $(2,0,0)$,\linebreak
$(2,1,0)$, $(2,2,0)$, $(2,2,1)$, $(3,0,0)$, and $(4,0,0)$ in $2\pi/L$ units.
Of these, only $(0,0,0)$, $(1,0,0)$ and $(2,0,0)$ are used to calculate the form factors; the rest allow us to calculate the
dispersion relation of the~$D^\ast$.
For $(1,0,0)$ and $(2,0,0)$, two different orientations of the momenta are considered, namely, parallel and perpendicular to the $D^\ast$ polarization.

The outline of the analysis of the two-point functions, explained in detail in the following subsections, is as follows. 
First the zero-momentum correlators are fit using phenomenological guidance for the prior central values.
For the ground state, we set a prior similar to the physical mass of the mesons, and the excited states differ by $\Delta E=0.5$~GeV.
The prior widths are large enough to accommodate significant departures from these assumptions.
These choices for the central value and widths of the priors are such that they have no influence on the fit result for the ground states.
In fact, we consider several variations of the energy priors to verify that their only function is to guarantee the stability of the 
fits without influencing the ground-state fit parameters.
The results of the zero-momentum fits are used to construct priors for the dispersion-relation fits.
In particular, the ground-state energies are expected to follow the continuum dispersion relation, and the overlap factors for the local operators 
should be approximately constant, barring, in both cases, discretization effects.
Using data for a variety of momenta, we fit the ground-state energies to a dispersion-relation expression that includes discretization terms,
see Eq.~(\ref{DispRel}). The resulting fit is used to calculate a prior for the energy of the ground state of the two nonzero-momentum correlators.

\subsubsection{Two-point function fits}
\label{twopointfits}

For the two-point function fits, we employ the form
\begin{align}
    C^{\textrm{2pt}}_{Y_a\to Y_b}(t) &=
        \sum_{i=0,1} (-1)^{i(t+1)} \mathcal{Z}_{i,a}\mathcal{Z}_{i,b}\left(e^{-E_i t} + e^{-E_i (L_t-t)}\right) \nonumber\\
    & \hspace{-1cm} + \sum_{i=2}^{2N-1} (-1)^{i(t+1)} \mathcal{Z}^2_{i,ab}\left(e^{-E_i t} + e^{-E_i (L_t-t)}\right),
\end{align}
where the state oscillates in time for odd~$i$, but not for even~$i$.
This kind of fit is denoted as $N+N$, meaning we include $N$ nonoscillating and $N$ oscillating states.
Both oscillating and nonoscillating excited states are fitted as the logarithm of the energy difference $\Delta E_i = E_i - E_{i-2}$
in order to avoid the collapse of two energy levels.
In the higher states, we never interrelate energies of oscillating and nonoscillating states, i.e., the
fitted $\Delta E_i$ always refer to the difference between two states of the same type. %(non)oscillating states.
The overlap factors in Eq.~\eqref{2ptDecomp} are included in the fit function via
\begin{equation}
    \mathcal{Z}_j = \sqrt{Z_j / 2E_j}.
    \label{zFacPoint}
\end{equation}
For the $\mathcal{Z}$ factors of the ground states, we also use a logarithm, forbidding the possibility
$\mathcal{Z}\le0$.

We perform joint fits of all available correlators for a given combination of meson and momentum.
That gives us three correlators corresponding to the $d$-$d$, $1S$-$1S$, and the crossed average between the
$d$-$1S$ and the $1S$-$d$ operators. % use hyphens, not minus signs
In the cases where we distinguish between different orientations of the polarization of the $D^\ast$ meson with respect to its momentum,
the total number of correlators increases to six.
The fitter uses the covariance matrix of the whole set of data, where the fit parameters are constrained with Gaussian priors.
The prior central values for the energy levels in the fit functions for the zero-momentum correlators are guided by the experimental
values for the meson mass in question and an empirical analysis of the data. The prior width of the physical (oscillating) ground state is
chosen to be $140$ MeV ($520$ MeV). In the fit functions for the nonzero-momentum correlators used for the dispersion relation, the ground-state
energy prior central values are set equal to $\sqrt{M^2 + \bm{p}^2}$, where $M$ is the posterior ground state energy from the fit to the corresponding
zero-momentum correlator.
The width of the prior is enlarged to encompass the expected discretization errors $O(\alpha_s a^2p^2)$.
The prior central value for the energy difference between two neighboring oscillating or nonoscillating states is taken to be $0.5$~GeV.
Their widths vary with the ensemble, but they are always larger than $0.2$~GeV.
The fit functions for the nonzero-momentum correlators employed in the three-point function analysis, namely momenta $2\pi(1,0,0)/L$ and $2\pi(2,0,0)/L$,
use the dispersion-relation results as priors for the ground-state energy. 

The energy levels are constrained to be the same across smearings, but the overlap factors are different, and they are represented with different
parameters.
For the ground states of the crossed average, we do not fit the $\mathcal{Z}_j$ amplitudes, but we impose the exact constraint
\begin{equation}
    \mathcal{Z}^2_{d,1S} = \mathcal{Z}_{1S}\mathcal{Z}_d.
    \label{ZCons}
\end{equation}
The $\mathcal{Z}_j$ amplitudes of the excited states of the crossed average are treated as separate fit parameters as they may describe a mix of excited states.
Our $\mathcal{Z}_j$ amplitudes are also allowed to depend on the orientation of the momentum, when applicable, and Eq.~\eqref{ZCons} applies independently to each orientation.
The priors for the $\mathcal{Z}_j$ factors of the ground states follow a log-normal distribution.
Their central values are estimated following an empirical examination of the data, and their widths are large enough to accommodate significant departures
from those original choices, roughly within one order of magnitude. In particular, the width of the physical ground state amplitude prior is set to $0.5$ for all ensembles,
and the width of the posterior is usually 20 times smaller. The prior is enlarged for the nonzero momentum correlators by a factor $\approx (1+2\alpha_s p^2)$. In the case of the
oscillating ground state, the width of the amplitude is set to $1.2$ for the zero momentum correlators and $2.0$ for the nonzero momentum ones, with a typical posterior width of $0.5$.
In contrast, we use a Gaussian distribution for the excited-state priors. The width is fixed to be $3.5$ for all ensembles, whereas the width of the resulting posteriors
is typically an order of magnitude smaller.
We test thoroughly that the fit results for the ground states are largely unaffected by the choice of priors and prior widths, as long as the fit remains stable.

\begin{figure*}
  \centering
    \includegraphics[width=0.45\linewidth]{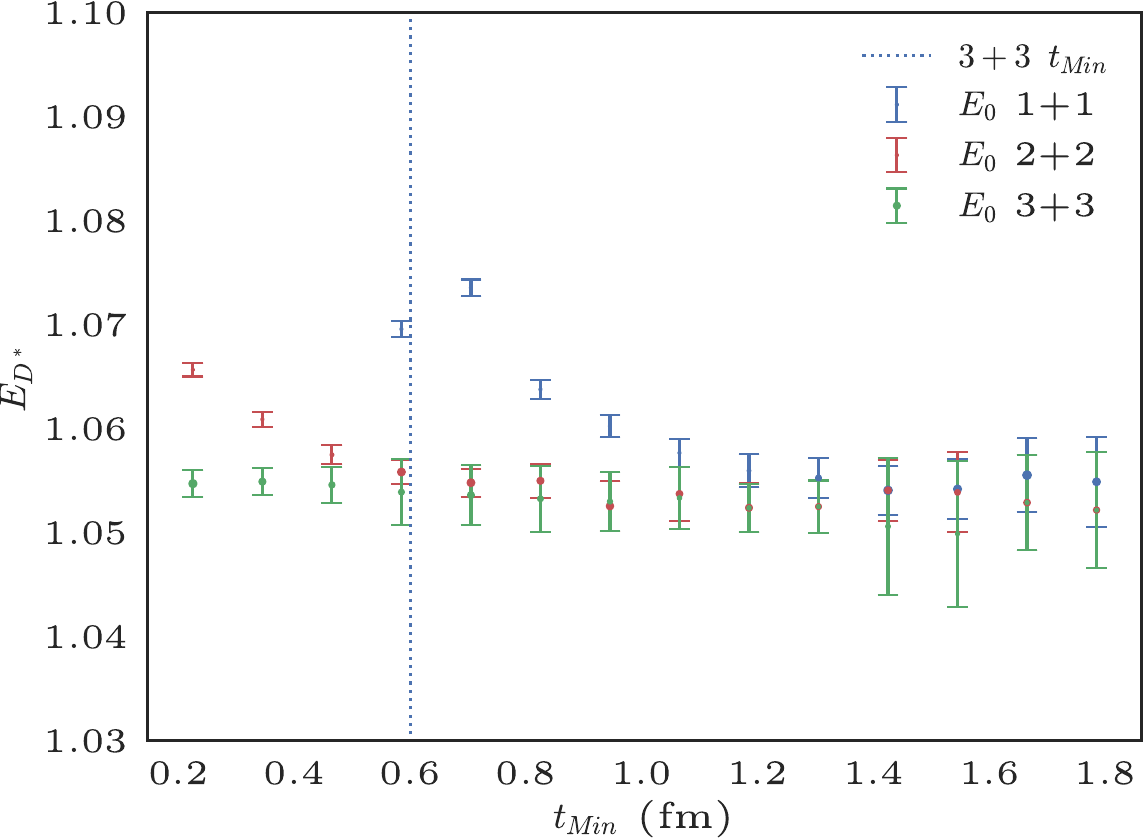} \hfill
    \includegraphics[width=0.45\linewidth]{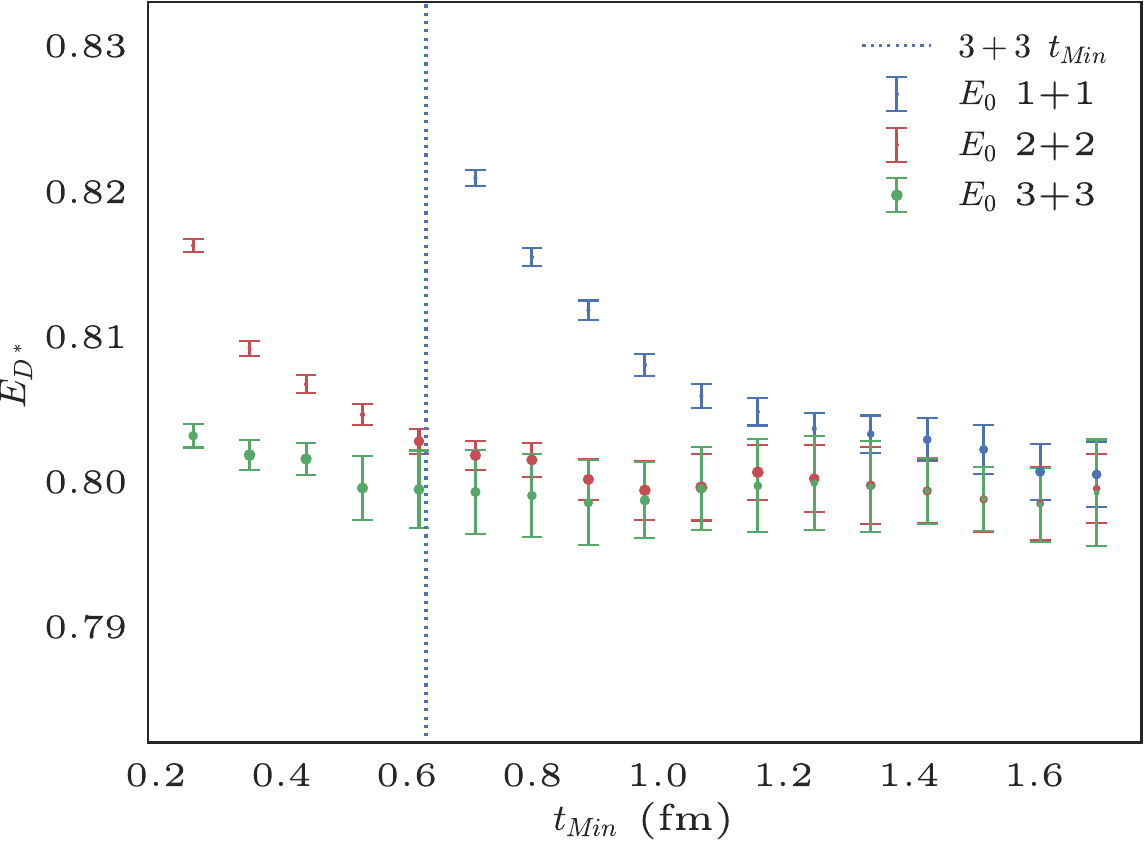}  \\[0.5em]
    \includegraphics[width=0.45\linewidth]{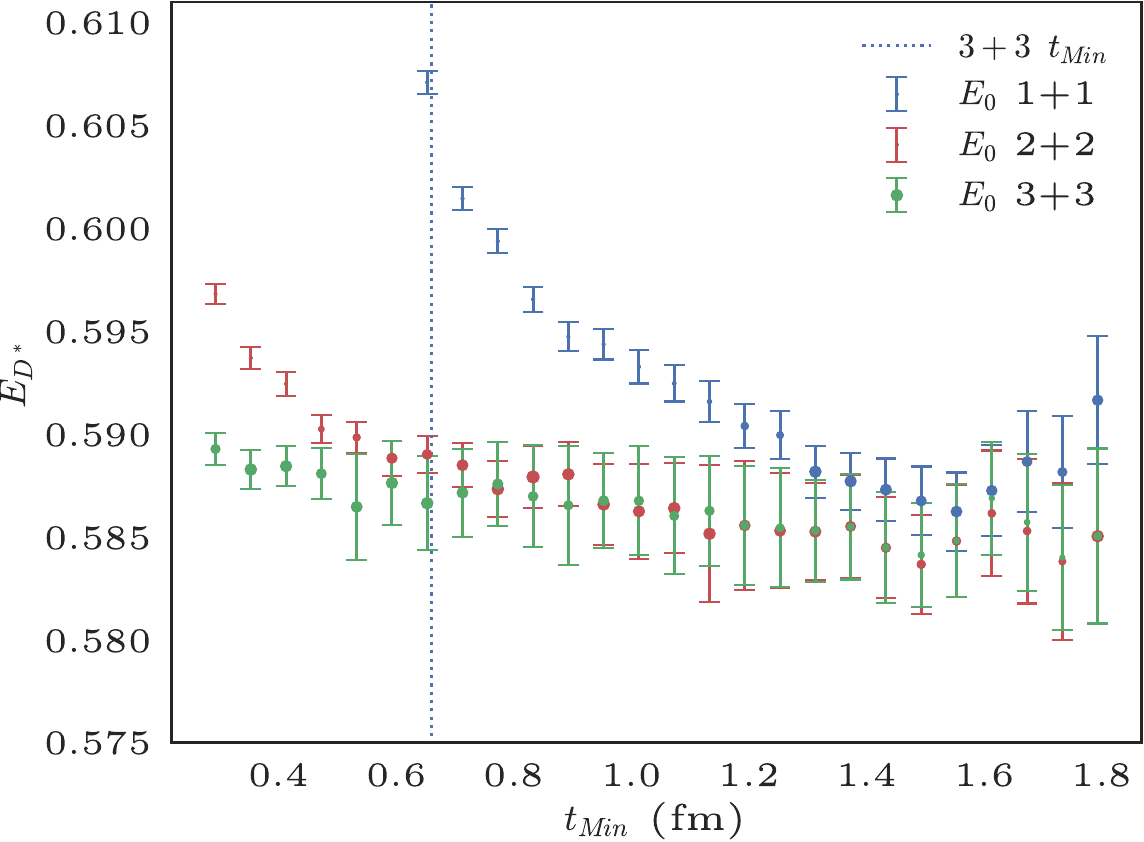} \hfill
    \includegraphics[width=0.45\linewidth]{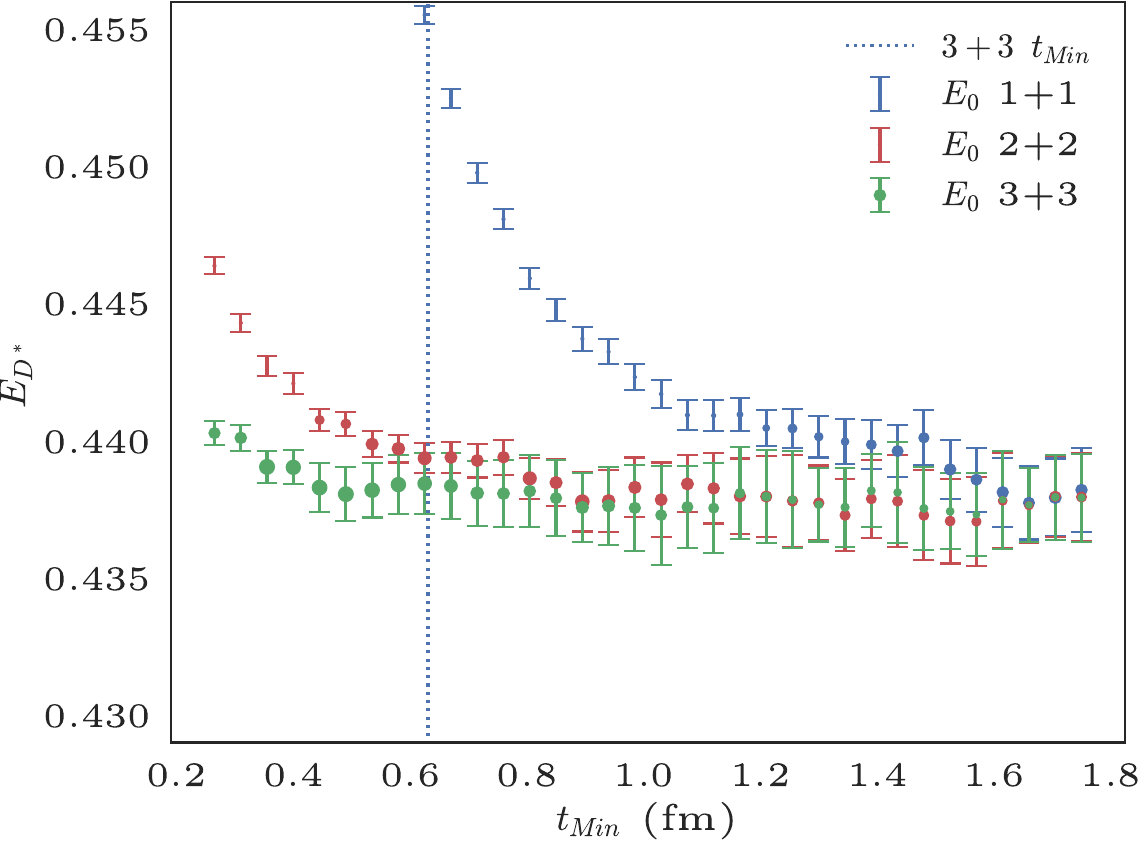}
    \caption{Stability plots for the $D^\ast$-meson energy in lattice units with zero momentum (i.e., the mass) for four different ensembles with $m_l/m_s = 0.2$ and approximate lattice
             spacings 0.12 fm (upper left), 0.09 fm (upper right), 0.06 fm (lower left) and 0.045 fm (lower right). The size of each point is proportional to the $p$ value, and visible point markers
             represent reasonable fits. The $y$ axis represents the energy of the ground state in lattice units on each ensemble, with a range that roughly correspond to the same interval in physical
             units. The $t_{\textrm{Min}}$ chosen for the analysis, which corresponds to $3+3$ correlator fits, is marked by a vertical dotted line. The point chosen is
             the closest one to the line. It is evident that on some ensembles smaller choices of $t_{\textrm{Min}}$ are still in the stability region, which means that our common $t_{\textrm{Min}}$
             is conservatively chosen.}
   \label{StabTest}
\end{figure*}

The fit ranges are chosen following a systematic procedure: $t_{\textrm{Max}}$ is chosen such that the correlator points for
$t<t_{\textrm{Max}}$ have fractional errors smaller than $\approx20$--30\%.
In this way, the covariance matrix is not contaminated by excessive noise, but the value of $t_{\textrm{Max}}$ becomes ensemble dependent. 
However, the correlator fits are generally insensitive to variations of $t_{\textrm{Max}}$ within this constraint.
In contrast, $t_{\textrm{Min}}$ is chosen to have the same value in physical units for all ensembles and momenta.
We do this because we expect the degree to which excited states influence the fit depends on their physical separation from the ground state.
We apply the following four criteria to select the best $t_{\textrm{Min}}$ value: (1) When including all ensembles and momenta, the
$p$~value must follow a sufficiently flat distribution for 15 ensembles,%
~% \footnote{Since with 15 ensembles the statistics are not very high, we require that there are no unreasonable peaks in the 
% distribution.} 
(2) the $2+2$ and the $3+3$ fits must agree on the
nonoscillating ground-state energy and overlap factors, (3) the fit result must be stable under small variations of the fit range,
and (4) the product $\mathcal{Z}_d\sqrt{2E} = Z_d$ for the overlap factor of the ground state should be approximately independent of
the momentum, barring discretization effects.
When these conditions are all fulfilled, we consider that the systematic errors due to the omission of still further excited states
have been included in the statistical fit error.
This usually leaves us with a small range of possible values for $t_{\textrm{Min}}$, we chose among those the one that complies best
with all these conditions. The selected values are listed in Table~\ref{TableFitRanges}.
An example of the level of agreement that is reached between our $2+2$ and $3+3$ two-point correlator fits is shown in
Table~\ref{TableFitResults}.

Previous experience \cite{Bailey:2014tva,Lattice:2015tia,Lattice:2015rga,Bailey:2015dka,Bazavov:2016nty,Bazavov:2019aom},
which also applies to this study, has shown that it is better to impose the four criteria introduced above on a set of fits,
rather than choosing, on a case-by-case basis, the fit with the smallest $\chi^2/$dof, the smallest error, or some other notion
of ``best'' fit. The case-by-case approach amplifies meaningless statistical fluctuations, which can introduce problems in
subsequent steps of the analysis (here, the chiral-continuum extrapolation).
Figure~\ref{StabTest} shows the stability of 1+1, 2+2, and 3+3 fits on ensembles at four lattice spacings, denoting the common $t_\text{Min}$.
It illustrates that we could have chosen smaller values of $t_\text{Min}$ on some ensembles, if we had adopted ensemble-by-ensemble criteria. Hence, our common $t_\text{Min}$ value is conservatively chosen.

\begin{table}
    \caption{Fit ranges in physical units for the two-point function fits used in the analysis of the form factors.
    As the number of included states increases, $t_{\text{Min}}$ is reduced to include information from the rapidly decaying excited states in the fit.
    For the coarsest ensembles and momentum $2\pi(4,0,0)/L$, the fit ranges do not yield enough points to perform a fit.
    In those cases the $2\pi(4,0,0)/L$ point is simply dropped.}
    \label{TableFitRanges}
  \begin{tabular}{cccc}
    \hline
    \hline
 \MulRow{2}{*}{$p$ $(2\pi/L)$} &      \MulCol{3}{c}{Two-point}       \\
                                & $2+2$ $t_{\text{Min}}$ (fm) & $3+3$ $t_{\text{Min}}$ (fm) & $t_{\text{Max}}$ (fm) \\ 
    \hline
            (0,0,0)             &    \MulRow{3}{*}{1.021}     &    \MulRow{3}{*}{0.631}     &         3.301         \\
            (1,0,0)             &                             &                             &         3.241         \\
            (2,0,0)             &                             &                             &         2.431         \\
    \hline
            (1,1,0)             &    \MulRow{8}{*}{1.021}     &    \MulRow{8}{*}{0.631}     &         2.836         \\
            (1,1,1)             &                             &                             &         2.551         \\
            (2,1,0)             &                             &                             &         2.251         \\
            (2,1,1)             &                             &                             &         2.161         \\
            (2,2,0)             &                             &                             &         1.921         \\
            (2,2,1)             &                             &                             &         1.801         \\
            (3,0,0)             &                             &                             &         1.681         \\
            (4,0,0)             &                             &                             &         1.201         \\
    \hline
    \hline
  \end{tabular}
\end{table}

%\begingroup
%\squeezetable
\begin{table*}
  \scriptsize
  \setlength\tabcolsep{2pt}
  \caption{Results for the $D^\ast$-meson two-point correlator fits on the $a\approx0.12$~fm, $m_l=0.14m_s$ ensemble.
      We compare the nonoscillating ground-state energy and overlap factors for the $2+2$ and $3+3$ state fits.
      We include here all fits that distinguish between the different orientations of the momentum.
      In the analysis we use the $3+3$ state fit result.}
  \label{TableFitResults}
  \begin{tabular*}{\textwidth}{cS[table-format = 1.4(2)]
                                S[table-format = 1.4(2)]
                                S[table-format = 1.3(2)]
                                S[table-format = 1.3(2)]
                                S[table-format = 1.4(2)]
                                S[table-format = 1.4(2)]
                                S[table-format = 1.3(2)]
                                S[table-format = 1.3(2)]
                                S[table-format = 1.4(2)]
                                S[table-format = 1.4(2)]
                                S[table-format = 3.1]@{}l@{}l
                                S[table-format = 3.1]@{}l@{}l}
    \hline
    \hline
 \MulRow{2}{*}{$p$ $(2\pi/L)$} &  \MulCol{2}{c}{$E_{D^\ast}$}  & \MulCol{2}{c}{$\mathcal{Z}^\bot_{1S}$} & \MulCol{2}{c}{$\mathcal{Z}^\bot_d$} & \MulCol{2}{c}{$\mathcal{Z}^\parallel_{1S}$} & 
                                \MulCol{2}{c}{$\mathcal{Z}^\parallel_d$} & \MulCol{6}{c}{$\chi^2/\text{dof}$} \\
                               &    {$2+2$}    &    {$3+3$}    &    {$2+2$}    &    {$3+3$}    &    {$2+2$}    &    {$3+3$}    &    {$2+2$}    &    {$3+3$}    &    {$2+2$}    &    {$3+3$}    & \MulCol{3}{c}{$2+2$} & \MulCol{3}{c}{$3+3$} \\
    \hline
           (1,0,0)             &   1.0837(21)  &   1.0841(19)  &   2.527(37)   &   2.531(39)   &   0.2777(43)  &   0.2767(46)  &   2.682(39)   &   2.674(42)   &   0.2912(45)  &   0.2897(48)  &    101.8 & / & 92    &     120.5 & / & 95   \\
           (2,0,0)             &   1.1854(49)  &   1.1890(50)  &   1.555(61)   &   1.579(63)   &   0.253(10)   &   0.259(10)   &   1.842(76)   &   1.881(83)   &   0.296(12)   &   0.302(12)   &     47.3 & / & 50    &      62.7 & / & 60   \\
    \hline
    \hline
  \end{tabular*}
\end{table*}
%\endgroup

\subsubsection{The dispersion relation}
The calculation of the dispersion relation serves two purposes: first, we can estimate a good prior for the two-point functions that
enter in the analysis of the form factors; second, by checking the size of the deviations from the continuum dispersion relation, we
can test whether the discretization errors due to the heavy quarks are under control.
The dispersion relation which includes discretization effects can be written as 
\begin{align}
    a^2E^2(\bm{p}) &= (aM_1)^2 + \frac{M_1}{M_2}(a\bm{p})^2 \nonumber \\
        &\phantom{=} + \frac{1}{4}\left[\frac{1}{(aM_2)^2} - \frac{aM_1}{\left(aM_4\right)^3}\right](a^2\bm{p}^2)^2 \nonumber \\
        &\phantom{=} - \frac{aM_1w_4}{3}\sum_{i=1}^3(ap_i)^4 + O(p_i^6),
    \label{DispRel}
\end{align}
where $M_1$ is the rest mass, $M_2$ is the kinetic mass, and $M_4$ is a further mass-like quantity.
A key observation of Ref.~\cite{ElKhadra:1996mp} is that the matching of the relativistic Wilson action via HQET or NRQCD to continuum QCD
removes discretization effects that grow uncontrollably with $aM$. In Eq.~\eqref{DispRel} discretization effects are described by the
coefficients of the $(a\bm{p})^n$ terms, parameterized by $w_4$, $M_1$, $M_2$ and $M_4$, for which explicit expressions
are given in Ref.~\cite{ElKhadra:1996mp}.
We tune the kinetic mass $M_2$ to match the experimentally observed mass, according to the nonrelativistic
interpretation of the clover action~\cite{ElKhadra:1996mp}.\footnote{In the Fermilab formulation, one can adjust $M_1=M_2$ by introducing an
asymmetry parameter into the lattice action, but this is not necessary for valence quarks. Adjusting $M_4=M_2$ requires a more improved action~\cite{Oktay:2008ex}.}

The leading $O(a^2)$ discretization effects are due to $M_1/M_2\sim 1$. Expectations for this ratio can be inferred from perturbation theory for
the quark masses~\cite{Mertens:1997wx} and by tracing contributions to the binding energy~\cite{Kronfeld:1996uy,Bernard:2010fr}.
On this basis, we expect $M_1/M_2$ to be $1+O(\alpha_s,(am_{0c})^2)$, and we would like to test whether the
leading deviation from the continuum dispersion relation, $E^2=\bm{p}^2+M^2$, grows as $O(\alpha_s a^2p^2)$.
We can check whether our nonzero momentum fits show deviations of order $O(\alpha_s a^2p^2)$ from the continuum dispersion relation,
and we can also fit the energies from our correlator fits to Eq.~\eqref{DispRel}, considering the coefficients
in front of the powers of momenta as fit parameters.
These results are used to guide the prior central values for the ground-state energies of the two-point correlators with nonzero momentum that are part of the 
three-point analysis which yields the form factors. 
In order to make this prior independent of the form-factor data, we exclude the $p=2\pi(1,0,0)/L$,
$2\pi(2,0,0)/L$ momenta from the dispersion-relation calculation.
As explained in Sec.~\ref{SubSec:2pt}, data for different polarizations of the $D^\ast$ meson are available only for
$p=2\pi(1,0,0)/L$ and $2\pi(2,0,0)/L$.
Therefore, these are the only momenta for which we obtaine the form factors at nonzero recoil. 

\begin{figure}[h]
    \centering
    \includegraphics[width=\linewidth]{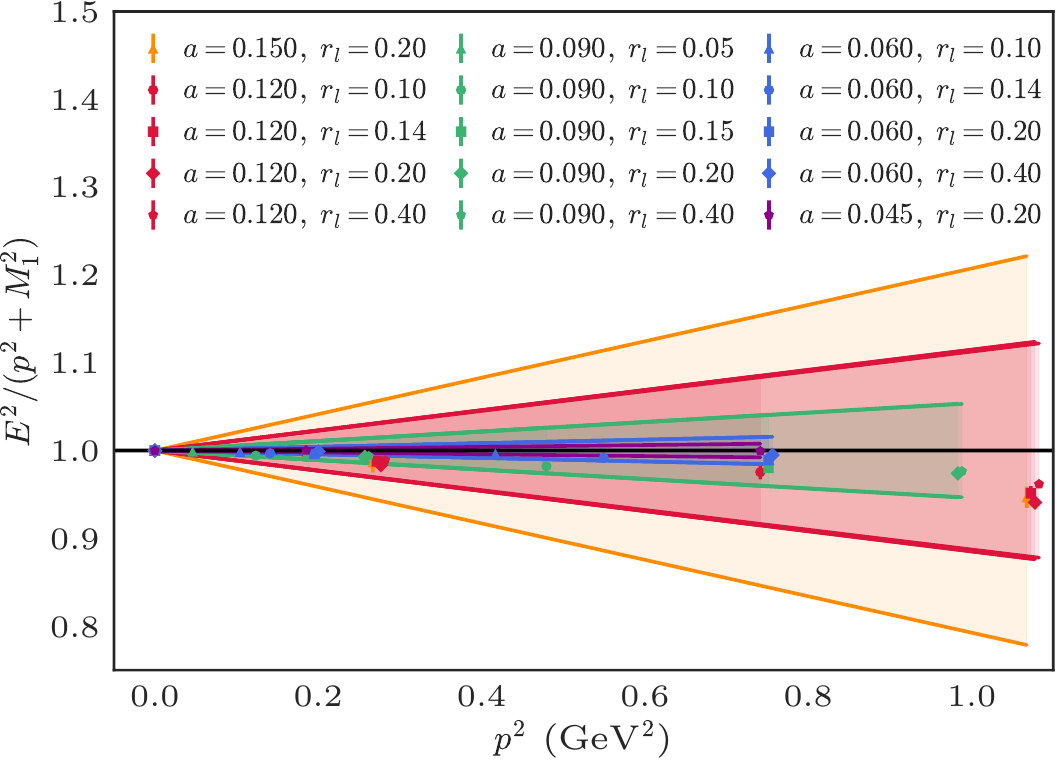}
    \caption{Ratio between the fitted energies and the expected energies from the continuum dispersion relation.
        The color encodes the lattice spacing, whereas the different symbols encode the light-to-strange mass ratio.
        The cones show the expected size of the discretization errors per lattice spacing for our lattice formulation.
        Hence, each cone follows the equation $\alpha_s a^2 p^2$, where $\alpha_s$ is different for each lattice spacing,
        and it is calculated following Refs.~\cite{Brodsky:1982gc,Lepage:1992xa}.
        The cones are color coded to match points with the same lattice spacing.
        All plotted points for each particular lattice spacing lie within their respective cones, which in turn means that
        the discretization errors are well within the expected size.}
    \label{PlotDispRel}
\end{figure}

In Table~\ref{TableFitDispResults}, we show the results of our two-point correlator fits that enter in the dispersion-relation fit
for a particular ensemble.
There is good agreement between the $2+2$ and the $3+3$ state fits, indicating that the systematic error from the omission of
higher states is negligible.
Results for other ensembles show a similar behavior.
Figure~\ref{PlotDispRel} compares the continuum dispersion relation with our data.
The data points show small discretization errors, which tells us that, indeed, these errors are under control.

%\begingroup
\begin{table*}
    \caption{Results for the $D^\ast$-meson two-point correlator fits on the $a=0.12$~fm, $m_l=0.14m_s$ ensemble.
        We compare the nonoscillating ground-state energy and overlap factors for the $2+2$ and $3+3$ state fits.
        We include here all fits that did not distinguish between the different orientations of the momentum.
        Momentum $2\pi(4,0,0)/L$ did not have enough degrees of freedom left for the $3+3$ state fit, and hence it is not shown.
        In the analysis we use the $3+3$ state fit result.}
    \label{TableFitDispResults}
  \begin{tabular*}{\textwidth}{@{\extracolsep{\fill}}
                               cS[table-format = 1.4(2)]
                                S[table-format = 1.4(2)]
                                S[table-format = 1.3(2)]
                                S[table-format = 1.3(2)]
                                S[table-format = 1.4(2)]
                                S[table-format = 1.4(2)]
                                S[table-format = 2.1]@{}c@{}l
                                S[table-format = 2.1]@{}c@{}l}
    \hline
    \hline
\MulRow{2}{*}{$p$ $(2\pi/L)$} & \MulCol{2}{c}{$aE_{D^\ast}$} & \MulCol{2}{c}{$\mathcal{Z}_{1S}$} & \MulCol{2}{c}{$\mathcal{Z}_d$} & \MulCol{6}{c}{$\chi^2/\text{dof}$} \\
                              &    {$2+2$}   &    {$3+3$}    &     {$2+2$}    &     {$3+3$}      &    {$2+2$}     &    {$3+3$}    & \MulCol{3}{c}{$2+2$} & \MulCol{3}{c}{$3+3$} \\ 
    \hline
           (0,0,0)            &  1.0419(17)  &  1.0405(17)   &    3.079(25)   &    3.052(24)     &   0.2865(26)   &   0.2835(30)  &    54.5 & / & 55     &    56.2  & / & 59    \\
           (1,1,0)            &  1.1205(31)  &  1.1215(26)   &    2.217(44)   &    2.215(44)     &   0.2786(56)   &   0.2782(51)  &    36.8 & / & 36     &    47.2  & / & 40    \\
           (1,1,1)            &  1.1569(36)  &  1.1576(30)   &    1.939(45)   &    1.928(43)     &   0.2769(61)   &   0.2760(53)  &    30.3 & / & 27     &    35.5  & / & 31    \\
           (2,1,0)            &  1.2191(43)  &  1.2202(35)   &    1.478(40)   &    1.477(41)     &   0.2682(64)   &   0.2683(55)  &    10.5 & / & 21     &    19.5  & / & 25    \\
           (2,1,1)            &  1.2513(47)  &  1.2536(38)   &    1.324(40)   &    1.323(39)     &   0.2666(67)   &   0.2663(56)  &    16.5 & / & 18     &    23.5  & / & 22    \\
           (2,2,0)            &  1.3089(65)  &  1.3100(49)   &    1.052(49)   &    1.050(44)     &   0.2611(68)   &   0.2607(57)  &    13.8 & / & 12     &    15.4  & / & 16    \\
           (2,2,1)            &  1.3378(72)  &  1.3402(58)   &    0.989(53)   &    0.959(44)     &   0.2598(68)   &   0.2596(58)  &    12.0 & / & 9      &    15.5  & / & 13    \\
           (3,0,0)            &  1.3168(79)  &  1.3168(79)   &    0.822(73)   &    0.83(12)      &   0.2523(68)   &   0.2513(59)  &     3.6 & / & 6      &     9.0  & / & 10    \\
    %      (4,0,0)            &  1.366(11)   &  1.3646(75)   &    0.763(41)   &    0.693(51)     &   0.2463(44)   &   0.2438(38)  &    1.67 & / & 2      &     0.85 & / & 3     \\
    \hline
    \hline
  \end{tabular*}
\end{table*}
%\endgroup

\subsection{Three-point functions}
With our previously defined interpolating operators, we can also construct three-point correlators by sandwiching a current between
two meson states,
\begin{equation}
C^{J^\mu}_{X_a\to Y_b}(\bm{p}, t) = \left\langle\mathcal{O}_{Y_b}(\bm{0}, T)J^\mu(\bm{p}, t)\,\mathcal{O}_{X_a}^\dagger(-\bm{p},0)\right\rangle.
\end{equation}
Using the same notation as in Eq.~\eqref{2ptDecomp}, we can write the spectral decomposition of the three-point correlators for a
particular source-sink separation $T$ as
\begin{align}
C^{J^\mu}_{X_a\to Y_b}(\bm{p}, t) &= \sum_n s_n(t)\,s_m(T - t)\sqrt{Z_{Y_b,n}(\bm{p})}\frac{e^{-E_n(\bm{p})t}}{2E_n(\bm{p)}} \nonumber \\
 & \hspace{-1.5cm}\left\langle Y_b, n, \bm{p}|J^\mu| X_a, m, \bm{0}\right\rangle \sqrt{Z_{X_a,m}(\bm{0})}\frac{e^{-M_m(T-t)}}{2M_m},\label{3ptDecomp}
\end{align}
where we choose $t < T \ll L_t$, such that wraparound terms with $t\to L_t - t$ and $T - t \to L_t - (T - t)$ in the exponent are completely negligible, at most $\sim 10^{-16}$,

In our three-point functions, we always use a Richardson $1S$ smearing for the $B$ meson, but the $D^\ast$ meson operator is either
$1S$-smeared or point $d$.  This gives a variety of possibilities for constructing ratios of correlators. For $x_f$ we use % $\bm{x}_f$ we use
\begin{equation}
    x_f(\bm{p}, t, T) = \frac{C^{V_j}_{D^\ast_{1S}\to D^\ast_a}(\bm{p}_{\bot,\parallel},t,T)}
        {C^{V_4}_{D^\ast_{1S} \to D^\ast_a}(\bm{p}_{\bot,\parallel},t,T)},
    \label{ratXf}
\end{equation}
where $a = d,1S$ and the orientation of the momentum can be arbitrary, as long as it is the same for the correlator in the numerator
and denominator.  These combinations cancel the leading overlap factors and exponentials.  The same cancelation can be achieved
in~$X_V$,
\begin{equation}
    X_V(\bm{p}, t, T) = \frac{C^{V_j}_{B_{1S}\to D^\ast_a}(\bm{p}_\bot,t,T)}{C^{A_j}_{B_{1S}\to D^\ast_a}(\bm{p}_\bot,t,T)}.
    \label{ratXv}
\end{equation}
In these two ratios we can find the desired matrix element in the limit $t \gg 0$ and $T - t \gg 0$, with $t < T$.
The double ratio and the other two single ratios can be expressed in the same way
\begin{align}
    X_0    (\bm{p}, t, T) &=
        \frac{C^{A_4}_{B_{1S}\to D^\ast_a}(\bm{p}_\parallel,t,T)}{C^{A_j}_{B_{1S}\to D^\ast_a}(\bm{p}_\bot,t,T)}
        \sqrt{\frac{Z_{D^\ast,a}(p_\bot)}{Z_{D^\ast,a}(p_\parallel)}},
    \label{ratR0}\\
    X_1    (\bm{p}, t, T) &=
        \frac{C^{A_j}_{B_{1S}\to D^\ast_a}(\bm{p}_\parallel,t,T)}{C^{A_j}_{B_{1S}\to D^\ast_a}(\bm{p}_\bot,t,T)}
        \sqrt{\frac{Z_{D^\ast,a}(p_\bot)}{Z_{D^\ast,a}(p_\parallel)}},
    \label{ratR1}\\
    R_{A_1}(\bm{p}, t, T) &=
        \frac{C^{A_j}_{B_{1S}\to D^\ast_a}(\bm{p}_\bot,t,T) \, C^{A_j}_{D^\ast_a\to B_{1S}}(\bm{p}_\bot,t,T)}
             {C^{V^4}_{D^\ast_a\to D^\ast_{1S}}(\bm{0},t,T) \, C^{ V^4 }_{B_{1S}\to B_{1S}}(\bm{0},t,T)} \nonumber \\
        & \hspace{-1cm} \frac{Z_{D^\ast,a}(p_\bot)}{\sqrt{Z_{D^\ast,a}(0)\,Z_{D^\ast,1S}(0)}}
                        \frac{M^2_{D^\ast}}{E^2_{D^\ast}(\bm{p})} e^{-(E_{D^\ast}(\bm{p}) - M_{D^\ast})T},
    \label{ratRA1}
\end{align}
but in this case the computed ratios depend on extra factors that must be removed before extracting the matrix elements.  The
overlap factors are removed per jackknife bin using the results of the two-point correlator fits.  In this way we can propagate
correlations from one fit to the other.  The $M_{D^\ast}/E_{D^\ast} = 1/w$ factor in Eq.~\eqref{ratRA1} is removed using the value
of the recoil parameter, as extracted from Eq.~\eqref{wDef0} per jackknife bin.

The double ratio $R_{A_1}$ deserves further comment. First, we reanalyze the ze\-ro-mo\-men\-tum correlators~\cite{Bailey:2014tva} using
the criteria given above. That also implies the double ratio $R_{A_1}(\bm{p}=\bm{0})$ is constructed only for the $a=1S$ smearing.
Second, we did not generate three-point functions at nonzero momentum of the form\linebreak $C^{A_j}_{D^\ast_a\to B_{1S}}(\bm{p}_\bot,t,T)$,
so we use the time reversal operation $\mathcal{T}$ to obtain the missing correlator,
\begin{equation}
    C^{A_j}_{B_{1S}\to D^\ast_a}(\bm{p}_\bot,t,T) \xrightarrow{\mathcal{T}} C^{A_j}_{D^\ast_a\to B_{1S}}(\bm{p}_\bot,T-t,T).
\end{equation}
%The resulting ratio is not exactly symmetric because the denominator of Eq.~\eqref{ratRA1} is not symmetrized.

\subsubsection{Three-point function fits}
The three-point functions are also affected by the oscillating states introduced by the staggered regularization.
So are the ratios constructed with such three-point correlators, but the dependence on the oscillating states is not as clean as in
the case of the two-point functions.
Our ratios do not show any noticeable oscillatory behavior in source and sink, but states that oscillate at both ends introduce a
nonnegligible overall shift on the ratio central value that depends on the sink time $T$ as $(-1)^T$.
In order to remove this contribution, we smooth the data following Refs.~\cite{Bernard:2008dn,Bailey:2014tva,Lattice:2015rga,Bailey:2008wp},
namely, we calculate the three-point correlators at two different values of the sink time $T$, and then we compute the following weighted
average to suppress this unwanted shift in most ratios:
\begin{align}
\bar{R}(t,T)\equiv\frac{1}{2}R(t,T) + \frac{1}{4}R(t,T+1) + \frac{1}{4}R(t+1,T+1), \label{avratio} \\ 
\textrm{with}\quad R=X_0,X_1,X_V,x_f\textrm{ and }R_{A_1}(\bm{p}=0).\nonumber
\end{align}
The contribution of the oscillating shift is then greatly suppressed.

\begin{figure*}
   \centering
   \includegraphics[width=\linewidth]{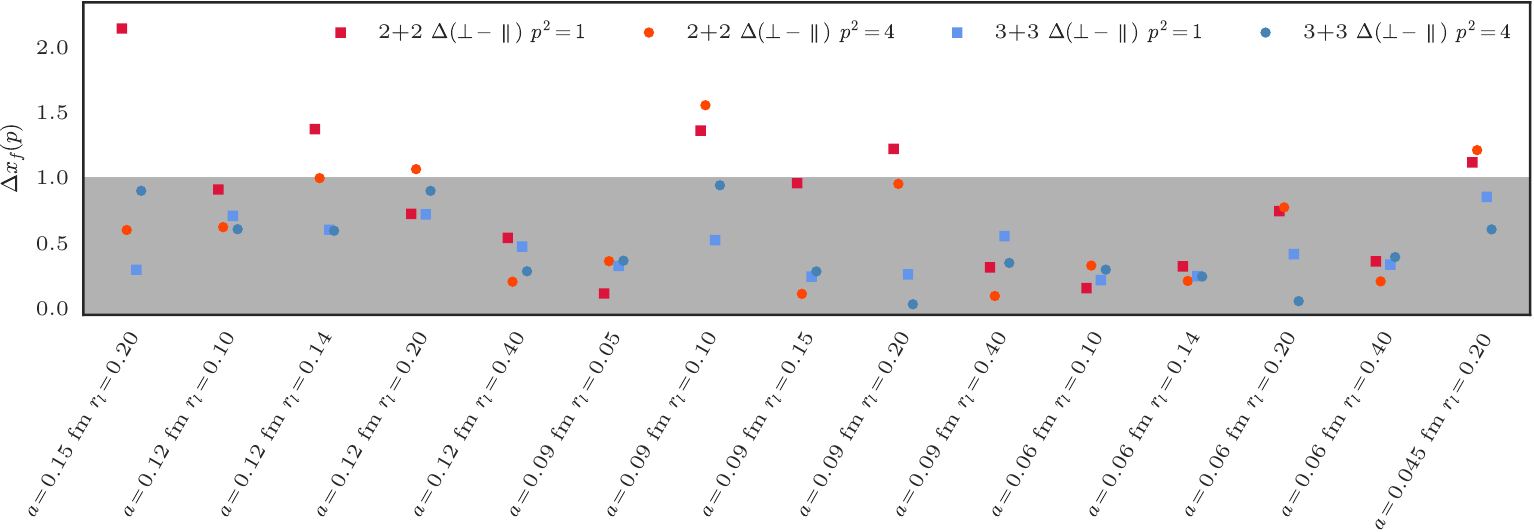}
   \caption{Difference in sigmas between $x_f$ calculated with parallel and perpendicular polarizations, for momenta
       $p^2=1,4$ and one ($2+2$) and two ($3+3$) excited states at source and sink in the ratio. The gray band corresponds to 
       differences $\le 1\sigma$. It is clear that an increase in the number of excited states reduces the difference between 
       polarizations to $1\sigma$ or less, whereas with a single excited state the difference can reach as high as $2\sigma$ in 
       some cases. Since we do not have any other way to keep in check the systematics due to excited states for other ratios, we 
       take a conservative approach and use the $3+3$ state fits for the rest of the analysis.}
   \label{xfComp}
\end{figure*}

The double ratio at nonzero momentum, $R_{A_1}(\bm{p}\neq0)$, requires the explicit removal of the sink-dependent exponentials in order to avoid bias,
\begin{align}
\bar{R}_{A_1}(\bm{p}\neq 0,t,T)\equiv &                  \frac{1}{2}R_{A_1}(\bm{p},t,  T  )e^{(E_{D^\ast}(\bm{p}) - M_{D^\ast})T} \nonumber \\
                                      & \hspace{-2cm} + \frac{1}{4}R_{A_1}(\bm{p},t,  T+1)e^{(E_{D^\ast}(\bm{p}) - M_{D^\ast})(T+1)} \nonumber\\
                                      & \hspace{-2cm} + \frac{1}{4}R_{A_1}(\bm{p},t+1,T+1)e^{(E_{D^\ast}(\bm{p}) - M_{D^\ast})(T+1)}.
\label{FixRatio}
\end{align}
These exponentials are removed using the energy and mass values coming from the two-point correlator fits per jackknife bin.
The ratio averages defined in Eqs.~(\ref{avratio}) and (\ref{FixRatio}) suppress the contributions from the unwanted oscillations to a fraction of the statistical errors.
Therefore, we henceforth employ only the averaged ratios in our analysis and omit the bar for simplicity.
The data are then processed through a single elimination jackknife, and the extra overlap factors and exponentials are removed by
using the values obtained in the two-point correlator fits per jackknife bin.
Then the ratios are fitted to the functional form:
\begin{align}
    R(\bm{p}, t, T) &= K\Big(1 + A_1 e^{-\Delta E^1_{X}t} + A_2 e^{-\Delta E^2_{X}t} \nonumber \\
    & \phantom{=} + B_1 e^{-\Delta E^1_{Y}(T-t)} + B_2 e^{-\Delta E^2_{Y}(T-t)}\Big),
    \label{ratAnsatz}
\end{align}
where $K$ is the matrix element we want to extract, and the extra terms take into account the presence of excited states, assuming their contribution is small.
The labels $X$, $Y$ represent mesons at source and sink, respectively, and $\Delta E^j_{X,Y}$ represents the energy difference between the ground state and the
$j^{\textrm{th}}$ excited state.
The second excited states at source and sink (included in the $A_2$ and $B_2$ terms) are necessary to remove systematic errors due to unaccounted excited states.
In order to check this point, we computed the ratio $x_f$ with different polarizations of the $D^\ast$ meson.
We expect the extracted matrix elements from different polarizations to agree, except for discretization effects that should be
reduced as the lattice spacing decreases.
Nonetheless, our results show a difference between the analysis with a single excited state at source and sink and the analysis with
two excited states at each end.
The addition of extra excited states not only increases the error, as expected, but also brings the central values calculated with
different polarizations closer.
Overall there is a large reduction in the difference between the cases with polarization parallel and perpendicular to the momentum.
This behavior depends only mildly on the lattice spacing, as can be checked in Fig.~\ref{xfComp}.

The two available sets of correlators with smearing operators $a=d$, $1S$ are fit simultaneously, so that they share $\Delta E$ and
$K$, but each smearing has its own $A_{1,2}$ and $B_{1,2}$.  We employ a loose prior for $K$ with a central value roughly set by the
ratios at $\approx T/2$, where the excited states are more suppressed, and with a width large enough to accommodate
significant variations: in the case of the double ratio $R_{A_1}$ the width of the prior is set to $0.1$, whereas the other ratios use
$0.05$. In all cases, the width of the prior encompasses all available correlator points for the single ratios. Except for the double ratio,
the excited states at source and sink carry different signs, hence we ensure the prior covers the central value of the matrix element.
Typically, the posterior is almost an order of magnitude narrower that the prior, although in the least precise cases th the posterior
width is $\approx60\%$ of the prior.
The priors for the $\Delta E$ of the first (second) excited states are taken from the two-point
function fits with $3+3$ states, but we increase the error by a factor of three (eight) to allow $\Delta E$ to differ from the two-point
correlator values.  This increase takes into account the fact that the $t_{\textrm{Min}}$ in the ratio fits is much smaller than in
the two-point function fits, and the excited-state pattern might be different as well. The priors for $A_{1,2}$ and $B_{1,2}$ are
taken to be $0(2)$ and $0(1)$ in the smeared and point source cases respectively.\footnote{We observe that the local operators have a
smaller overlap with the excited states, and we reduce the width of their corresponding priors for $A_{1,2}$ and $B_{1,2}$
accordingly.}
As stated above, our priors are conservative enough that significant variations of their central values and/or widths do not
result in a relevant change in the posterior for the matrix element.

The fit ranges are chosen following criteria similar to the two-point function case.  We use the same value of $t_{\textrm{Min}}$
and $t_{\textrm{Max}}$ in physical units for all ensembles and all ratios, except for the double ratio, where we use
$t_{\textrm{Max}}=T-t_{\textrm{Min}}$ to account for the fact that the states on source and sink are exactly the same.  In this
case, we take the same $t_{\textrm{Min}}$ as for the rest of the ratios.  We choose the fits that show stability over small
variations of the fit range and result in a reasonably flat distribution for the $p$~values.

\subsection{Calculation of the recoil parameter}
\label{wCalcSec}

\begin{figure*}
    \centering
    \includegraphics[width=\linewidth]{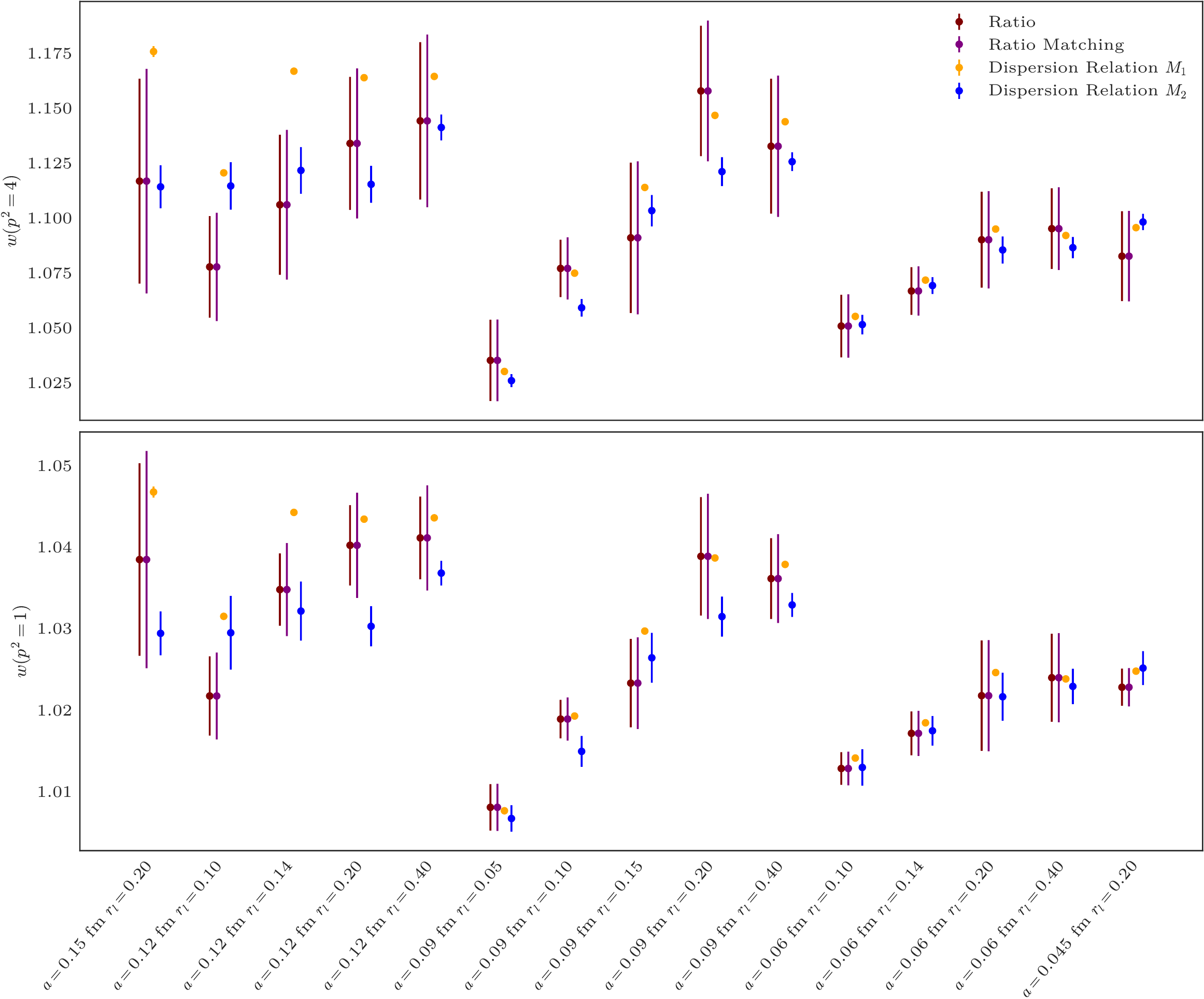}
    \caption{Comparison of different definitions of the recoil parameter.
        The horizontal axis labels the ensemble, whereas the vertical axis shows the recoil value $w$, calculated from the 
        different definitions.
        The ratio method is shown before (``Ratio") and after (``Ratio matching") including the matching errors, whereas the points 
        labeled as ``Dispersion relation $M_{1,2}$" use Eq.~(\ref{wParmNew}) and the fitted rest ($M_1$) or kinetic ($M_2$) mass of 
        the $D^\ast$ meson. 
        The three-momentum is set to $p=2\pi(1,0,0)/L$ [$p=2\pi(2,0,0)/L$] in the lower (upper) panel.}
    \label{wCompPlot}
\end{figure*}

As in previous work~\cite{Bailey:2012rr,Lattice:2015rga}, we use the ratio $x_f$ to define the recoil parameter, following
Eq.~\eqref{wDef0}. The disadvantage of this method is that it introduces systematic errors due to the renormalization of the
currents.  One could also use the continuum dispersion relation to define the recoil parameter:
\begin{equation}
    w = \sqrt{\frac{M_{D^\ast}^2 + \bm{p}^2}{M_{D^\ast}^2}},
    \label{wParmNew}
\end{equation}
where $\bm{p}$ is the three-momentum of the $D^\ast$ meson, and the mass is either $M_1$ or $M_2$.
The different choices for the different definition of the mass are expected to result in slightly different discretization errors
that are resolved in our chiral-continuum extrapolation, so the choice should not affect the final results.
As shown in Fig.~\ref{wCompPlot}, the error in the $x_f$ method encompasses the differences in the rest- and the kinetic-mass
versions of Eq.~(\ref{wParmNew}).
In this work, we take a conservative approach and define $w$ via Eq.~\eqref{wDef0}, but we
note that all choices lead to results for the form factors, $|V_{cb}|$, and $R(D^\ast)$ that are compatible within errors.

\subsection{Current renormalization and blinding}
\label{analysis:matching}

As outlined in Sec.~\ref{eFfMe}, the ratios described in Eqs.~(\ref{ratXf})--(\ref{ratRA1}) are constructed in such a way that the
flavor-diagonal renormalization factors $Z_{V^4_{cc}}Z_{V^4_{bb}}$ from Eq.~\eqref{cRat} cancel out.
Hence, what remains is only the computation of the different matching factors $\rho_X$ that enter in the ratios.
These factors can be calculated using perturbation theory, but the calculation becomes cumbersome for $w>1$.
In this article, we use the approximation $am_{2c}\to 0$, where $am_{2c}$ is the charm kinetic mass, which removes the dependence on $w$,
because a light quark cannot modify the dynamics of a heavy quark in the heavy quark limit (see \ref{ApHQCorr} and Refs.~\cite{Harada:2001fi,Harada:2001fj}).
Then we incorporate errors coming from these approximations.
The $w$ dependence introduces an error proportional to $w-1$, and the $am_{2c}\to 0$ approximation increases the error by
$O(\alpha_s am_{2c})$.

The calculation of the axial matching factor $\rho_{A_j}(1)$ then follows exactly the procedure of Ref.~\cite{Bailey:2014tva}.
The other ratios require further calculations that are detailed in \ref{ApMatch}.
The resulting values for all matching factors are gathered in Table~\ref{TableMatchingFactors}, where the errors shown
are from the VEGAS integration. 
The errors associated with the approximations we make to obtain those factors are discussed in \ref{ApMatch}.
They are added to those in Table~\ref{TableMatchingFactors} before carrying out the chiral-continuum extrapolation.

\begin{table*}
    \caption{Matching factors calculated at one loop in perturbation theory for all ensembles and form factors.
        The errors shown in the table do not include the systematic errors coming from the approximations employed in the
        calculation of the matching factors.
        They are nonetheless included in the chiral-continuum extrapolation.
        These errors and more details on the calculation are described in \ref{ApMatch}.
        The last two columns include the values of $\alpha_V$ and $m_{2c}a$ used in~\ref{ApMatch} to estimate the matching errors.}
    \label{TableMatchingFactors}
    \begin{tabular*}{\textwidth}{@{\extracolsep{\fill}}ccS[table-format = 1.6(2)]S[table-format = 1.6(2)]S[table-format = 1.7(2)]S[table-format = 1.4]S[table-format = 1.4]}
    \hline
    \hline
            $a$ (fm)          &   $am_l$/$am_s$  &  {$\rho_{A_j}$} & {$\rho_{A_4}/\rho_{A_j}$} & {$\rho_{V_j}/\rho_{A_j}$} & $\alpha_V$ & $m_{2c}a$ \\ 
    \hline                                                                                                                                          
         $\approx 0.15$       &   0.0097/0.0484  &   0.990864(48)  &        0.945800(10)       &       0.8835513(23)       &   0.3589   &   0.6000  \\
    \hline                                                                                                                                          
\MulRow{4}{*}{$\approx 0.12$} &     0.02/0.05    &   0.994199(25)  &        0.961710(9)        &       0.9133327(20)       &   0.3047   &   0.4259  \\
                              &     0.01/0.05    &   0.988811(22)  &        0.957865(9)        &       0.9096509(22)       &   0.3108   &   0.4460  \\
                              &    0.007/0.05    &   0.993665(41)  &        0.960041(9)        &       0.9101966(21)       &   0.3102   &   0.4469  \\
                              &    0.005/0.05    &   0.993640(23)  &        0.960038(9)        &       0.9101935(21)       &   0.3102   &   0.4471  \\
    \hline                                                                                                                                          
\MulRow{5}{*}{$\approx 0.09$} &   0.0124/0.031   &   0.998292(13)  &        0.971160(5)        &       0.9320012(15)       &   0.2582   &   0.3269  \\
                              &   0.0062/0.031   &   0.998077(13)  &        0.970701(6)        &       0.9310817(16)       &   0.2607   &   0.3317  \\
                              &  0.00465/0.031   &   0.997956(18)  &        0.970559(6)        &       0.9308184(16)       &   0.2611   &   0.3350  \\
                              &   0.0031/0.031   &   0.997901(12)  &        0.970405(7)        &       0.9305130(16)       &   0.2619   &   0.3361  \\
                              &  0.00155/0.031   &   0.997889(16)  &        0.970353(7)        &       0.9303983(16)       &   0.2623   &   0.3365  \\
    \hline                                                                                                                                          
\MulRow{4}{*}{$\approx 0.06$} &   0.0072/0.018   &   1.001973(7)   &        0.978328(4)        &       0.9460531(12)       &   0.2238   &   0.2343  \\
                              &   0.0036/0.018   &   1.002050(11)  &        0.978408(4)        &       0.9461304(11)       &   0.2245   &   0.2327  \\
                              &   0.0025/0.018   &   1.002013(13)  &        0.978356(5)        &       0.9460189(12)       &   0.2249   &   0.2336  \\
                              &   0.0018/0.018   &   1.002015(6)   &        0.978309(4)        &       0.9459167(12)       &   0.2253   &   0.2339  \\
    \hline                                                                                                                                          
         $\approx 0.045$      &   0.0028/0.014   &   1.005588(8)   &        0.984902(2)        &       0.9578549(7)        &   0.2013   &   0.1647  \\
    \hline
    \hline
    \end{tabular*}
\end{table*}
We also calculate the matching factors for the ensembles involved in the heavy-quark (HQ) mistuning corrections.
This is a departure from Ref.~\cite{Lattice:2015rga}, where the matching factors are applied after the HQ-mistuning correction.
In this way, we completely separate the HQ-mistuning corrections from the matching errors.

During the analysis, we blinded the form-factor data by multiplying the matching factor $\rho_{A_j}$ by an unknown, random number close to 1.
The ratios $\rho_{A_4}/\rho_{A_j}$ and $\rho_{V_j}/\rho_{A_j}$ are left unchanged.
Via the correlator ratios and Eqs.~\eqref{eq:matched-hA1}--\eqref{eq:matched-hV}, all form-factor values on all ensembles are thus multiplied by a common factor.
All stages of the analysis were tested either through independent fits performed by two co-authors or by using independent methods or codes.
Only after the full analysis was complete, including construction of the systematic error budget (Sec.~\ref{ErrorBudget}) and $z$~expansion
(Sec.~\ref{Results}), the blinding factor was removed from $\rho_{A_j}$, and the analysis scripts were rerun to extract the unblinded results for the form
factors.

\subsection{Heavy-quark-mass adjustment}

The bare masses for the $b$ and $c$ quarks are tuned such that the kinetic masses of the $B_s$ and $D_s$ meson on each ensemble are
equal to their physical values.
Nonetheless, the tuning procedure has errors that must be taken into account.
The procedure is explained in \ref{ApHQCorr} and outlined here.
As we generate configurations, values for the heavy-quark masses that give approximately the correct meson masses can be estimated.
These initial values are employed to compute the two- and three-point functions that we analyze in this work.
At the end of the data generation, the much larger statistical sample allows for a more precise determination of the $b$ and $c$ quark
masses by using the procedure detailed in Ref.~\cite{Bailey:2014tva}. We must then correct for this mismatch.

The correction is calculated nonperturbatively by studying the effect of a varying heavy-quark mass in a single ensemble on all correlation functions.
Since the heavy-quark mismatch is small, a linear fit in $1/m_Q$ for each flavor $Q=c$, $b$ and form factor is usually sufficient.
Once the functions that describe the evolution of the form factors with the quark masses are known, we can apply the correction to all other ensembles.
Table~\ref{HQcExample} in \ref{ApHQCorr} gathers all our adjustments.
The error in the correction comes from the error in the heavy-quark-mass tuning procedure and from the error of the linear fit to
find the heavy-quark mass dependence.

\subsection{Chiral-continuum extrapolation}
After applying the renormalization factors and the heavy-quark-mistuning corrections as described in previous subsections, the
resulting form factors still need to be corrected for the fact that they are calculated at nonzero lattice spacing and nonphysical
values for the light-quark masses.
An extrapolation to both the continuum limit and the physical value of the light-quark masses is thus necessary to extract values
that can be used in a physical calculation.
The extrapolation should be based on an appropriate effective field theory (EFT) description of lattice QCD.
The relevant EFT for the calculation at hand is rooted staggered chiral perturbation theory (rS$\chi$PT), which describes how the
form factors behave as the lattice spacing and the light-quark masses approach the desired limits, extended to include heavy-light
observables~\cite{Aubin:2005aq}.
The unquenched MILC configurations generated with 2+1 flavors of improved staggered fermions make use of the fourth-root procedure
for eliminating the unwanted four-fold degeneracy of staggered quarks. At nonzero lattice spacing, this procedure has small violations
of unitarity~\cite{Prelovsek:2005rf,Bernard:2006gj,Bernard:2006zw,Bernard:2007qf,Aubin:2008wk} and locality~\cite{Bernard:2006ee}.
Nevertheless, a careful treatment of the continuum limit, in which all assumptions are made explicit, argues that lattice QCD with
rooted staggered quarks reproduces the desired local theory of QCD as $a\to 0$~\cite{Shamir:2006nj,Shamir:2004zc}.
When coupled with other analytical and numerical evidence (see Refs.~\cite{Durr:2005ax,Sharpe:2006re,Kronfeld:2007ek} for reviews),
this gives us confidence that the rooting procedure is indeed correct in the continuum limit.
We then use the following functions obtained in $\text{SU}(3)$ rS$\chi$PT to lowest nontrivial order in the heavy-quark expansion to fit
the different form factors: 
\begin{align}
    h_Y(a, m, m_s, w) &= \Bigg(K_Y + \frac{\chi_Y(\Lambda_\chi)}{m_c^{k_Y}} + f_Y^w + f_Y^\text{NLO} \nonumber \\
    & \phantom{=} + f_Y^\text{NNLO}\Bigg) \times \left(1 + f_Y^\text{HQ}\right),
    \label{eq:fitfunction}
\end{align}
where $Y=A_1$, $A_2$, $A_3$, and $V$; $K_{A_2}=0$ but $K_Y=1$ otherwise; $k_{A_1}=2$ but $k_Y=1$ otherwise.
These expressions contain the correct dependence in $\chi$PT on the light- and strange-quark masses, the lattice spacing, and the
recoil parameter $w$ at next-to-leading order (NLO). This result expands on the one in Ref.~\cite{Laiho:2005ue} by adding the missing
recoil dependence in the relevant places.
The terms
\begin{align}
    f_Y^{\textrm{NLO}} &= \frac{g^2_{D^\ast D\pi}}{48\pi^2 f_\pi^2 r_1^2} \, \mathrm{logs}^Y_\text{SU(3)}(a, m, m_s, w, \Lambda_\chi) \nonumber \\
    &\phantom{=} + c_{m_1,Y}x_l + c_{a_1,Y}x_{a^2},
    \label{NLOChCt}
\end{align}
introduce nonanalytic dependence on the light and the strange-quark masses through the chiral logarithms\linebreak
$\mathrm{logs}^Y_\text{SU(3)}$.
Those terms also include the leading taste-breaking discretization effects from the light-quark sector.
The explicit expression of the logarithms for each form factor is given in \ref{ApChiral} and includes a dependence
on the recoil parameter $w$.
The coefficient of the chiral logarithms comes from $\chi$PT and it is known, but the current determinations of the coupling
$g_{D^\ast D\pi}$ are not very accurate, hence we fit the coupling with a Gaussian prior $0.53\pm0.08$, compatible with experimental
data~\cite{Anastassov:2001cw,Lees:2013uxa,Lees:2013zna} and lattice-QCD results
\cite{Detmold:2011bp,Can:2012tx,Becirevic:2012pf,Flynn:2015xna,Bernardoni:2014kla,Detmold:2012ge}.
We fix the pion decay constant appearing in the chiral logs in Eq.~(\ref{NLOChCt}), and elsewhere in the fit function to the
three-flavor FLAG 2019 average with the error increased by the estimated 0.7\% charm sea-quark contribution
$f_\pi=130.2\pm 1.2$~MeV~\cite{Aoki:2019cca}.

The other terms in Eq.~\eqref{NLOChCt} introduce analytic NLO corrections in the light-quark masses through\linebreak
$x_l=2B_0m/(8\pi^2f_\pi^2)$ and in the lattice spacing through $x_{a^2} = [a/(4\pi f_\pi r_1^2)]^2$, where $B_0$ is the
low-\-en\-er\-gy-\-con\-stant (LEC) of $\chi$PT that relates the light- and strange-quark masses with the meson masses.
The value of $B_0$ for each lattice spacing is the same as in the earlier analysis at zero
recoil~\cite{Bailey:2014tva}, which uses exactly the same ensembles, and is given in \ref{ApChiral}.
We take into account truncation errors by including the term
\begin{equation}
    f_Y^{\textrm{NNLO}} = c_{c,Y} x_l x_{a^2} + c_{m_2,Y} x_l^2 + c_{a_2,Y} x_{a^2}^2,
    \label{NNLOChCt}
\end{equation}
which describes the dependence on the light-quark\linebreak masses and the lattice spacing $a$ at next-to-next-to-leading order (NNLO), not including logarithmic terms.
According to $\chi$PT power-counting, these analytical terms are expected to have coefficients of $O(1)$, so we take them as fit
parameters with priors $0\pm 1$.
We don't include analytical terms in the strange-quark mass because we do not have data at different values of $m_s$ and nonzero
recoil, and our $h_{A_1}(1)$ result using only zero-recoil data agrees within errors with our previous result from
Ref.~\cite{Bailey:2014tva}.
Also, $\chi$PT predicts a much milder dependence on $m_s$ than on the light-quark masses.

We allow a simple NLO analytical dependence on $w$ to describe the behavior in our small-recoil range through the term
\begin{equation}
    f_Y^w = -  \rho_Y^2(w-1) + \kappa_Y(w-1)^2,
    \label{wChCt}
\end{equation}
where the fit parameters $\rho_Y$ and $\kappa_Y$ are related to the slope and curvature of the form factor $h_Y$ respectively.
We can reasonably expect the slope of the form factors to be roughly $1$, but in order to accommodate substantial deviations from
this value, we set the prior of $\rho_Y$ to $1\pm 2$.
The priors of $\kappa_Y$ are chosen to be $0\pm3$, and the posteriors are compatible with zero within a fraction of a sigma. 

The constant term $\chi_Y(\Lambda_\chi)$ in Eq.~(\ref{eq:fitfunction}) is a LEC of the chiral effective theory, and it is suppressed
for $h_{A_1}$ by a factor of $1/m_c^2$ due to Luke's theorem~\cite{Luke:1990eg}, whereas the other form factors receive
contributions of order $O(1/m_c)$ in the heavy-quark power counting.
The dependence of this LEC on the chiral scale $\Lambda_\chi$ cancels against the dependence of the nonanalytical terms in
Eq.~(\ref{NLOChCt}).
We set the prior of this LEC to $0\pm 1$, except for $h_{A_1}$ where we use $0.0\pm 0.2$ to reflect the suppression due to Luke's
theorem.

The last term in Eq.~\eqref{eq:fitfunction} accounts for the heavy quark discretization errors,
\begin{equation}
    f_Y^\textrm{HQ} = \beta^{\alpha_s a}_Y \alpha_s a\Lambda_{\textrm{QCD}} + \beta_Y^{a^2} a^2\Lambda^2_{\textrm{QCD}} +
        \beta_Y^{a^3} a^3\Lambda^3_{\textrm{QCD}},
    \label{HQChCt}
\end{equation}
with $\beta^p_Y$ the coefficient of the term of order $O(p)$ corresponding to the form factor $h_Y$, and
$\Lambda_\textrm{QCD}=0.6$~GeV for normalization purposes.
In previous articles,\footnote{See, for instance, Ref.~\cite{Bazavov:2011aa}.} we have employed the universal functions described in\linebreak
Refs.~\cite{ElKhadra:1996mp,Oktay:2008ex}.
But in those cases there was only one heavy meson.
Here we need to deal with the $B$ and the $D^\ast$, and our data are not accurate enough to
distinguish the different terms described by the universal functions.
In order to avoid terms that mimic the effect of others, we consider it a better strategy to implement generic discretization error
terms, which would account for the same dependence as the universal functions.
A side effect of this approach is that the heavy- and NLO light-quark discretization effects become
mixed together through terms with the same dependence on the lattice spacing.
To avoid this, we drop the $O(a^2)$ term from Eq.\eqref{HQChCt}, which has the same dependence as the $O(a^2)$ term already present
in Eq.~\eqref{NLOChCt}, and we enlarge the prior of the latter assuming that both corrections are independent, i.e., with a
quadrature sum.
The priors for the $\beta^p_Y$ coefficients are set to $0\pm 1$, but the $O(a^2)$ coefficients of Eqs.~\eqref{NLOChCt}
and~\eqref{HQChCt} have different normalizations.
For this reason, the final prior for $c_{a_2,Y}$ becomes $0.0\pm 6.1$.
This approach does not allow us to distinguish cleanly the origin of the discretization errors, but it accounts for the correct
dependence and size of discretization errors.
However, absorbing the $O(a^2)$ term from Eq.~\eqref{HQChCt} in Eq.~\eqref{NLOChCt} may have further effects, since the correction in
Eq.~\eqref{HQChCt} is applied to the chiral-continuum fit function in Eq.~\eqref{eq:fitfunction}. It is possible that this procedure does not
account for discretization effects in the shapes of the form factors, which would give rise to higher order terms of the form $a^2(w-1)$ and
$a^2(w-1)^2$. We test for such effects by performing two alternate chiral-continuum fits. In the first, we add the terms $x_{a^2}(w-1)$ and
$x_{a^2}(w-1)^2$ to Eq.~\eqref{wChCt}, where the priors for the coefficients of these terms are chosen as $0(1)$. We find that the results
of this fit differ by at most $0.1\sigma$ in their central values from our base fit, while the statistical fit uncertainty is unchanged.
In the second variation, we keep the $O(a^2)$ term in Eq.~\eqref{HQChCt}. In this case, the central values are consistent with those of our
base fit within $0.25\sigma$, again with an unchanged uncertainty. Hence, we conclude that such discretization effects are already
accounted for in our base fit.

Since each ensemble is statistically independent from the others, there are no correlations among them.
On the other hand, we keep track of correlations both between different form factors within the same
ensemble and within the same form factor calculated at different momenta, by combining the jackknife data 
of all form factors into a large, block-diagonal dataset.
Our large statistics allow us to resolve the full covariance matrix without resorting to thinning procedures or
singular-value-decomposition cuts on its eigenvalues.
Nonetheless, we use the shrinkage procedure described in Refs.~\cite{Ledoit2003,Schaefer2005,Ledoit2017,Simone2017} to ensure the
small eigenvalues of the covariance matrix have the correct behavior, and we find that our results do not change with respect to the
analysis without shrinkage.

The systematic errors coming from the heavy-quark mistuning corrections and those introduced by the matching factors are built into
our chiral-continuum extrapolation by constructing the combined covariance matrix,
\begin{equation}
    C_{ij} = C_{ij}^{\rm stat} + \delta_i^{(\rho)}\delta_j^{(\rho)} + \delta_i^{(\kappa)}\delta_j^{(\kappa)}, \label{cCovCor}
\end{equation}
where the first term includes the statistical covariance, and the second and the third ones account for the matching factor and
heavy-quark-mass mistuning correction errors respectively.
The $i,j$ indices run over all form factors, ensembles, and momenta. With $\delta^{(\rho,\kappa)}_i$ we represent either
the shift in the $i^\text{th}$ datum due to a correction the heavy-quark mass ($\kappa$) or the propagated error of the form factor
from the errors in the matching factors ($\rho$) as calculated in Eqs.~\eqref{eq:ZQA1w}--\eqref{eq:ZXVw}.
As a result, the systematic errors introduce new correlations between all data points.
In fact, Eq.~\eqref{cCovCor} assumes the worst case scenario that the matching systematic errors and the errors coming from the
heavy-quark mistuning are 100\% anticorrelated.

\begin{figure*}
  \centering
    \includegraphics[width=0.45\linewidth]{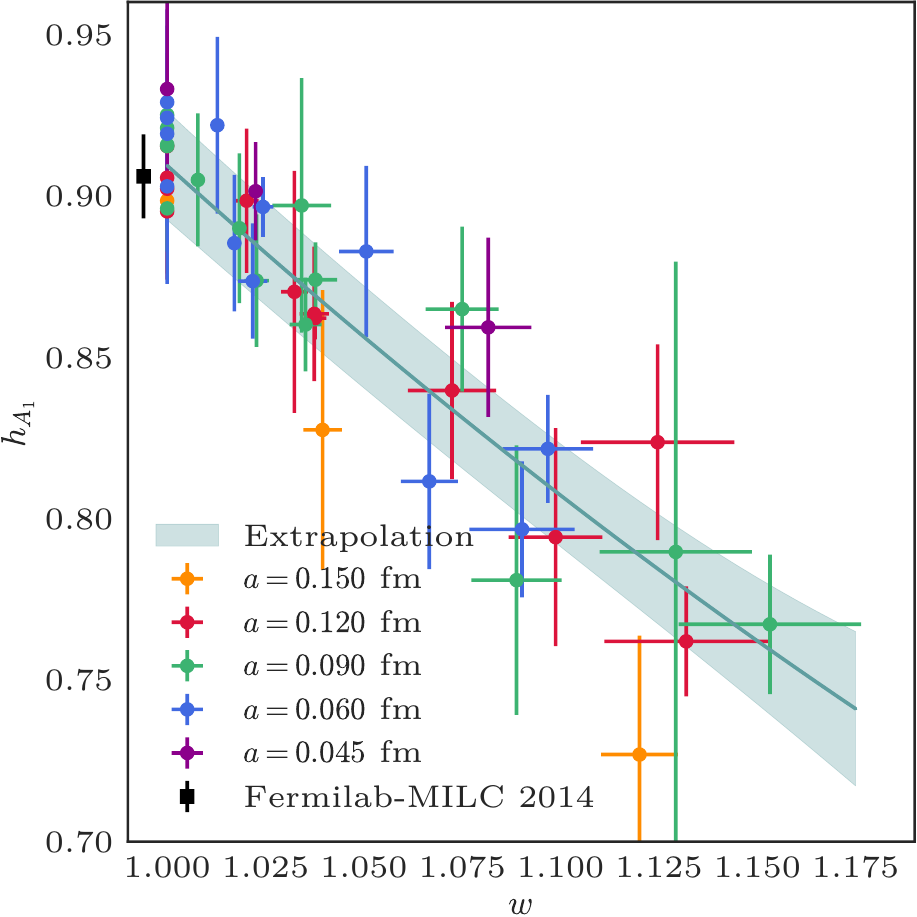} \hfill
    \includegraphics[width=0.45\linewidth]{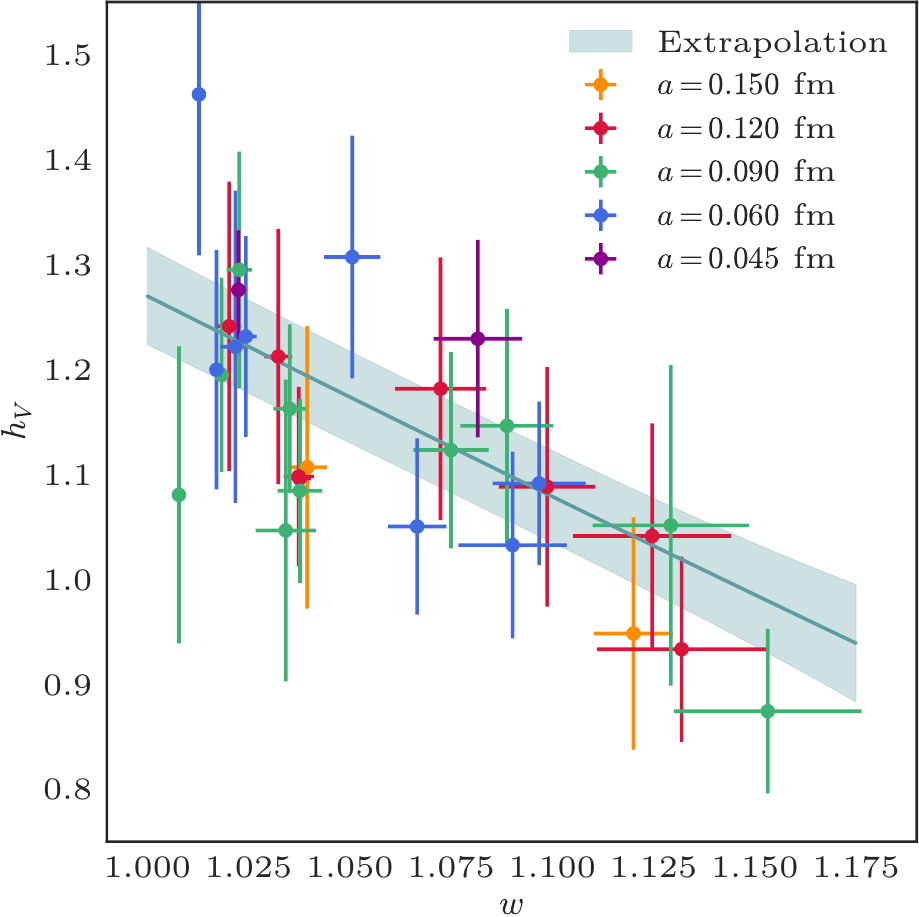}  \\[0.5em]
    \includegraphics[width=0.45\linewidth]{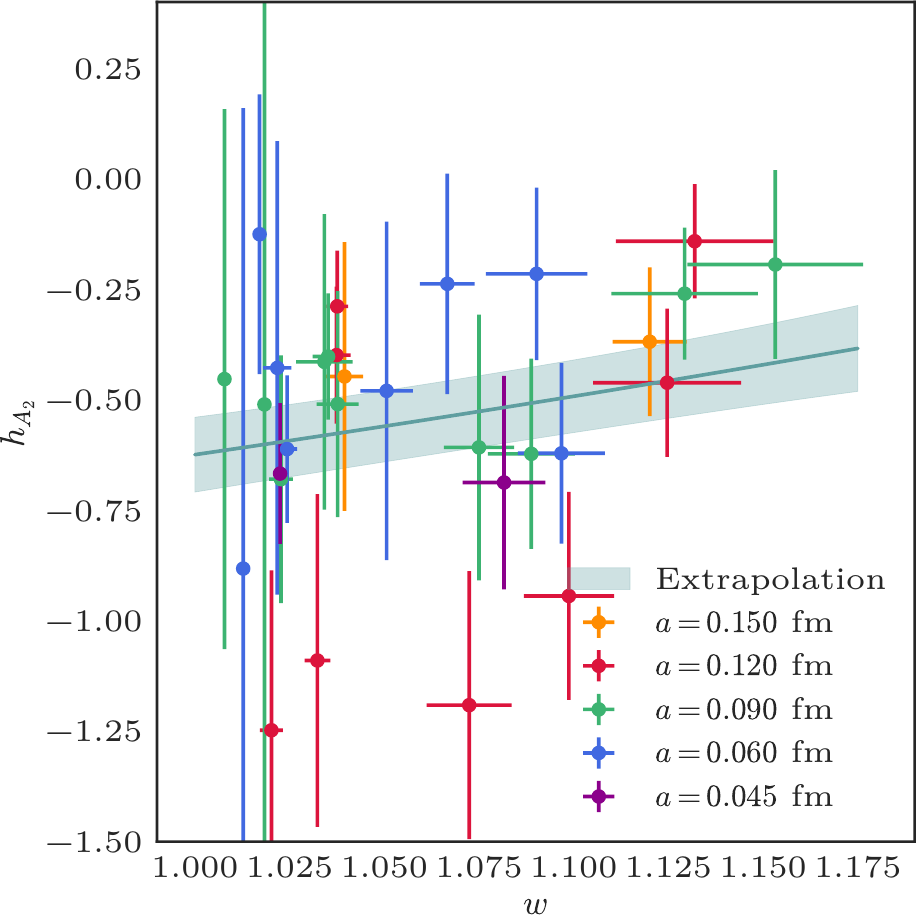} \hfill
    \includegraphics[width=0.45\linewidth]{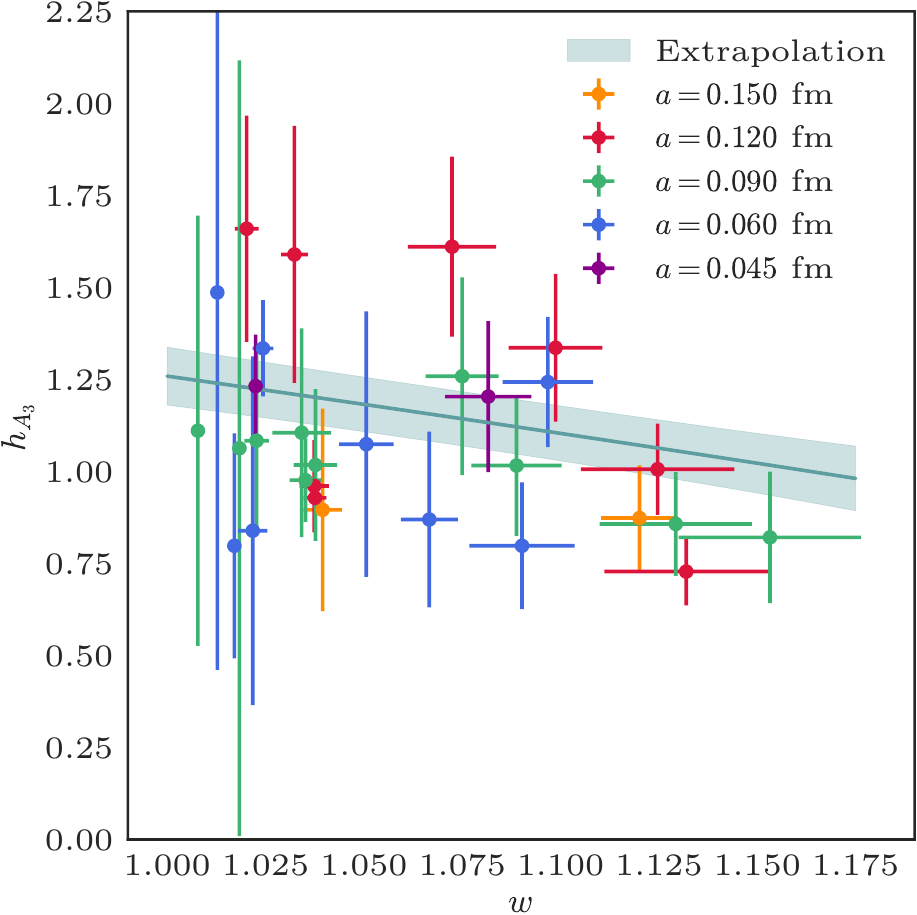}
    \caption{Chiral-continuum extrapolation for the form factors $h_{A_1}$ (top left), $h_V$ (top right), $h_{A_2}$ (bottom left),
        and $h_{A_3}$ (bottom right).
        The color encodes different lattice spacings, whereas the band shows the result of the fit.
        The upper-left plot for $h_{A_1}$ also shows the zero-recoil result from Ref.~\cite{Bailey:2014tva}.
        Note that correlations in $w$ and between form factors play an important role in controlling the final uncertainty.
    }
   \label{PlotFormFactors}
\end{figure*}

The extrapolation results for the four form factors are shown in Fig.~\ref{PlotFormFactors}.
As one can see, $h_{A_1}$, which is protected by Luke's theorem, receives small corrections from~1 at $w=1$.
The other form factors do not enjoy this privilege, and for them the plots show large corrections from the HQET limit.
Figure~\ref{PlotFormFactors} also shows the result of the previous Fermilab-MILC calculation at zero recoil $w=1$ for
comparison~\cite{Bailey:2014tva}.
The agreement is good, although the errors have increased, mainly due to more conservative choices in this work, which stem from the
data at nonzero recoil requiring an extra excited state at source and sink in the ratio calculations, which resulted in larger
errors.
For consistency, we employed the same approach at zero recoil as well.
The deaugmented $\chi^2$/dof of the chiral-continuum extrapolation is $85.2/95$.

\section{Systematic errors}
\label{ErrorBudget}

This section provides specific information on our estimates of every source of systematic error in the determination of the form factors $h_X$.
Even though only $h_{A_1}$ contributes to the decay amplitude at zero recoil, all form factors are nonzero at $w=1$, and their
errors need not be suppressed at small recoil.
Even if the errors of $h_V$, $h_{A_2}$, and $h_{A_3}$ become large, however, their contribution to the decay amplitude and, hence,
the resulting uncertainty in the decay amplitude is still suppressed at small recoil.

\begin{figure*}
  \centering
    \includegraphics[width=0.49\linewidth]{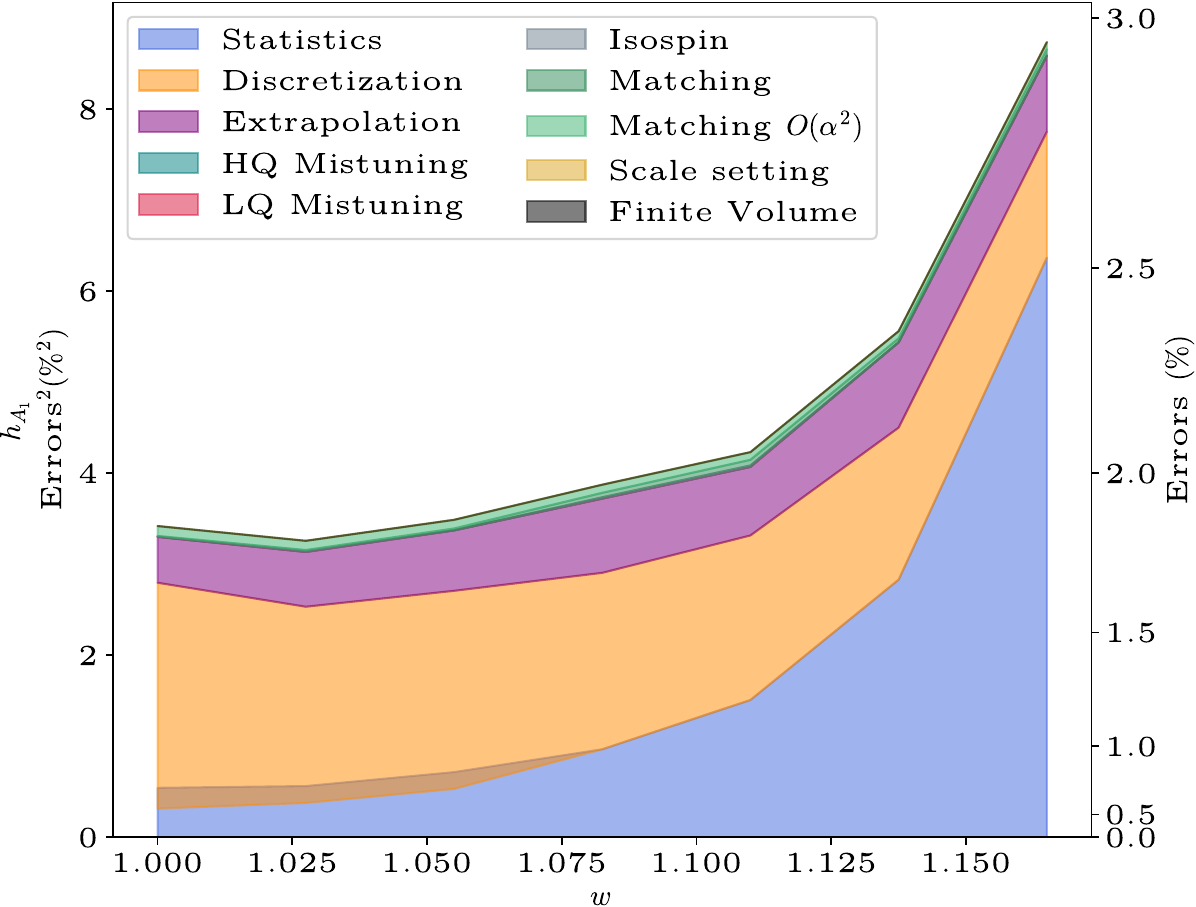} \hfill
    \includegraphics[width=0.49\linewidth]{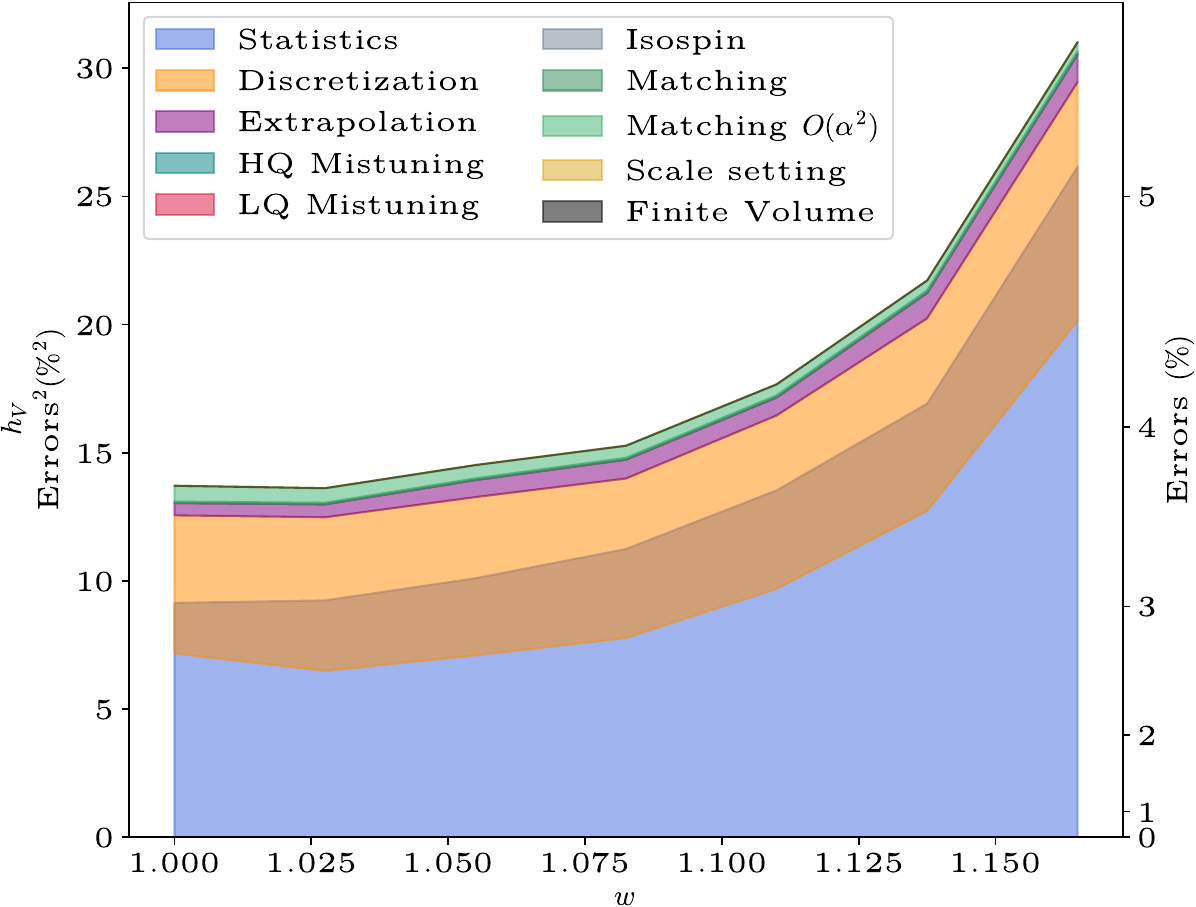}  \\[0.5em]
    \includegraphics[width=0.49\linewidth]{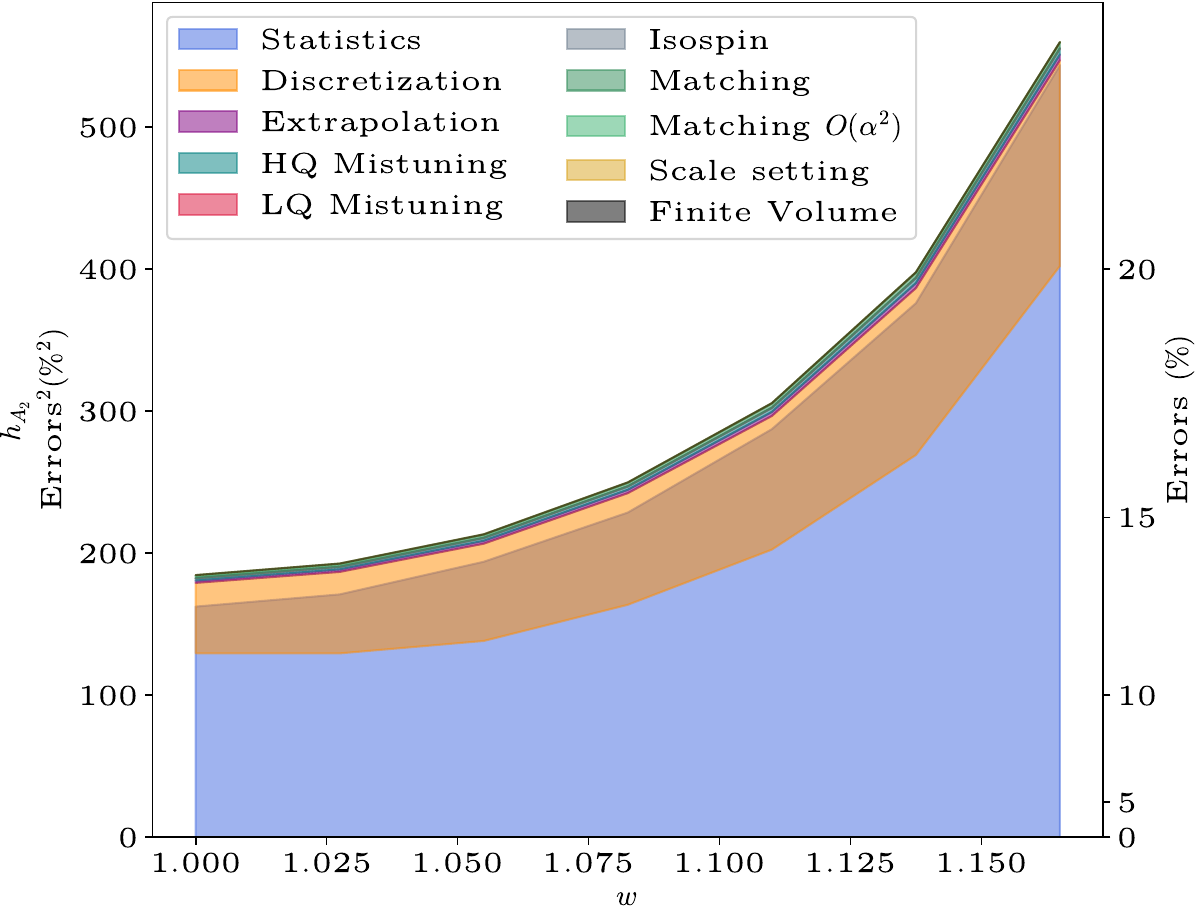} \hfill
    \includegraphics[width=0.49\linewidth]{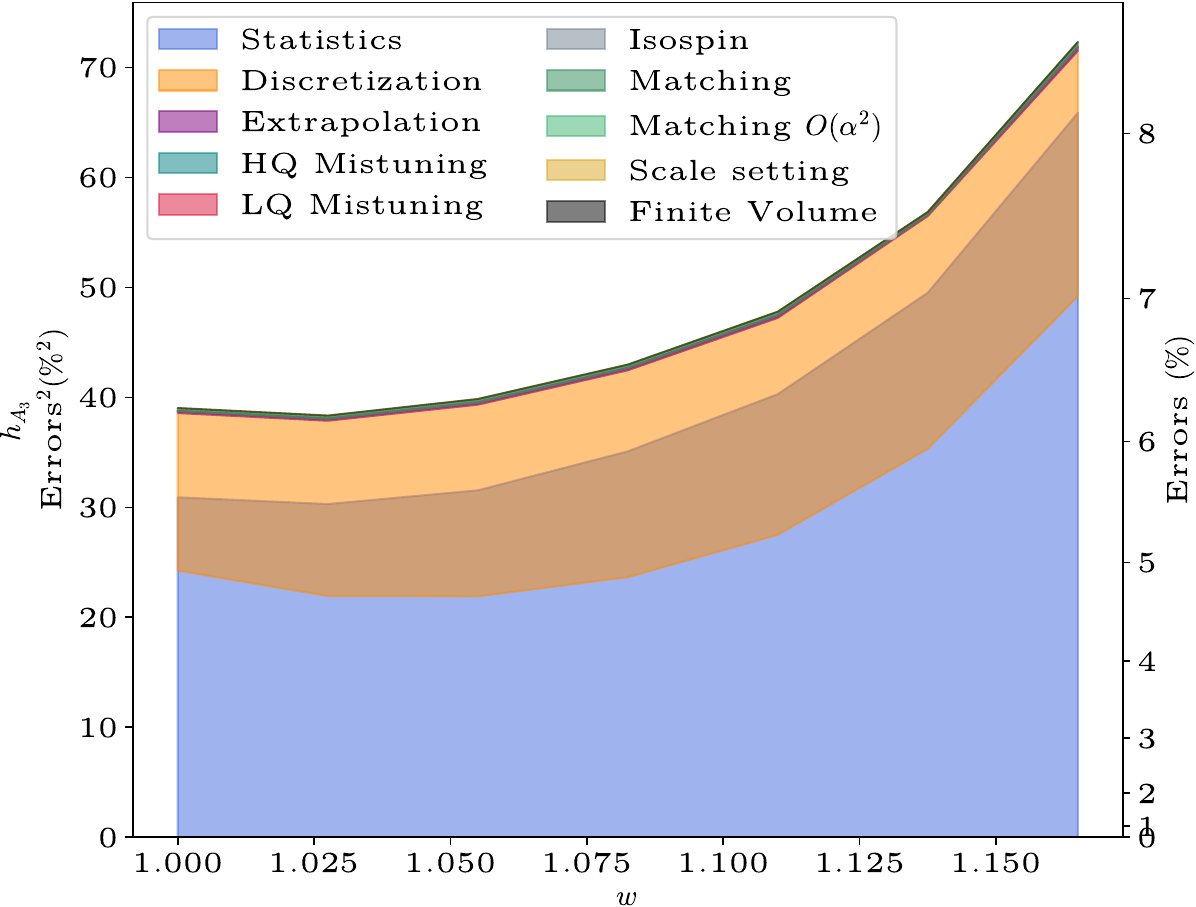}
    \caption{Contributions to the total error of the form factors $h_{A_1}$ (top left), $h_V$ (top right), $h_{A_2}$ (bottom left),
        and $h_{A_3}$ (bottom right) as a function of the recoil parameter~$w$. 
        The two largest contribution come from statistics, in blue, and quark discretization effects, in orange.
        These two contributions overlap in the brown band in the plot, because of the common term of order $a^2$
        in Eqs.~\eqref{NLOChCt} and~(\ref{HQChCt}). 
        The remaining contributions do not overlap.
        Note the differences in vertical scales.}
    \label{PlotSysErrors}
\end{figure*}

Some general features of the uncertainties in the form factors can be understood via HQET.
The form factor $h_{A_1}$ is protected by Luke's theorem~\cite{Luke:1990eg} and, indeed, we find HQET corrections of a few percent.
The form factors $h_V$ and $h_{A_3}$, which are not protected by Luke's theorem, receive HQET corrections at the $\sim 30\%$ level.
The form factor $h_{A_2}$ starts in HQET with terms of order $\alpha_s$ and $1/m_c$, which is roughly consistent with our data,
$h_{A_2}\sim-\frac{1}{2}$.
Figure~\ref{PlotSysErrors} shows the error budget for the different form factors in the continuum as a function of the recoil
parameter.
The relative uncertainty in each form factor follows the same pattern as the HQET corrections: small for $h_{A_1}$, moderate for
$h_V$ and $h_{A_3}$, and large for $h_{A_2}$.
In the last case, the relative uncertainty is large, because the overall value of $h_{A_2}$ is smaller than the others.

Our chiral-continuum extrapolation ansatz to NNLO incorporates errors from statistics, choices in the chiral-continuum
extrapolation, discretization effects,\linebreak $O(am_c \alpha_s)$ matching errors, and heavy-quark parameter mistuning.
Thus, they are all entangled in the fit, and it is not straightforward to extract each particular contribution.
In addition, our treatment of the heavy-quark discretization errors includes a term identical to one of gluon and light-quark
discretization errors.
We can, however, roughly estimate each contribution by making modifications to the fit.
In this spirit, we define the statistical contribution to the error as the error obtained in a NLO fit without mistuning correction
or matching-factor errors included.
We have a specific way to deal with the matching factors, which is explained below.
The contribution coming from the chiral-continuum extrapolations is estimated by comparing the fit errors with and without NNLO
terms.

There are more contributions to the final error that have been taken into account: light-quark mass mistuning, scale setting,
isospin effects, and finite-volume effects.
The final error is taken to be the quadrature sum of these uncertainties with that of the chi\-ral-\-con\-tinuum extrapolation error,
which (again) includes statistical, chiral-continuum extrapolation, discretization, heavy-quark mistuning, and matching errors, as
shown in Table~\ref{TableErrorBudget}.
In the rest of this section, we discuss each source of uncertainty one by one, explaining how they enter this error budget.

\begin{table}
    \scriptsize
    \setlength\tabcolsep{2pt}
    \caption{Error budget for all form factors at $w=1.11$.
        The first row shows the combined error coming from our chiral-continuum fit, which encompasses the statistical errors, the
        matching errors, the systematics due to our chiral-continuum extrapolation, errors coming from HQ-mistuning corrections, and
        discretization errors.
        The next several rows show estimates of the individual contributions (in parentheses as a reminder that they are contained
        in the first row).
        Since the terms that describe the discretization errors come from both the (N)NLO terms in the extrapolation and the HQ
        discretization terms, there is an overlap between the statistical errors (determined as those of a chiral-continuum 
        extrapolation at NLO without any matching or heavy-quark mistuning errors taken into account) and the discretization 
        errors, and the sum in quadrature of the numbers in parenthesis does not equal the first row.
        The remaining rows show other contributions, which are added to the first row in quadrature to obtain the total error in 
        the last row. Dashes represent terms so small that were not included in the final computation of the error.}
    \label{TableErrorBudget}
    \begin{tabular}{lcccc}
    \hline\hline
                      Source           & $h_V(\%)$ & $h_{A_1}(\%)$ & $h_{A_2}(\%)$ & $h_{A_3}(\%)$ \\
    \hline
    Chiral-continuum fit error         &    4.2    &      2.0      &     17.4      &      6.9      \\
    \qquad(Statistics)                 &   (3.7)   &     (1.2)     &    (16.9)     &     (6.3)     \\
    \qquad(Chiral-continuum extr.)     &   (0.8)   &     (0.9)     &     (1.7)     &     (0.5)     \\
    \qquad(LQ and HQ discretization)   &   (2.6)   &     (1.3)     &     (9.7)     &     (4.4)     \\
    \qquad(HQ mistuning)               &   (0.0)   &     (0.0)     &     (1.7)     &     (0.0)     \\
    \qquad(Matching $O(am_c\alpha_s)$) &   (0.3)   &     (0.2)     &     (1.7)     &     (0.5)     \\
    LQ mistuning                       &    0.0    &      0.0      &      0.1      &      0.0      \\
    Matching $O(\alpha_s^2)$           &    0.7    &      0.3      &      0.5      &      0.3      \\
    Scale setting                      &    0.0    &      0.0      &      0.3      &      0.1      \\
    Isospin effects                    &    0.1    &      0.1      &      0.4      &      0.2      \\
    Finite volume                      &    ---    &      ---      &      ---      &      ---      \\
    \hline
    Total error                        &    4.3    &      2.0      &     17.4      &      6.9      \\
    \hline\hline
  \end{tabular}
\end{table}

\subsection{Statistics and stability of the correlator fits}

In principle, the determination of masses, energies, and form factors depends on choices made in fitting the two- and three-point
correlation functions, but we argue that the associated uncertainties are encompassed in the statistical component of the first line
of Table~\ref{TableErrorBudget}.
We have analyzed the two-point functions with both 2+2 and 3+3 states in the fit.
Only when the two results agree within statistical errors do we select a particular fitting range.
In this way, the influence of excited states is reduced below the statistical uncertainty of the $B$~masses and $D^*$ energies.
For the three-point-function ratios, we find that excited states play a more important role, as can be seen for the example of $x_f$
in Fig.~\ref{xfComp}.
When fitting the three-point functions, we therefore include extra states at the source and sink in order to control this potential
source of systematic error.

Bias can arise from the choice of fitting ranges.
To avoid the problems that can come from choosing different fitting ranges for different ensembles, we impose the same
$t_\text{Min}$ in physical units for all two-point correlator fits.
Our $t_\text{Max}$ is chosen differently and varies from ensemble to ensemble,
but the impact of a different $t_\text{Max}$ is much smaller, because these points have much larger errors.
We refer to the reader to Sec.~\ref{twopointfits}, where all the details are explained.
For the form-factor ratio fits, we employ the same range in physical units for all ensembles and most ratios.
For the double ratio $R_{A_1}$ and $x_f$, where the same pattern of states is expected at source and sink, we use a symmetric fit range with
$t_\text{Max}=T-t_\text{Min}$.
Our fits also take into account all correlations between the data points fitted, and, as pointed out in Sec.~\ref{sec:corfunc},
we find that autocorrelations in our data are negligible.

For each correlator fit, we compute a $p$~value from the deaugmented $\chi^2$ and number of degrees of freedom, as explained in~\ref{sec:corfunc}.
We then verify that these $p$~values follow an approximately uniform distribution.

\subsection{Stability of the chiral-continuum extrapolation}

\begin{figure*}
    \centering
    \includegraphics[width=0.45\linewidth]{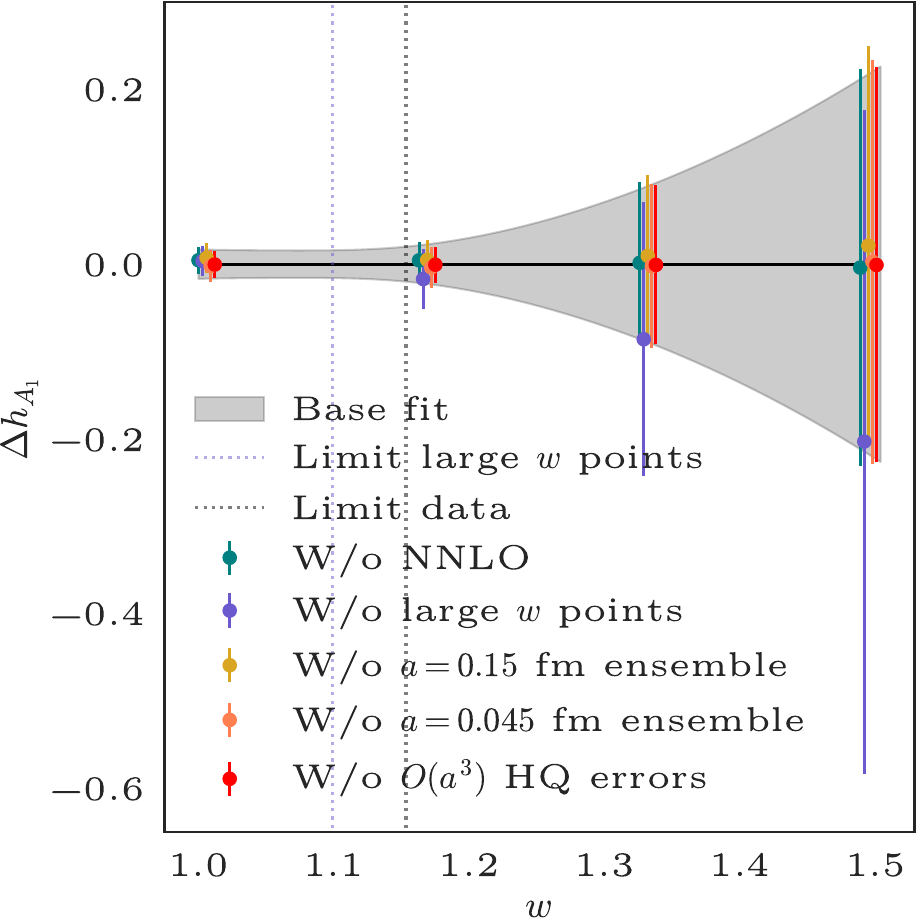} \hfill
    \includegraphics[width=0.45\linewidth]{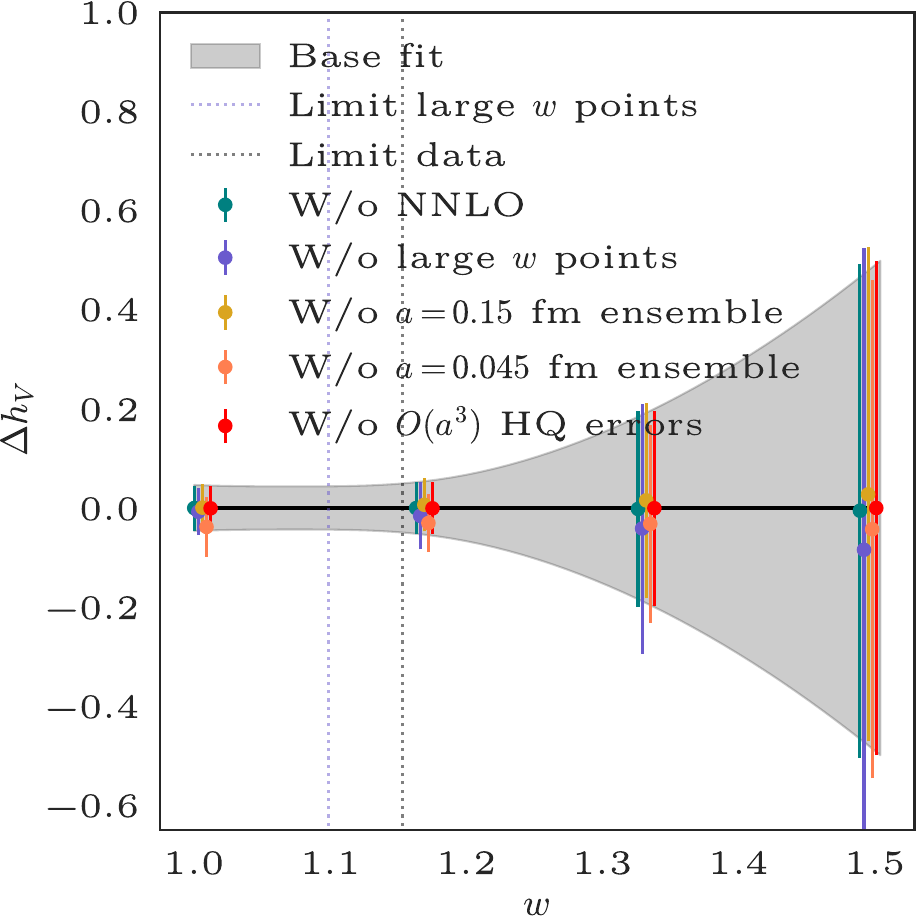}  \\[0.5em]
    \includegraphics[width=0.45\linewidth]{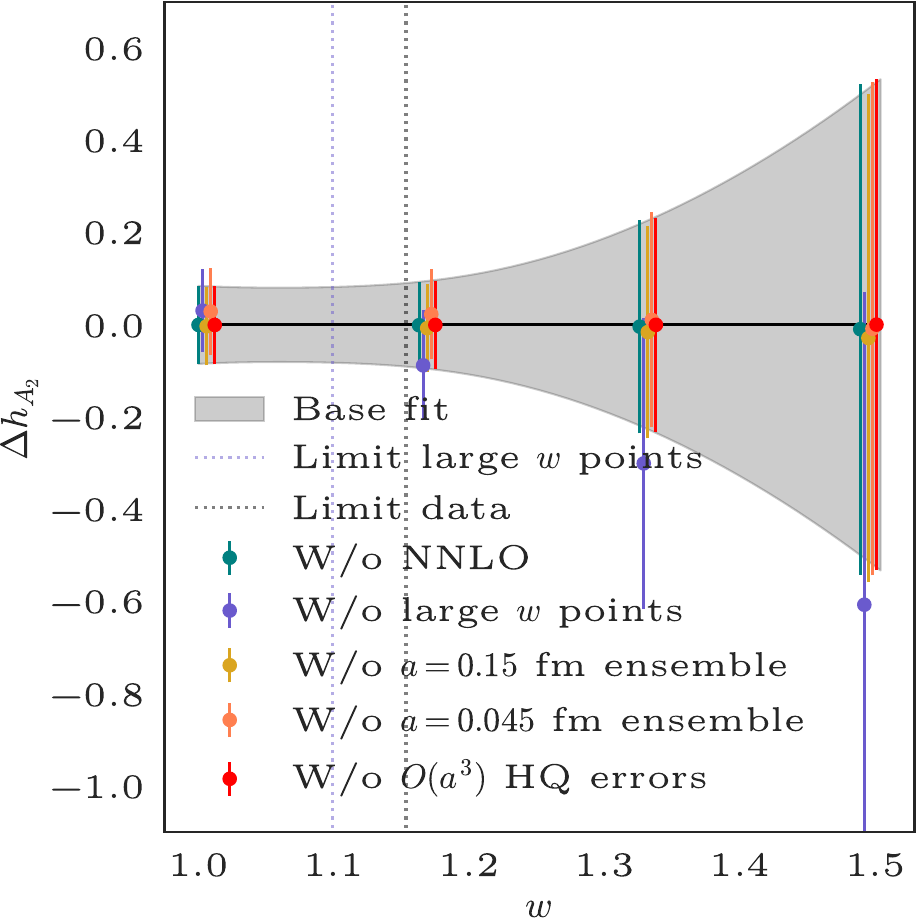} \hfill 
    \includegraphics[width=0.45\linewidth]{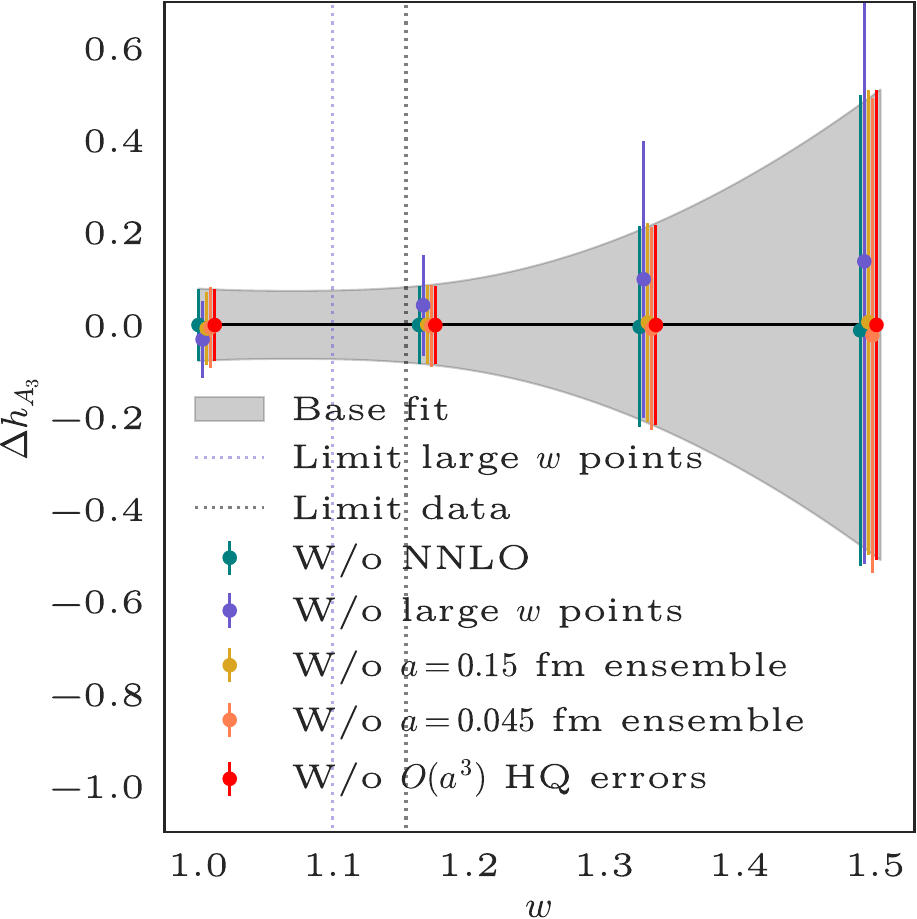}
    \caption{Changes in chiral-continuum-extrapolated values as a result of selected variations in the analysis for $h_{A_1}$ (top
        left), $h_V$ (top right), $h_{A_2}$ (bottom left), and $h_{A_3}$ (bottom right) as a function of the recoil parameter.
        The gray band shows the error band of the base fit.
        The points at $w=1.00$, $1.17$, $1.34$, and $1.50$ show the deviation between the form
        factors calculated from the base fit and each variation of the chiral-continuum fit.
        The dotted vertical lines show the maximum $w$ for which lattice data are available and the limit $w=1.10$ used in one of
        our stability tests.}
    \label{StabFit}
\end{figure*}

To assess the stability of the chiral-continuum extrapolation, we repeat the fit for several different functional forms.
Compared with the base fit, we omit, in turn, the NNLO terms, data at $w>1.10$, the coarsest ensemble, the finest ensemble, and
heavy-quark discretization terms of order $a^3$.
The results are very stable under modifications, as shown in Fig.~\ref{StabFit}.
The most dramatic changes occur when we remove the $w>1.10$ data, leading to shifts of around one standard deviation in the worst
case.
Obviously, the large-recoil extrapolation is affected when removing data at larger recoil, but we find that the $z$~expansion,
discussed below in Sec.~\ref{zExpSec}, stabilizes the final results for $|V_{cb}|$ and $R(D^\ast)$ in this respect.
Table~\ref{chi2Stab} shows that the quality of fit remains good, with $\chi^2/\text{dof}\lesssim1$, in all cases.

\begin{table*}
    \caption{Values of $\chi^2$/dof for several variations in the chiral-continuum extrapolation.}
    \label{chi2Stab}
    \begin{tabular*}{\textwidth}{@{\extracolsep{\fill}}ccccccc}
    \hline\hline
                      & Base     & W/o NNLO  & W/o large $w$ & W/o $a=0.15$ fm & W/o $a=0.045$ fm & W/o HQ $O(a^3)$ \\
    \hline
     $\chi^2$/dof     & 85.2/95  &  86.0/107 &    71.1/75    &     79.4/86     &     81.6/86      &     85.3/99     \\
    \hline\hline
    \end{tabular*}
\end{table*}

\subsection{Discretization errors}

The improved action in the ensemble simulations has light-quark and gluon discretization errors of order $\alpha_s a^2$.
Simple pow\-er-\-counting arguments suggest that the dis\-cret\-i\-za\-tion errors range from $\sim 0.5\%$ in the finest ensemble to $\sim 10\%$
in the coarsest one.
The chi\-ral-\-con\-tin\-u\-um extrapolation describes, however, such terms via the lattice-spacing dependence of the chiral logarithms and
the analytical terms proportional to~$a^2$.
Moreover, the form factors do not seem to be sensitive to the lattice spacing.
Hence, we expect the chiral-continuum extrapolation to take all these errors into account, and no further systematic uncertainties
are added to the final result.

In order to take into account the discretization errors coming from the lattice treatment of the heavy quarks, we include extra
terms in the chiral-continuum extrapolation as shown in Eq.~\eqref{HQChCt}.
These terms are motivated by the HQET description of cutoff effects~\cite{Harada:2001fj,Kronfeld:2000ck}, which uses HQET to
derive the mismatch between the lattice gauge theory at hand and continuum QCD.
The result is a set of functions that depend on the heavy-quark mass, and that can account for discretization effects of different
sizes (in our case, order $\alpha_sa$, $a^2$ and $a^3$).

In our analysis, we would like to introduce these functions for both the $B$ and the $D^\ast$ mesons.
Our data do not, however, distinguish between the contributions of the two mesons, so we instead use a single generic term for both
mesons.
Eq.~\eqref{HQChCt} shows the terms used in the end: $\alpha_sa$, $a^2$, and $a^3$.
Since the $a^2$ term is already included in the light-quark discretization errors, it would be superfluous to include it again here.
The downside of this approach is that it is impossible for us to disentangle light- and heavy-quark discretization errors in the
error budget, and as such, we report them together.

One can estimate the size of these individual effects from variations of the chiral-continuum extrapolation with and without the
terms in Eq.~\eqref{HQChCt}, and also removing the $O(a^2)$ term coming from NLO corrections in Eq.~\eqref{NLOChCt}.
Heavy- and light-quark discretization errors turn out to be the largest contribution to the total error in our analysis, and the
inclusion of the terms listed in Eq.~\eqref{HQChCt} is key in order to account for the heavy-quark systematic errors.

\subsection{Matching errors}

The matching factors are calculated at one-loop order in perturbation theory.
\ref{ApMatch} explains how we estimate the uncertainties, listed in Eqs.~(\ref{eq:ZQA1w})--(\ref{eq:ZXVw}).
They are included in the chiral-continuum extrapolation through Eq.~\eqref{cCovCor}.
We can estimate the error introduced by the uncertainty in the matching up to order $am_c\alpha_s$ by removing the contribution of
the matching factors to Eq.~\eqref{cCovCor}. The effect of higher order contributions is estimated by including an overall factor
$(1 + r^{h_X}_2 \alpha_s^2 + r^{h_X}_3 \alpha_s^3)$ multiplying Eq. \eqref{eq:fitfunction} in the fit and checking the shift in the central
value of the form factor. The priors for the $r^{h_X}_{2,3}$ coefficients are set to $0(1)$; the posteriors have central values and widths close
to those of the priors. We see no impact in including $O(\alpha_s^3)$ terms, but the $O(\alpha_s^2)$ contribute to the final error at the subpercent
level. We collect all observed differences in the corresponding line in Table~\ref{TableErrorBudget}.

\subsection{Heavy quark mistuning}
The form factors are adjusted for the differences between the simulated masses of the heavy quarks and the physical ones before the
chiral-continuum extrapolation.
The correction procedure is detailed in \ref{ApHQCorr}.
The largest correction is about $1\sigma$, but in general the correction is negligible.
Equation~\eqref{cCovCor} includes the contribution of the mistuning in the chiral-continuum extrapolation, therefore, we do not need
to add any further error.
Switching off these corrections gives small variations in the results in our chiral-continuum extrapolation, as shown in
Table~\ref{TableErrorBudget} and Fig.~\ref{PlotSysErrors}.

\subsection{Light quark mistuning}

The endpoint for the light quark masses in the chiral-continuum extrapolation is set to
$r_1m_l=0.003612(126)$ \cite{Bazavov:2010hj}.
We can determine the uncertainty in the form factors coming from a mistuning in the light quark mass by varying $r_1m_l$ within
$1\sigma$ and monitoring its effect on the form factors.
The resulting uncertainty is shown in Table~\ref{TableErrorBudget} and Fig.~\ref{PlotSysErrors}.

\subsection{Scale setting}

In order to determine the relative lattice spacing, we use the distance scale $r_1/a$ defined from the force between static
quarks~\cite{Sommer:1993ce,Bernard:2000gd}, which has been extensively computed~\cite{Bazavov:2009bb}.
Absolute scale setting is taken from the chiral $f_\pi$ analysis of the MILC collaboration~\cite{Bazavov:2011aa}, leading to
$r_1=0.3117(22)~\text{fm}$.
The form factors are dimensionless, so uncertainties from scale setting appear indirectly through the tuning of the heavy-quark
masses, the setting of the light-meson masses in the chiral logarithms, and in the approach to the continuum limit.

We estimate the systematic error associated with $r_1/a$ and $r_1$ by propagating their errors to the final result.
We find that the form factors change only slightly when we vary $r_1$ or $r_1/a$ by $\pm1\sigma$, and we include an extra error
associated to this variation as shown in Table~\ref{TableErrorBudget} and Fig.~\ref{PlotSysErrors}.

\subsection{Isospin effects}
The whole calculation of the form factors has been done assuming isospin symmetry.
The main effect of isospin breaking is to modify the endpoint of the chiral extrapolation through a change in the pion mass.
This effect could bring the endpoint of the extrapolation closer to the $D\pi$-threshold cusp described by the chiral logs.
We estimate the errors introduced by this approximation by varying the endpoint of the extrapolation in the 
pion mass from $m_{\pi^0}$ to $m_{\pi^+}$, by modifying the value of $r_1m_l$ from $0.003612$ to $0.004065$.
While the pion-mass difference is mainly due to isospin breaking QED effects, here we are using it as proxy for the valence quark mass difference, to which we do not have direct access.
%Although the difference between the charged and the neutral pion masses is mainly an isospin breaking QED effect, we are unconcerned with the origin
%of the mass differences, but with the fact that such a difference exists.
%Taking as a lower (higher) limit $r_1m_u=0.002236$ ($r_1m_d=0.004988$), we observe a shift of a fraction of one percent in all form
%factors.
%Our values for $r_1 m_{u,d}$ come from the ratio $m_u/m_d = 0.4529(^{+157}_{-82})$~\cite{Basak:2018yzz}, and the $r_1m_l$ value
%calculated using the asqtad ensembles employed in this work.
Following the resulting difference, we assign an error ranging from 0.0\% to 0.5\%, depending on the form factor and the value of the recoil parameter, as
shown in Fig.~\ref{PlotSysErrors} and Table~\ref{TableErrorBudget}. This increase in the error has no impact in the final result for the form factors.

As an alternative way of estimating these effects, we have also tried to move the endpoint of the extrapolation to isospin symmetric points
with $m_l = m_u$ and $m_l = m_d$. The difference of the values of the form factors between these two endpoints overestimates the isospin breaking errors,
because it includes sea-quark effects that cancel out at first order. The estimate of the isospin breaking errors is larger with this method, but it is
still negligible. Hence we can safely assume that isospin effects are insignificant at our current level of precision.

\subsection{Finite-volume effects}
To estimate finite-volume effects in our heavy-light $\chi$PT description of the form factors, we replace the loop integrals by
discrete sums.
Following Refs.~\cite{Arndt:2004bg} and~\cite{Laiho:2005ue}, we estimate the correction to the integrals in the formulas appearing
in $B\to D^\ast$ at zero recoil to be smaller than $0.01\%$.
This comes from the fact that the contribution of the chiral logarithms to the form factors is quite small.
We have not calculated the corrections at nonzero recoil, and one also expects an increase in the error close to the cusp of the
chiral logs.
Given that $M_\pi L > 4$ on most ensembles, and $M_\pi L\ge 3.7$ always, there is no reason to expect such a large increase in the
error as to make the finite-volume corrections sizable. Hence, we do not assign any additional error due to them.

\section{Determination of \boldmath \texorpdfstring{$|V_{cb}|$}{|Vcb|} and \texorpdfstring{$R(D^\ast)$}{R(D*)}}
\label{Results}
After calculating the form factors, we can reconstruct the decay amplitude using Eq.~\eqref{probAmp} and use experimental data to
extract $|V_{cb}|$.  Similarly, the form factors lead directly to $R(D^\ast)$ via Eqs.~\eqref{RDstDef}, \eqref{DiffDecayRate},
and~\eqref{eq:BR}.  There is a problem: the form factors are obtained only at small values of the recoil parameter, and an
extrapolation to large $w$ with the chi\-ral-\-continuum fit formula would greatly increase the error.  To bring the large $w$
behavior under control we use a standard, model-independent parametrization based on unitarity and analyticity to extrapolate
the form factors to the large recoil region.

Historically the CLN parametrization~\cite{Caprini:1997mu} has been widely used for this process.  However, recent developments have
called into question the reliability of CLN fits, given the high accuracy of the latest experiments and
calculations~\cite{Bigi:2017njr,Grinstein:2017nlq}.  Apart from using outdated data to derive the coefficients of the expansions,
the main criticism of CLN in its most common usage is the lack of error estimates for theoretical ingredients.  Even though the
original CLN article~\cite{Caprini:1997mu} includes equations defining the covariance matrix of the slope and the curvature of the
reference form factor, the final expressions omit this information.  Finally, the strong unitarity constraints, based
on heavy-quark symmetry, play an important role in the CLN parametrization.  Instead of introducing these constraints as an
additional assumption, we would rather perform a fit without imposing them at the outset and then use them after the fits as a
consistency check.

To perform the $z$~expansion we therefore use the completely general BGL
parametrization~\cite{Boyd:1995sq,Boyd:1995cf,Boyd:1997kz}.  Nonetheless, we compare our results with an updated version of CLN in
Sec.~\ref{CLNCompSec}.  For a review on the status of the parametrizations for heavy-to-heavy decays, see
Ref.~\cite{Gambino:2020jvv}.

\subsection{\textit{z} expansion with the BGL parametrization}
\label{zExpSec}

The $z$ expansion is based on a conformal map that takes $w$ to a variable $z$, which remains small over the
physical region for the decay process, namely,
\begin{equation}
    z = \frac{\sqrt{w+1} - \sqrt{2N}}{\sqrt{w+1} + \sqrt{2N}}.
    \label{zParm}
\end{equation}
The value $N=1$ is most commonly used, because it fixes the point $z=0$ at zero recoil, but a symmetric range has been advocated,
with the claim that it reduces errors in the expansion~\cite{Caprini:1997mu,Boyd:1997kz}.  With $N=1$, the maximum recoil point for
massless leptons $w_\text{Max}\approx1.503$ becomes $z_\text{Max}\approx0.056$, so indeed $z\ll1$ in any case, and any reasonable
expansion in $z$ with coefficients of order~$1$ converges with a few terms.  The conformal map given in Eq.~\eqref{zParm} also
pushes the branch cut in the $w$ plane onto the unit circle, $|z|=1$, and subthreshold poles onto the real axis near $z=-1$.  Since
the nearest threshold is very far ($|z|\sim1$) from the valid kinematic range ($0\le z\lesssim 0.056$) and higher-energy thresholds
even farther, there is no advantage in using an alternative parametrization, such as the one proposed in Ref.~\cite{Bourrely:2008za},
that would take care of the behavior at such values of $z$.

The BGL parametrization does not apply directly to the $h_X$ form factors, but to combinations with definite
spin-parity~\cite{Gambino:2020jvv}:
\begin{align}
  g =& \frac{h_V}{M_B \sqrt{r}}, \label{eq:BGL:g}  \\
  f =& M_B \sqrt{r}(1+w) h_{A_1}, \\
  \mathcal{F}_1 =& M_B^2 \sqrt{r} (1+w) \Big[(w-r)h_{A_1} \nonumber \\
     &\phantom{=}\hspace*{3em} - (w-1)\left(rh_{A_2} + h_{A_3}\right)\Big], \\
  \mathcal{F}_2 =& \frac{1}{\sqrt{r}} \Big[(1+w)h_{A_1} + (rw-1)h_{A_2} + (r-w)h_{A_3}\Big], \label{eq:BGL:F2}
\end{align}
where $r = M_{D^\ast}/M_B$.  These form factors are proportional to the helicity amplitudes $H_--H_+$, $H_++H_-$, $H_0$, and $H_S$,
respectively [cf., Eqs.~(\ref{eq:helicity:T})--(\ref{eq:helicity:S})].  Thus, the form factor $\mathcal{F}_2$ is important only with
massive leptons, in particular in the determination of $R(D^\ast)$.

The BGL parametrization expresses the dependence of the form factors on $z$ as
\begin{equation}
f_i(z) = \frac{1}{P_i(z)\,\phi_i(z)}\sum_{j=0}^\infty a_{i,j} z^j,
\label{pBGL}
\end{equation}
where the functions $P_i(z)$ are called \emph{Blaschke factors}, and the $\phi_i$ are known as \emph{outer functions}.  As discussed
below, wise choices of $\phi_i(z)$ make the coefficients of the expansion $a_{i,j}$ of order~$1$ and ensure rapid convergence of the
series.  The Blaschke factors are given by
\begin{equation}
P_i(z) = \prod_p\frac{z-z_p}{1 - zz_p},
\end{equation}
with
\begin{equation}
z_p(M_p, N) = \frac{\sqrt{\left(1 + r\right)^2 - \frac{M_p^2}{M_B^2}} - \sqrt{4Nr}}%
{\sqrt{\left(1 + r\right)^2 - \frac{M_p^2}{M_B^2}} + \sqrt{4Nr}}.
\label{eq:pole-position}
\end{equation}
They include the explicit poles with mass $M_p$ below the $BD^\ast$ threshold and with the appropriate quantum numbers.
Table~\ref{InnerParms} shows the poles we use for the BGL form factors.  Although some analyses employ four $1^-$ resonances, the
fourth one is very far from $z>0$ and its value uncertain.  We therefore follow Ref.~\cite{Bigi:2017jbd} and use only three.

The $z$-expansion coefficients of the BGL form factors are then defined via
\begin{align}
      g       =& \frac{1}{P_{1^-}(z)\,\phi_g              (z)}\sum_{j=0}^\infty a_j z^j,\label{gConst}  \\
      f       =& \frac{1}{P_{1^+}(z)\,\phi_f              (z)}\sum_{j=0}^\infty b_j z^j,\label{fConst}  \\
\mathcal{F}_1 =& \frac{1}{P_{1^+}(z)\,\phi_{\mathcal{F}_1}(z)}\sum_{j=0}^\infty c_j z^j,\label{F1Const} \\
\mathcal{F}_2 =& \frac{1}{P_{0^-}(z)\,\phi_{\mathcal{F}_2}(z)}\sum_{j=0}^\infty d_j z^j,\label{F2Const} 
\end{align}
where the Blaschke factors' subscripts denote the $J^P$ of the $l\nu$ final state.

\begin{table}
    \centering
    \caption{Poles for the Blaschke factors, taken from Ref.~\cite{Bigi:2017jbd} and references therein.
         For $J^P=1^-$, $0^-$ ($1^+$), the first two (first only) resonances are well determined from either
         a lattice calculation or experimental measurements. All other masses are based on model estimates.}
    \label{InnerParms}
    \begin{tabular}{@{\extracolsep{\fill}}ccl}
        \hline\hline
        Form factor          & $J^P$ & \multicolumn{1}{c}{Masses $M_p$ (GeV)} \\
        \hline
        $g$ \Tstrut          & $1^-$ & 6.329, 6.920, 7.020        \\
        $f$, $\mathcal{F}_1$ & $1^+$ & 6.739, 6.750, 7.145, 7.150 \\
        $\mathcal{F}_2$      & $0^-$ & 6.275, 6.842, 7.250 \Bstrut \\
        \hline\hline
    \end{tabular}
\end{table}

Setting $N=1$, we choose the outer functions to be~\cite{Boyd:1995sq,Boyd:1995cf,Boyd:1997kz}
\begin{align}
    \phi_g               =&      16r^2      \sqrt{\frac{n_I}{3\pi\tilde{\chi}^T_{1^-}(0)}}
        \frac{(1+z)^2(1-z)^{-\frac{1}{2}}}{\left[(1+r)(1-z) + 2\sqrt{r}(1+z)\right]^4},
    \label{gOuter}  \\
    \phi_f               =& \frac{4r}{M^2_B}\sqrt{\frac{n_I}{3\pi      {\chi}^T_{1^+}(0)}}
        \frac{(1+z)  (1-z)^{ \frac{3}{2}}}{\left[(1+r)(1-z) + 2\sqrt{r}(1+z)\right]^4},
    \label{fOuter}  \\
    \phi_{\mathcal{F}_1} =& \frac{4r}{M^3_B}\sqrt{\frac{n_I}{6\pi      {\chi}^T_{1^+}(0)}}
        \frac{(1+z)  (1-z)^{ \frac{5}{2}}}{\left[(1+r)(1-z) + 2\sqrt{r}(1+z)\right]^5},
    \label{F1Outer} \\
    \phi_{\mathcal{F}_2} =&   8\sqrt{2}r^2  \sqrt{\frac{n_I}{ \pi\tilde{\chi}^L_{1^+}(0)}}
        \frac{(1+z)^2(1-z)^{-\frac{1}{2}}}{\left[(1+r)(1-z) + 2\sqrt{r}(1+z)\right]^4},
    \label{F2Outer} 
\end{align}
where the undefined symbols under the square root are given in Table~\ref{OuterParms}, along with the numerical values we use for
$M_B$ and $M_D^\ast$.
In testing the effect of this pole and various other choices for the pole positions, we find that although the numerical values of
the $z$-expansion coefficients depend on the details, the final curves for the form factors are largely independent of such choices.

With these outer functions, the coefficients of the expansion satisfy the following unitarity
constraints~\cite{Boyd:1995sq,Boyd:1995cf,Boyd:1997kz},
\begin{equation}
\sum_{j=0}^\infty a_j^2 \lesssim 1,\quad\sum_{j=0}^\infty \left(b_j^2 + c_j^2\right) \lesssim 1,\quad\sum_{j=0}^\infty d_j^2 \lesssim 1,
\label{UniCons}
\end{equation}
where the $\lesssim$ symbols reflect the fact that the values for the $\chi$ factors in Table~\ref{OuterParms} are not exact,
so the bounds are not precisely known.  These constraints provide information from unitarity and analyticity, which can be used
together with the output of the chiral-continuum extrapolation.

In this analysis, we do not impose the unitarity constraints, Eq.~\eqref{UniCons}, on the BGL coefficients,
but we check that the final results comply with the constraints within errors.
We note that uncertainties in the values of $\chi_{J^P}^{T,L}$ provided in Table~\ref{OuterParms} complicate strict unitarity constraints on
the $z$~expansion coefficients. Still, we find that an implementation of the unitarity constraints using hard cutoffs (following,
for instance, Ref.~\cite{Ferlewicz:2020lxm}) leaves the fit results essentially unchanged.

\begin{table}
    \centering
    \caption{Inputs for the outer functions, taken from Ref.~\cite{Bigi:2017jbd} and references therein.
    The $\chi$ parameters are calculated in perturbative QCD through $O(\alpha^2_s)$, and depend on
    charm and bottom quark mass inputs, see Ref.~\cite{Bigi:2016mdz}.}
    \label{OuterParms}
      \begin{tabular}{cc}
        \hline
        \hline
            Input     \Tstrut          &         Value         \\
\hline
$M_{D^\ast}$ (GeV)    \Tstrut          &         2.010         \\
$M_B$        (GeV)                     &         5.280         \\
$n_I$                                  &          2.6          \\
$\chi^T_{1^+}(0)$         (GeV$^{-2}$) & $3.894 \times10^{-4}$ \\
$\tilde{\chi}^T_{1^-}(0)$ (GeV$^{-2}$) & $5.131 \times10^{-4}$ \\
$\tilde{\chi}^L_{1^+}(0)$\Bstrut       & $1.9421\times10^{-2}$ \\
        \hline
        \hline
      \end{tabular}
\end{table}

There are two kinematic relations between the form factors, one at zero recoil and another at maximum recoil,
\begin{align}
\mathcal{F}_1(1)           =& M_B(1-r)f(1),                                           \label{KinF1} \\
\mathcal{F}_2(w_\text{Max}) =& \frac{1+r}{M_B^2(1+w_\text{Max})(1-r)r}\mathcal{F}_1(w_\text{Max}). \label{KinF2}
\end{align}
These constraints follow trivially from the HQET basis of form factors (the $h_X$), but the BGL parametrization does not
automatically impose them.\footnote{It is interesting to note that the CLN parametrization does include the constraint at zero
recoil given in Eq.~\eqref{KinF1}.  The constraint at maximum recoil, Eq.~\eqref{KinF2}, is not imposed in CLN, and indeed
does not hold unless the original CLN expressions are modified.}
Equation~\eqref{KinF1} is straightforward to implement for the BGL parametrization with $N=1$ in Eq.~\eqref{zParm}, since it amounts
to a relationship between the $b_0$ and the $c_0$ coefficients of the expansion,
\begin{equation}
\frac{1-r}{\sqrt{2}(1+\sqrt{r})^2} b_0 = c_0.
\end{equation}
On the other hand, Eq.~\eqref{KinF2} can be imposed by adding an extra data point that enforces the constraint.  Alternatively, we
can remove $d_0$ and write it as a function of the remaining $d_j$ and $c_j$. The two approaches give compatible
results.  In our final value, we choose not to impose the second constraint. The results of the chiral-continuum extrapolation
trivially build in both constraints, so we would expect any well-behaved expansion to keep this property, even when extrapolating to
the whole recoil range.  Indeed, the zero-recoil constraint is satisfied to very high accuracy, even if we do not impose it.  This
happens because the $z$~expansion is quite constrained by the lattice-QCD values. For the maximum-recoil constraint, we just check for
compatibility within errors.

\subsubsection{Synthetic data}
\label{sec:synth}

The output of the chiral-continuum extrapolation is not a set of points, but a set of functions that express the form factors at any
value of the recoil parameter.  In order to make the results amenable to a BGL fit with experimental data, we use \emph{synthetic
data}, together with their covariance, based on selected central values of the chiral-continuum fit, evaluated at zero lattice
spacing and physical quark masses.  The selected values with correlations are included in the ancillary files, as explained in
\ref{ApFullRes}.  We choose data points at three $w$ values, $\{1.03, 1.10, 1.17\}$, as representative of the span of our
lattice-QCD data, but we have checked that varying these values does not change significantly the curves generated for the form factors
from the $z$~expansion, as long as there are no $w$ values too close to $w=1$.  This robust behavior is not surprising, since the
full covariance matrix is available, and hence the same amount of information is provided.  The covariance matrix of the lattice
data is well defined for all recoil values, but at $w=1$ the kinematic constraint given in Eq.~\eqref{KinF1} is exactly
satisfied. Therefore, the form factors are no longer independent as $w\to1$, and the covariance matrix becomes singular.
The singularity can be avoided by choosing any value slightly larger than $w=1$.  Also, we cannot push $w$ much higher than $1.17$
without going outside the region where lattice-QCD data are available, because the uncertainty grows rapidly.
We have selected three points per form factor because in the continuum limit there are only twelve free independent functions in our
chiral-continuum extrapolation, three per form factor.  Adding more points does not increase the accuracy of the $z$-expansion fits,
because the new points are not independent.

We first carry out BGL fits to our lattice-QCD form factors. The constant and linear coefficients are well determined by the data,
and no prior constraints are used for them. The quadratic and cubic coefficents are constrained with priors $0(1)$ to stabilize the fit.
Keeping terms up to quadratic (linear) order in $z$ and imposing the kinematic relation given in Eq.~\eqref{KinF1}, leaves just one (three)
degree(s) of freedom. Since unitarity constrains the size of the coefficients, we can include cubic terms with the unitary-inspired 
priors to check stability of the results against truncation effects.  Table~\ref{SynthResults} gathers the coefficients for these three versions of the $z$~expansion, with
the $\chi^2/\text{dof}$ and unitarity sums.  All fits satisfy the unitarity constraints within errors.
The unitarity sums are computed by taking the median of the distributions obtained from the Gaussian posteriors, along with the confidence levels, $\pm 34.1\%$, for the uncertainties.
We note that the distributions of the squares are not Gaussian when the errors on the posteriors are large. This is the case for coefficients that are not well determined by the underlying lattice data.
The kinematic constraint at $z=0$, Eq.~\eqref{KinF1}, is satisfied to very high accuracy, even when it is not imposed.
On the other hand, the constraint at $z_\text{Max}$, Eq.~\eqref{KinF2}, is satisfied only within approximately one standard
deviation, unless, of course, it is imposed.  
%%% remove repetition, as this point is made below. 

%The quadratic and unitarity-constrained cubic $z$~expansions are in essentially perfect agreement, indicating that the expansions have saturated.

\begin{table}
  \scriptsize
  \setlength\tabcolsep{2pt}
  \caption{Results of linear, quadratic, and unitarity-constrained cubic $z$ expansions using only lattice-QCD data.
           The coefficient $c_0$ is fixed by the constraint given in Eq.~\eqref{KinF1}, and it is shown for convenience.}
   \label{SynthResults}
    \begin{tabular}{cS[table-format = +1.6(2)]S[table-format = +1.6(2)]S[table-format = +1.6(5)]}
     \hline\hline
     \Tstrut               &     {Linear}     &     {Quadratic}     &      {Cubic}     \\
     \hline
     $a_0$                 &    0,0330(12)    &      0,0330(12)     &    0,0330(12)    \\
     $a_1$                 &    -0,157(52)    &      -0,156(55)     &    -0,155(55)    \\
     $a_2$                 &                  &       -0,12(98)     &     -0,12(98)    \\
     $a_3$                 &                  &                     &    -0,004(1000)  \\
     \hline
     $b_0$                 &   0,01229(23)    &     0,01229(23)     &   0,01229(23)    \\
     $b_1$                 &    -0,002(10)    &      -0,003(12)     &    -0,003(12)    \\
     $b_2$                 &                  &        0,07(53)     &      0,05(55)    \\
     $b_3$                 &                  &                     &    -0,01(100)    \\
     \hline
     $c_0$                 &  0,002059(38)    &    0,002059(38)     &  0,002059(38)    \\
     $c_1$                 &   -0,0057(22)    &     -0,0058(25)     &   -0,0057(25)    \\
     $c_2$                 &                  &      -0,013(91)     &     -0,02(10)    \\
     $c_3$                 &                  &                     &      0,10(95)    \\
     \hline
     $d_0$                 &    0,0508(15)    &      0,0509(15)     &    0,0509(15)    \\
     $d_1$                 &    -0,317(59)    &      -0,328(67)     &    -0,327(67)    \\
     $d_2$                 &                  &       -0,02(96)     &     -0,02(96)    \\
     $d_3$                 &                  &                     &  -0,0006(10000)  \\
     \hline
%$\chi^2/\text{dof}$\Tstrut &{0.83/5\Hs{6.5mm}}&   {0.64/3\Hs{5mm}}  & {0.64/3\Hs{1cm}} \\
$\chi^2/\text{dof}$\Tstrut &{0.83/5\Hs{6.5mm}}&   {0.63/1\Hs{5mm}}  & {0.61/-3\Hs{1cm}} \\
$\sum_i^N a_i^2$           &   0,026\epm{.9}{0.019}{0.013}  & 0,47\epm{1}{1.51}{0.36} & 1,4\epm{1.6}{2.3}{1.0} \\
$\sum_i^N (b_i^2 + c_i^2)$ & 2,4\epm{1.2}{1.8}{0.6}\cnt{-4} & 0,13\epm{1}{0.44}{0.10} & 1,6\epm{1.6}{2.2}{1.1} \\
$\sum_i^N d_i^2$\Bstrut    &   0,103\epm{.9}{0.041}{0.034}  & 0,54\epm{1}{1.40}{0.38} & 1,4\epm{1.6}{2.2}{1.0} \\
%$\sum_i^N a_i^2$           &     0,026(16)    &       0,04(24)      &      0,04(24)    \\
%$\sum_i^N (b_i^2 + c_i^2)$ &  0,000193(69)    &      0,005(70)      &      0,01(18)    \\
%$\sum_i^N d_i^2$\Bstrut    &     0,103(37)    &      0,110(61)      &     0,110(52)    \\
     \hline\hline
    \end{tabular}
\end{table}
The cubic coefficients  ($a_3$, $b_3$, $c_3$, $d_3$) shown in Table~\ref{SynthResults} are not well 
determined by our lattice data, resulting in unitarity sums, with central values $>1$, while still consistent with unitarity within error.
However, the cubic fit provides useful information on truncation effects.
We see that the lower-order coefficients in the cubic fit are in very good agreement with the coefficients in the quadratic fit.
Further, we find, that the cubic coefficients have little effect on the decay rate and the form factors, as expected, since $|z|\ll1$.
In contrast, and as Table~\ref{SynthResults} shows, the linear expansion leads to coefficients with nearly the same central values as
in the quadratic and cubic fits, but with slightly smaller errors, suggesting an underestimation of the truncation error.
We conclude that the errors coming from the truncation of the series in Eqs.(\ref{gConst})--(\ref{F2Const}) are already included in
the uncertainties on the coefficients of the quadratic fit, which we choose for the main result of the $z$ expansion.
The full correlation matrix of the quadratic fit, which can be used to reconstruct the output of the $z$~expansion,
is included in the ancillary files in binary format, as explained in \ref{ApFullRes}.
Form factors from this information can be used in phenomenology with no assumption about the presence of new physics.
Below we discuss $z$ fits incorporating shape information from experiment, which are more precise but possibly contaminated
by new physics in the light semileptonic channel.

Experimentalists~\cite{Waheed:2018djm,Dey:2019bgc} usually adjust the order of the $z$ expansion to allow unconstrained fits to the
form factors without violating unitarity.
Following this criterion we find that we can remove the $a_2$ and the $d_2$ coefficients, and obtain the same fit results as in our
quadratic fit without including any $0(1)$ priors for the higher order coefficients.
This ensures that our utilization of priors for some coefficients indeed does not influence the fit results, and that their only function
is to stabilize the fit.

\subsubsection{Functional method}

A functional method can also be used to fit the result of the chiral-continuum extrapolation to the BGL
parametrization~\cite{Lattice:2015tia}.  The method exploits the fact that the chiral-continuum fit functions are linear in the fit
parameters.  The covariance in the fit parameters is then easily converted to the covariance of the values of the resulting fitted
form factors (at zero lattice spacing and physical quark masses) at any pair of recoil parameters $(w,w')$.  Through the BGL
parametrization, this covariance is then converted to a covariance in the form factors values at any $z$~pair, $(z,z')$.  The method
has the esthetic property that information from the best fit continuum form factors is spread over the entire physical region in
$z$, rather than at a few arbitrarily chosen discrete points.

We have compared form factors from the functional approach with those from the synthetic data.  They show no discernible difference
in the form factors, implying that the systematic errors associated with the choice of synthetic data
from the chiral-continuum extrapolation are very small.  Since the functional fits do not provide any new insight, and they make it
difficult to combine data from several sources, we focus on the synthetic-data results in the rest of the paper.

\subsection{Determination of \texorpdfstring{\boldmath$|V_{cb}|$}{|Vcb|}}
\label{sec:Vcb}

The lattice-QCD form factors can be used in conjunction with experimental data to perform a joint fit to the BGL parametrization,
with an additional fit parameter for the relative normalization, which is nothing but $|V_{cb}|$.  In these fits, the low-recoil
behavior is determined by lattice QCD, and the large-recoil behavior by experiment.  As experimental input, we use the 2018 raw
dataset from Belle~\cite{Waheed:2018djm} and the synthetic data generated from the 2019 BaBar analysis~\cite{Dey:2019bgc}.
These data are combined with the lattice-QCD synthetic data.
We do not use Belle's 2017 tagged dataset~\cite{Abdesselam:2017kjf}, because it is still unpublished.

Experiments extract the fully differential decay rate, not only with respect to the recoil parameter, but also to all angular
variables in the decay chain $B\to D^\ast\ell\nu$, $D^\ast\to D\pi$~\cite{Bigi:2017njr,Waheed:2018djm,Dungel:2010uk},

\bOneCol
\begin{align}
    \frac{d\Gamma}{dw\,d\cos{\theta_v}\,d\cos{\theta_\ell}\,d\chi} =& \left|V_{cb}\right|^2 \left|\eta_\text{EW}\right|^2
        \frac{3G_F^2M_B^5}{1024\pi^4} r^3 \sqrt{w^2-1}(1-2wr+r^2) \times \nonumber \\
     & \left[\left(1-\cos{\theta_\ell}\right)^2\sin^2{\theta_v}H^2_+(w) +
         \left(1+\cos{\theta_\ell}\right)^2\sin^2{\theta_v}H^2_-(w) +
         4\sin^2{\theta_\ell}\cos^2{\theta_v}H_0^2(w)\right. \nonumber \\
     & - 2\sin^2{\theta_\ell}\sin^2{\theta_v}\cos{2\chi}H_+(w)H_-(w) -
     4\sin{\theta_\ell}\left(1-\cos{\theta_\ell}\right)\sin{\theta_v}\cos{\theta_v}\cos{\chi}H_+(w)H_0(w) \nonumber \\
     & \left.+ 4\sin{\theta_\ell}\left(1+\cos{\theta_\ell}\right)\sin{\theta_v}\cos{\theta_v}\cos{\chi}H_-(w)H_0(w)
     \vphantom{\left(1+\cos{\theta_\ell}\right)^2} \right]
     \mathcal{B}(D^*\to D\pi),
     \label{fDecAmp}
\end{align}
\eOneCol
\noindent where $\mathcal{B}(D^*\to D\pi)$ is the branching fraction of the daughter $D^*$ decay; further, $\theta_v$, $\theta_\ell$, and
$\chi$ are the polar angle of the $D$ in the $D^*$ rest frame, the polar angle of the charged lepton in the rest frame of the
virtual $W$ meson, and the angle between the $\ell\nu$ and $D\pi$ planes, respectively.
As with Eq.~(\ref{DiffDecayRate}), for neutral $B^0$ decays the right-hand side of Eq.~(\ref{fDecAmp}) should have an
additional factor $(1+\alpha\pi)$ for the Coulomb attraction in the final state (see for example, Refs.~\cite{deBoer:2018ipi,Cali:2019nwp}). 
Other electromagnetic corrections are expected to be smaller, and we will neglect them, keeping only $|\eta_\text{EW}|^2$ and the 
Coulomb factor.\footnote{Both the Belle and BaBar experiments use the PHOTOS package to account for low-energy EM radiation;
to the best of our knowledge, the EM interactions between charged particles in the final state, described by the Coulomb factor, are not included in PHOTOS.}
Following our previous estimates of EM effects~\cite{Bailey:2014tva,Lattice:2015rga}, as well as the HFLAV procedure~\cite{Amhis:2016xyh,Amhis:2019ckw},
we use $\eta_{\text{EW}} = 1.0066(50)$ in our calculation.

Belle marginalizes on one variable at a time, integrating (binning) the rest.
BaBar's method consists of a full, four-dimensional analysis without integrating over any variable.
The collaboration claims such an analysis is needed to achieve correct results~\cite{Dey:2019bgc}.
Nonetheless, both the Belle and BaBar collaborations give compatible final values of $|V_{cb}|$ in their respective
publications~\cite{Waheed:2018djm,Dey:2019bgc}.

For Belle, we integrate Eq.~\eqref{fDecAmp} in the required bins, and the BGL expressions are introduced in the integrated results.
Also, we multiply the right-hand side of Eq.~\eqref{fDecAmp} by the Coulomb factor $(1+\alpha\pi)$, because these data are for
neutral $B$~mesons only.
We perform a combined fit to both the electron and muon modes, instead of averaging them.\footnote{The systematic correlation matrices
given in Ref.~\cite{Waheed:2018djm} do not include off-diagonal blocks for the correlated
systematic errors between electron and muon modes.  They can be reconstructed from the given data~\cite{Bobeth:2021lya}, but we do not
attempt such a reconstruction in our analysis.} For BaBar, we fit the lattice-QCD synthetic data with those in Ref.~\cite{Dey:2019bgc}.
BaBar publishes results for a BGL fit to their data that includes both neutral and charged $B$-meson decays.
According to Ref.~\cite{BaBar:2015zkb}, $35.1\%$ of the decays in this data set correspond to $B^0$ decays, and the remaining
$64.9\%$ correspond to $B^\pm$ decays. These fractions imply that a Coulomb factor of $(1+0.351\alpha\pi)$ should be applied to the
$B^0$ + $B^\pm$ BaBar dataset.

\begin{figure*}
    \centering
    \includegraphics[width=0.45\textwidth]{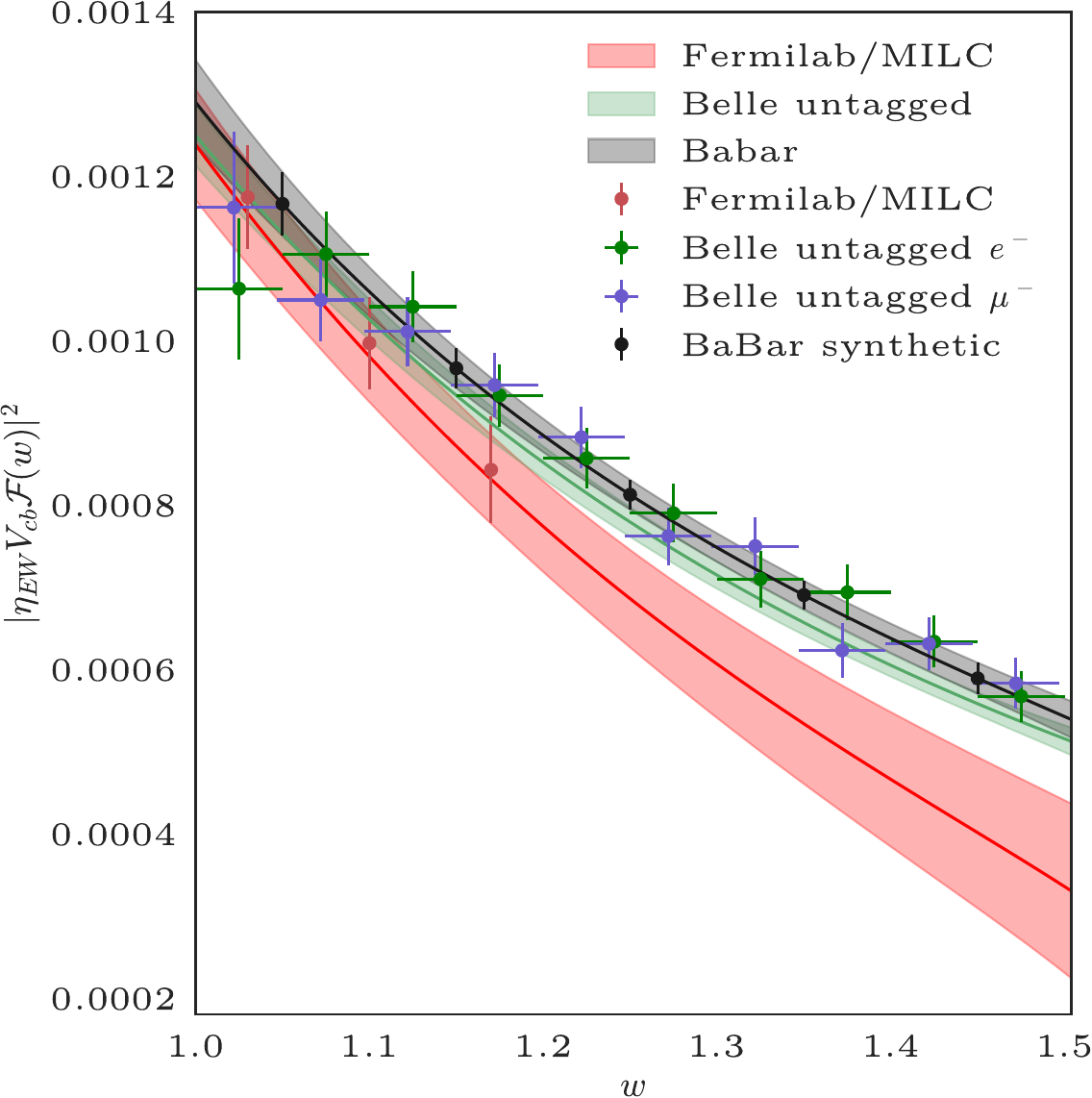} \hfill
    \includegraphics[width=0.45\textwidth]{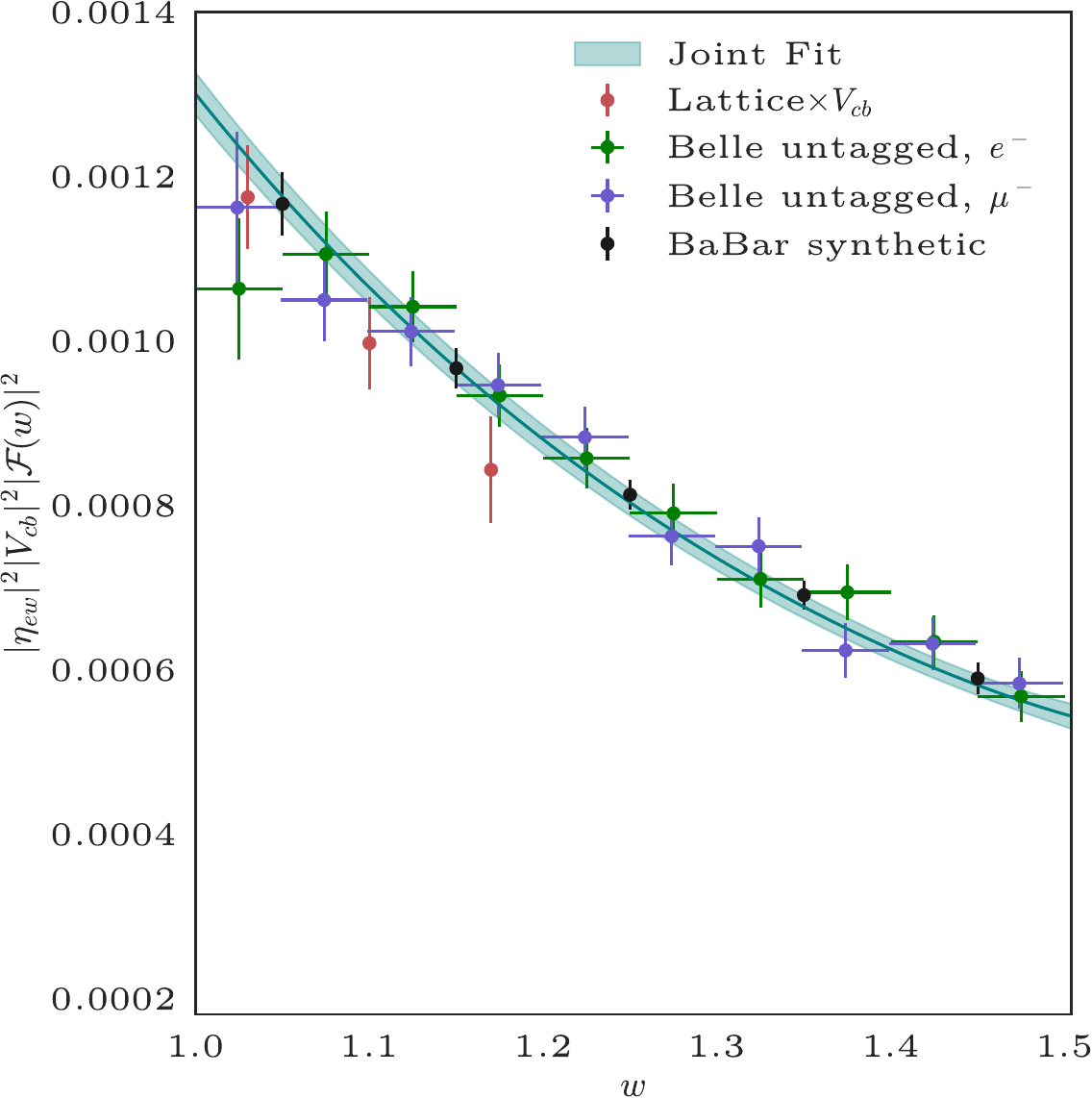}
         \caption{Results for separate fits to each dataset (left) and joint fit of all data (right).  On the left we compare
         the BaBar result (gray), the Belle result from the untagged dataset (green), and the lattice-QCD result coming from our
         synthetic data (red).
         To allow for an straightforward comparison of lattice and experimental data, the data points and bands have been normalized
         with the central value of $|V_{cb}|$ as obtained in our joint fit, and taking into account the Coulomb factor
         corresponding to each case.
         All results agree within $\approx2\sigma$ over the whole kinematic range.
         There is tension between the slope predicted by the lattice calculation and that of the experimental data.
         Since the lattice-QCD slope is well determined, correlations in the joint fit cause the central
         lattice-QCD values to fall slightly below the experimental values.}
   \label{DecayComp}
\end{figure*}

BaBar uses the previous Fermilab-MILC result of $h_{A_1}$ at zero recoil to extract $|V_{cb}|$~\cite{Bailey:2014tva,Dey:2019bgc}.
In order to minimize the influence of the lattice-QCD results for $h_{A_1}(1)$ from Ref.~\cite{Bailey:2014tva} in the joint fit, we
create synthetic data from Ref.~\cite{Dey:2019bgc} for $|\eta_{EW}|^2|V_{cb}|^2|\mathcal{F}(w)|^2$ at five recoil values away from
$w=1$, as shown in Fig.~\ref{DecayComp}. Since the BaBar collaboration employs a linear fit for all form factors, we exhaust
the number of degrees of freedom with just five points. Each one of these synthetic data points has a larger impact than each single
data point from Belle's untagged dataset, because the Babar synthetic data inherits its precision from the precision of the underlying,
full dataset. In the end, the green band for Belle in Fig.~\ref{DecayComp} is noticeably narrower than BaBar's, so it is expected that
the Belle untagged dataset has a larger impact in the final results for $|V_{cb}|$ and $R(D^\ast)$.

The kinematic constraint in Eq.~\eqref{KinF1} is included in these fits, and although there are no direct experimental measurements
that determine $\mathcal{F}_2$, experiments also have an impact on the $d_j$ coefficients through the correlations between this and
other form factors.  Our preferred fits, coming from quadratic $z$~expansions, are shown in Fig.~\ref{DecayComp} and Table~\ref{CompResults}.
As in the lattice-QCD-only fits, the fits in Table~\ref{CompResults} include $0(1)$ priors for the quadratic and higher (if applicable)
coefficients of each form factor, while leaving the rest of the coefficients unconstrained.
%As in the lattice QCD only z expansion and in our chiral-continuum extrapolation, theoretically motivated priors whose posteriors are very
%similar in width and central value are counted as data inputs, increasing the number of degrees of freedom.
Fig.~\ref{DecayComp} shows that the mean of the lattice estimate falls below the experimental curves, but the errors are large enough to make
the difference remain at $\approx 2\sigma$. The correlations between the different lattice, synthetic data points determine very precisely
the slope of the decay amplitude, forcing it to be noticeably larger than what we obtain from our fits to experimental data.
The full correlation matrix is provided in the ancillary files, as described in
\ref{ApFullRes}.  Form factors from this information can be used in phenomenology, under the assumption that only the
$\tau$ couples to new physics.

Our final result for $|V_{cb}|$ is obtained from the quadratic BGL fit to the lattice-QCD form factors and both experimental datasets
(see the column labeled ``Lat\-tice+\-both'' in Table~\ref{CompResults}), which yields
\begin{equation}
    \left|V_{cb}\right| = (38.40 \pm 0.78)\times 10^{-3},
    \label{eq:Vcb}
\end{equation}
and a $\chi^2/\text{dof} =126/84$. This relatively large $\chi^2/\text{dof}$ indicates tensions among the datasets:
a combined fit of Belle and BaBar data, using lattice-QCD input only for normalization results in a large $\chi^2/\text{dof}$ of $104/76$.
It is therefore to be expected that the combined fit would result in a similarly large $\chi^2/\text{dof}$.
Further, we note that our fit to the lattice-QCD form factors only has $\chi^2/\text{dof} < 1$, as
shown in Table~\ref{CompResults}, which also lists the results of joint fits of lattice-QCD form factors with each experimental dataset
separately, as well as with the combined Belle and BaBar data. We find that all joint lattice-QCD with experimental data fits have\linebreak
$\chi^2/\text{dof} > 1$, including the one leading to Eq.~\eqref{eq:Vcb}, but the central values of $|V_{cb}|$ do not differ
more than approximately one standard deviation among these fits, and the sizes of the errors are similar.
We also see a general agreement in the coefficients of the expansion, particularly in the important low-order ones.
\begin{table*}[h]
    \caption{Quadratic $z$~expansion results. The second column shows results from a fit only to synthetic lattice-QCD data (the
    same as the ``quadratic'' column in Table~\ref{SynthResults}), the third, from a joint fit to lattice QCD plus BaBar's synthetic
    data, the fourth, from lattice QCD plus Belle's untagged dataset, the fifth, lattice QCD plus both experiments,
    and the last, a combined fit of all experimental data using the value of $h_{A_1}(1)$ extracted from our chiral-continuum extrapolation
    as normalization. 
    %As with the $z$~fit to the chiral-continuum extrapolation, whenever a posterior is very similar to its prior, we count the prior as input data.
    The coefficient $c_0$ is fixed by the constraint given in Eq.~\eqref{KinF1}, and it is shown for convenience.}
  \label{CompResults}
    \begin{tabular*}{\textwidth}{@{\extracolsep{\fill}}cS[table-format = +1.5(2)]S[table-format = +1.5(2)]S[table-format = +1.5(2)]S[table-format = +1.5(2)]S[table-format = +1.5(2)]}
     \hline
     \hline
     \Tstrut                        &      {Lattice QCD}      &    {Lattice + BaBar}    &    {Lattice + Belle}    &     {Lattice + both}     &  {$h_{A_1}(1)$ + both}  \\
     \hline                                                                                                                           
     $a_0$ \Tstrut                  &        0,0330(12)       &        0,0331(12)       &         0,0325(11)      &         0,0321(10)       &         0,0262(42)      \\
     $a_1$                          &        -0,156(55)       &        -0,091(41)       &         -0,160(45)      &         -0,147(31)       &           0,03(15)      \\
     $a_2$                          &         -0,12(98)       &         -0,19(20)       &          -0,69(94)      &          -0,63(20)       &          -0,12(12)      \\
     \hline                                                                                                                           
     $b_0$                          &       0,01229(23)       &       0,01229(22)       &        0,01238(22)      &        0,01249(22)       &        0,01228(23)      \\
     $b_1$                          &        -0,003(12)       &        0,0104(72)       &          0,015(10)      &         0,0021(43)       &         0,0046(48)      \\
     $b_2$                          &          0,07(53)       &          0,44(17)       &          -0,30(24)      &           0,07(11)       &          -0,17(21)      \\
     \hline                                                                                                                      
     $c_0$                          &      0,002059(38)       &      0,002058(37)       &       0,002073(37)      &       0,002092(37)       &       0,002056(38)      \\
     $c_1$                          &       -0,0058(25)       &       -0,0010(11)       &         0,0010(17)      &        0,00062(86)       &        0,00163(92)      \\
     $c_2$                          &        -0,013(91)       &         0,022(50)       &          0,035(57)      &          0,060(26)       &          0,017(37)      \\
     $c_3$                          &                         &          0,24(77)       &          -0,34(76)      &          -0,94(48)       &          -0,66(62)      \\
     \hline                                                                                                                      
     $d_0$                          &        0,0509(15)       &        0,0517(15)       &         0,0522(15)      &         0,0531(14)       &                         \\
     $d_1$                          &        -0,328(67)       &        -0,220(55)       &         -0,180(49)      &         -0,201(42)       &                         \\
     $d_2$                          &         -0,02(96)       &          0,20(92)       &          -0,01(90)      &         0,0007(8980)     &                         \\
     \hline                                                                                                                  
    $\chi^2/\text{dof}$\Tstrut      &     {0.63/1\Hs{5mm}}    &    {8.50/4\Hs{5mm}}     &     {111/79\Hs{4mm}}    &     {126/84\Hs{4mm}}     &     {104/76\Hs{4mm}}    \\
    $\sum_i^N a_i^2$                & 0,47\epm{1}{1.51}{0.36} & 0,05\epm{1}{0.12}{0.05} &  0,7\epm{1.1}{2.0}{0.7} & 0,43\epm{1}{0.30}{0.21}  & 0,04\epm{1}{0.05}{0.03} \\
    $\sum_i^N (b_i^2 + c_i^2)$      & 0,13\epm{1}{0.44}{0.10} & 0,56\epm{1}{0.98}{0.39} & 0,54\epm{1}{1.00}{0.43} & 0,90\epm{.8}{1.17}{0.57} & 0,57\epm{1}{1.13}{0.55} \\
    $\sum_i^N d_i^2$\Bstrut         & 0,54\epm{1}{1.40}{0.38} & 0,46\epm{1}{1.35}{0.37} & 0,41\epm{1}{1.25}{0.34} & 0,41\epm{1}{1.25}{0.33}  &                         \\
     \hline                                                                                                                         
$|V_{cb}|\times 10^3$\Tstrut\Bstrut &                         &          39,1(10)       &         38,17(85)       &         38,40(78)        &         39,35(91)       \\
     \hline
     \hline
    \end{tabular*}
\end{table*}

Because most previous inclusive and exclusive determinations of $\left|V_{cb}\right|$ omit
the Coulomb factor, we also perform the BGL fits without it; the results are collected in Table~\ref{tab:noCoulomb}.
\begin{table*}[h]
    \caption{$\left|V_{cb}\right|$ results for our different BGL fits without including the Coulomb factor.}
    \label{tab:noCoulomb}
    \begin{tabular*}{\textwidth}{@{\extracolsep{\fill}}cS[table-format = +1.5(2)]S[table-format = +1.5(2)]S[table-format = +1.5(2)]S[table-format = +1.5(2)]S[table-format = +1.5(2)]}
     \hline
     \hline
     \Tstrut                   &    {Lattice + BaBar}    &    {Lattice + Belle}    &     {Lattice + both}     &  {$h_{A_1}(1)$ + both}  \\
     \hline
$|V_{cb}| \times 10^3$ \Tstrut &          39.3(11)       &         38.60(86)       &         38.74(78)        &         39.75(92)       \\
    $\chi^2/\text{dof}$\Bstrut &    {8.48/4\Hs{5mm}}     &     {111/79\Hs{4mm}}    &     {124/84\Hs{4mm}}     &     {103/76\Hs{4mm}}    \\
     \hline
     \hline
    \end{tabular*}
\end{table*}
Compared to the results for $\left|V_{cb}\right|$ in Table~\ref{CompResults}, the central values are shifted by the respective Coulomb factors. They are consistent with previous exclusive determinations,
for example $|V_{cb}|_\text{excl}=(39.9\pm0.9)\times10^{-3}$ from the PDG~\cite{Zyla:2020zbs}.  The long-standing tension with inclusive determinations thus remains:
$|V_{cb}|_\text{incl}=(42.2\pm0.8)\times10^{-3}$~\cite{Zyla:2020zbs}.

Since the Belle data are binned in different variables, there is a normalization constraint between the different bins, assuming
that they contain the same underlying data.  Then only 37 of the 40 bins are truly independent for each mode~\cite{Bobeth:2021lya},
because the sum of all bins for a particular variable should give the same total number of events.  Such constraints should be
reflected as zero eigenmodes, or ---with rounding errors--- very small eigenvalues in the $40\times40$ statistical correlation
matrices.  The correlation matrices provided in Ref.~\cite{Waheed:2018djm} are constructed using Monte Carlo simulations, and do not
resolve these constraints due to the underlying approximations.  We therefore investigate the effect of removing the last bin on
each one of the angular variables data and reconstructing its value from the total normalization.  We find that this procedure
correctly introduces the anticipated constraints between the bins, while the values of the reconstructed last bins are compatible
with those given Ref.~\cite{Waheed:2018djm}.  Hence, the expected zero eigenvalues in the statistical correlation matrices are
recovered.  With this procedure our combined Belle + lattice-QCD BGL fit does not yield any significant changes in the final values
and uncertainties for $|V_{cb}|$ and $R(D^\ast)$, but we observe a substantial decrease in $\chi^2/\text{dof}$ from 111/79 to 96/73.
In the case of our joint fit, which includes Belle, BaBar and lattice-QCD data, the $\chi^2/\text{dof}$ decreases from 126/84 to 109/78.
Nevertheless, the results for $|V_{cb}|$ and $R(D^\ast)$ quoted in this work use the Belle data and correlation matrices as given in Ref.~\cite{Waheed:2018djm}.

The BGL fit to the BaBar data~\cite{Dey:2019bgc} includes fewer coefficients than our BGL fit to the lattice-QCD form factors.  We
test for the presence of truncation errors by performing BGL fits to the BaBar data including higher coefficients with priors of
$0.0(5)$.  This increases the errors in the BaBar data points, most likely because the extra coefficients are completely
uncorrelated with the rest of the BaBar data.  Because the joint fit to all data is currently dominated by the Belle and lattice-QCD
data, the addition of extra coefficients in the BaBar expansion does not change our final results for $|V_{cb}|$ and $R(D^\ast)$ in
a meaningful way.  Hence, for the our final results quoted in this work, the synthetic data points from Ref.~\cite{Dey:2019bgc} are
generated without adding extra coefficients.

In the Belle and BaBar analyses the number of coefficients in the BGL $z$~expansion is limited to exclude those that cannot be
properly determined by their data, and thus avoiding apparent unitarity violations. This procedure, however, does not account for
possible truncation errors. Repeating the fits in Table~\ref{CompResults} without the $c_3$ and $d_2$ coefficients yields similar
results with reduced errors and much smaller sums for the unitarity constraints.

\subsection{Determination of \texorpdfstring{\boldmath$R(D^\ast)$}{R(D*)}} % DecayRD

\begin{figure}
    \centering
    \includegraphics[width=0.44\textwidth]{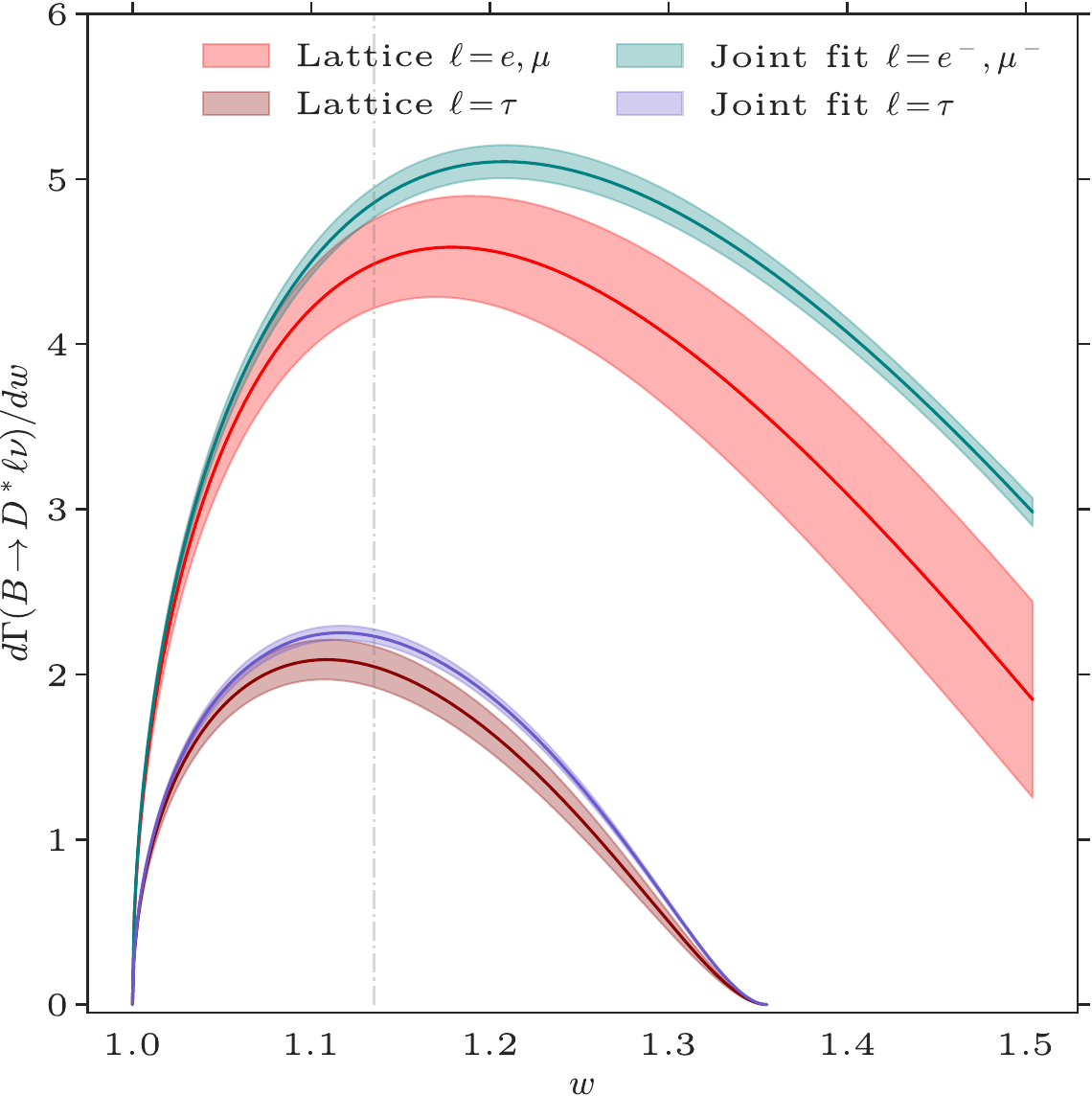} \hfill
    \includegraphics[width=0.46\textwidth]{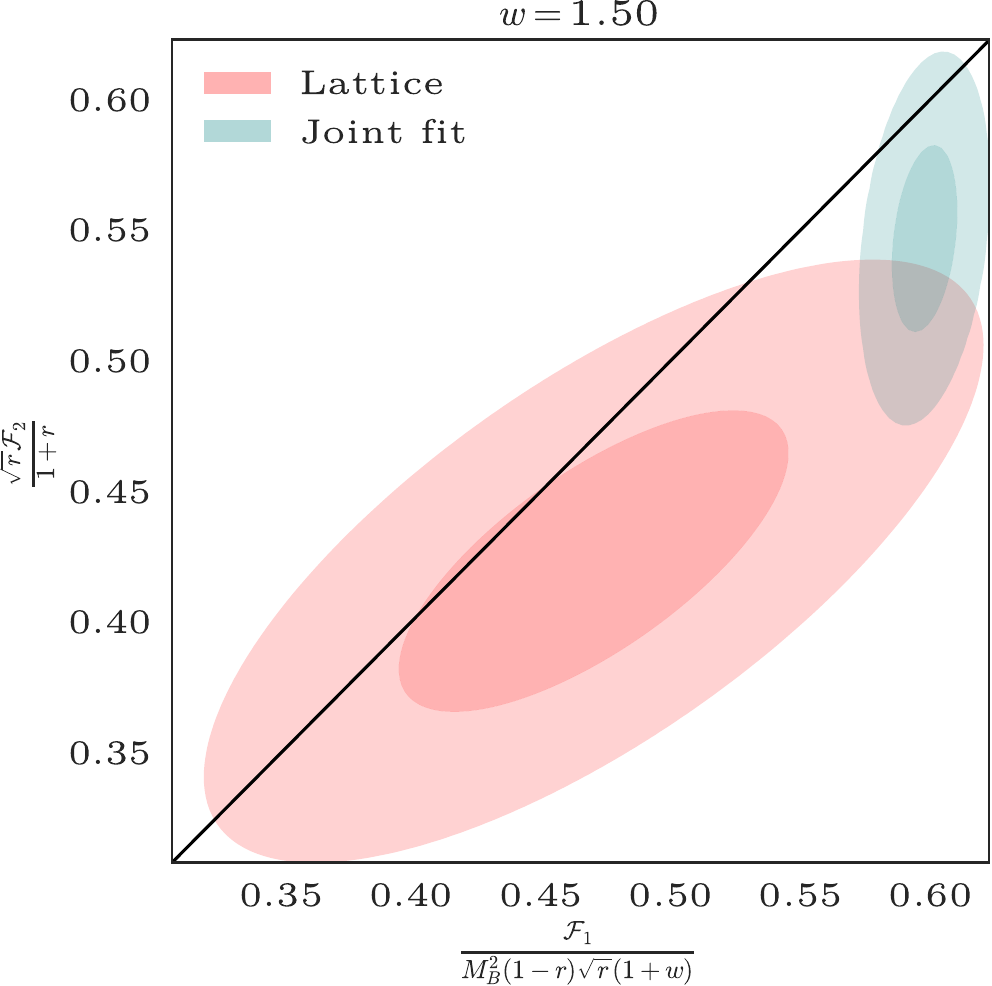}
    \caption{{\bf Top:} differential decay rate calculated using only lattice data (red and maroon) and lattice plus experimental
       data (green and blue).  The higher curves are for a massless lepton, whereas the lower curves are for the $\tau$.
       Although the pure-lattice curves are consistently below the experimental ones, especially at
       large recoil, both of them agree within $2\sigma$.  {\bf Bottom:} test of the kinematic constraint at maximum recoil
       Eq.~\eqref{KinF2}.  Shown is a contour plot up to $2\sigma$ of the form factors $\mathcal{F}_1$ and $\mathcal{F}_2$
       resulting from the lattice-data-only fit and the joint fit of lattice and experimental data.  The constraint is satisfied
       along the diagonal.  We see both fits satisfy the constraint within errors.}
   \label{wMaxConst}
\end{figure}

\begin{figure}
    \centering
    \includegraphics[width=\linewidth]{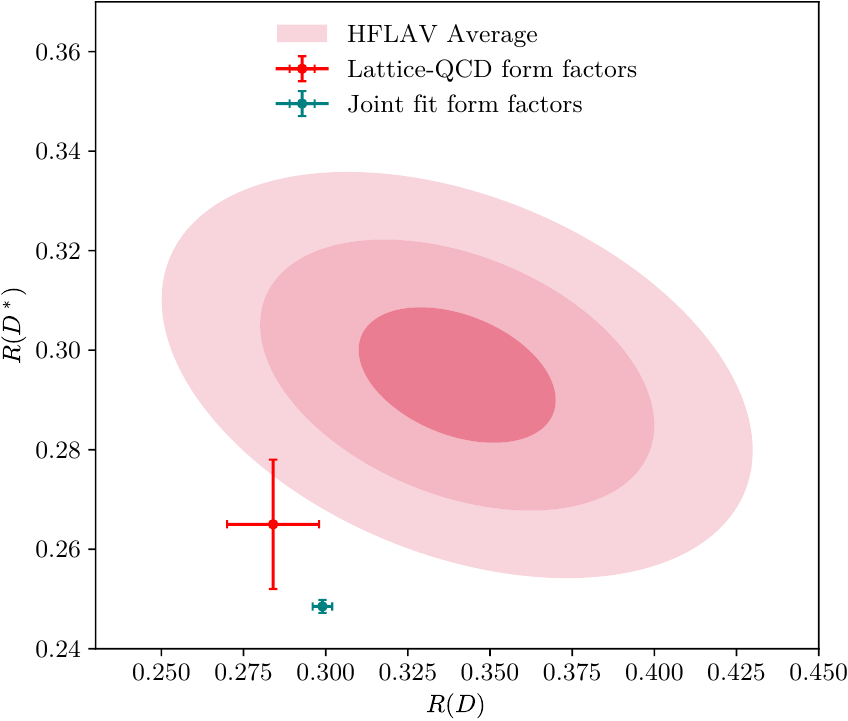}
    \caption{Current status of the evaluations of $R(D)$ and $R(D^\ast)$. The contours show the 2019 HFLAV experimental average~\cite{Amhis:2019ckw} at 1, 2, and 3$\sigma$.
             The red point with error bars uses our lattice-QCD prediction for $R(D^\ast)$ and $R(D)=0.284(14)$ calculated from lattice-QCD data~\cite{Lattice:2015rga}
             (without correlations, which are not available but could be important).
             The green point shows $R(D^\ast)$ from the joint fit yielding $|V_{cb}|$ with $R(D)=0.299(3)$ from HFLAV~\cite{Amhis:2019ckw}, which is similarly based on a
             joint fit to lattice-QCD and experimental data.} 
    \label{RDvsRDstPlot}
\end{figure}

From the fit results in Table~\ref{CompResults} we can calculate $R(D^\ast)$ through direct integration of the differential decay
rate over the whole kinematic range.  In Fig.~\ref{wMaxConst},
we show the differential decay rate as a function of the recoil parameter extracted using lattice-only data (red and brown curves),
compared with that of our joint fit.  The curves below (maroon and blue) show the differential decay rate for the $\tau$ case.  Our
final result for $R(D^\ast)$ from our purely lattice-QCD calculation is
\begin{equation}
    R(D^\ast)_{\text{Lat}} = 0.265 \pm 0.013 .
    \label{RDstLatVal}
\end{equation}
If we assume that new physics effects are visible only at large lepton masses (i.e., the $\tau$), we can use our joint fit of the
lattice and light-lepton experimental data to obtain a more precise SM value of $R(D^\ast)$.  We note that in our joint fit, the
curve corresponding to light leptons is determined mainly from experiment, and the one corresponding to the $\tau$ comes mainly from
the lattice data.  In that case, we obtain
\begin{equation}
    R(D^\ast)_{\text{Lat+Exp}} = 0.2484(13),
    \label{RDst-Vcbfit}
\end{equation}
where the Coulomb factor is included. Its removal does not change significantly neither the central value nor the error.
We emphasize, however, that Eq.~\eqref{RDstLatVal} is the SM prediction, relying only on lattice QCD, while Eq. \eqref{RDst-Vcbfit}
is also based on the shape information coming from experimental data. In any case, the correlated difference between the two results is $1.3\sigma$.
Our values also agree with previous theoretical determinations~\cite{Fajfer:2012vx,Bernlochner:2017jka,Bordone:2019vic,Gambino:2019sif,Jaiswal:2020wer}.
We note that more recent experimental measurements have found $R(D^\ast)$ to be consistently smaller than before, hence reducing the tension
between theory and experiment~\cite{Amhis:2019ckw}.  The current status of the $R(D)$-$R(D^\ast)$ determinations is summarized in
Fig.~\ref{RDvsRDstPlot}.

\subsection{Tests}

\subsubsection{Imposing the constraint at maximum recoil}

As we explained above, our preferred analysis does not impose the kinematic constraint in Eq.~\eqref{KinF2}, is trivially satisfied
in the HQET basis of form factors (the $h_X$) used in our chiral-continuum extrapolation. However, the BGL expansion does not naturally
incorporate it.  Maximum recoil is far from the region where lattice data are available, and there are no experimental data available
for this decay with a heavy lepton $\ell = \tau$.  Thus, to the extent that the BGL expansion does not match the HQET-basis form
factors precisely, we expect small deviations from Eq.~\eqref{KinF2} in the BGL fit.  Such deviations are tolerable because small
violations of the constraint do not have any physical consequences, as long as they are within errors.  Figure~\ref{wMaxConst} shows
that our fits, nonetheless, satisfy the maximum-recoil constraint to within approximately $1\sigma$.

Imposing the constraint in the fit model, we find new values for $|V_{cb}|=38.36(78)\times 10^{-3}$ and $R(D^\ast)=0.274(10)$,
which are compatible with the values obtained in our preferred analysis.  It is not surprising that the constraint does not alter the
value of $|V_{cb}|$.  After all, the CKM matrix element is extracted mainly from the behavior of the form factors at small recoil and
does not entail the form factor $\mathcal{F}_2$. The error on $R(D^\ast)$, on the other hand, is slightly reduced by the constraint.

A comparison of the BGL coefficients of the constrained analysis with those of our preferred one is shown in Table~\ref{CompConst}.
We do not find significant changes, and the coefficients in both analyses are compatible with each other within $1\sigma$, although
the differences increase with the order of the coefficients.  This behavior is expected, as the low-order coefficients are well
determined by the data, and the higher-order coefficients become more relevant at maximum recoil.  The new information does not
improve the quality of the fit in a significant way.  This is particularly clear in the pure lattice fit, where the $\chi^2$ almost
doubles, but the number of degrees of freedom increases just from three in the unconstrained fit to four in the constrained one.  It
appears that the constraint introduces small tensions with the BGL expansion.  Because of this, and since both the unconstrained and
the constrained fit give compatible results with only small differences, we choose the unconstrained fit as our preferred result.

\begin{table*}
    \caption{Comparison of results of the $z$ expansion fits with and without the kinematic constraint given in Eq.~\eqref{KinF2}.
             The largest differences appear in the higher-order coefficients.
             The coefficient $c_0$ is fixed by the constraint given in Eq.~\eqref{KinF1}, and it is shown for convenience.
             The Coulomb factors are included in the fits with experimental data input.}
    \label{CompConst}
    \begin{tabular*}{\textwidth}{@{\extracolsep{\fill}}cS[table-format = +1.6(2)]S[table-format = +1.6(2)]S[table-format = +1.6(2)]S[table-format = +1.6(2)]}
     \hline
     \hline
     \Tstrut                      &       \multicolumn{2}{c}{Lattice QCD alone}       &    \multicolumn{2}{c}{Lattice QCD + Experiment}    \\
                                  &    {Unconstrained}      &     {Constrained}       &     {Unconstrained}      &      {Constrained}      \\
     \hline
     $a_0$                        &      0,0330(12)         &      0,0330(12)         &        0,0321(10)       &        0,0320(10)       \\
     $a_1$                        &      -0,156(55)         &      -0,156(55)         &        -0,147(31)       &        -0,142(31)       \\
     $a_2$                        &       -0,12(98)         &       -0,12(98)         &         -0,63(20)       &         -0,60(19)       \\
     \hline
     $b_0$                        &     0,01229(23)         &     0,01228(23)         &       0,01249(22)       &       0,01250(22)       \\
     $b_1$                        &      -0,003(12)         &       0,001(11)         &        0,0021(43)       &        0,0031(42)       \\
     $b_2$                        &        0,07(53)         &       -0,23(41)         &          0,07(11)       &          0,05(10)       \\
     \hline
     $c_0$                        &    0,002059(38)         &    0,002058(38)         &      0,002092(37)       &      0,002094(37)       \\
     $c_1$                        &     -0,0058(25)         &     -0,0051(23)         &       0,00062(86)       &       0,00090(84)       \\
     $c_2$                        &      -0,013(91)         &      -0,076(58)         &         0,060(26)       &         0,061(26)       \\
     $c_3$                        &                         &                         &         -0,94(48)       &         -1,08(47)       \\
     \hline
     $d_1$                        &      0,0509(15)         &      0,0509(15)         &        0,0531(14)       &        0,0531(14)       \\
     $d_1$                        &      -0,328(67)         &      -0,331(67)         &        -0,201(42)       &        -0,187(41)       \\
     $d_2$                        &       -0,02(96)         &        0,12(95)         &        0,0007(8980)     &          0,85(67)       \\
     \hline
$\chi^2/\text{dof}$\Tstrut        &    {0.63/1\Hs{5mm}}     &    {1.36/2\Hs{5mm}}     &    {126/84\Hs{4mm}}      &      {127/85\Hs{4mm}}   \\
$\sum_i^N a_i^2$                  & 0,47\epm{1}{1.51}{0.36} & 0,47\epm{1}{1.51}{0.36} & 0,43\epm{1}{0.30}{0.21}  & 0,39\epm{1}{0.28}{0.20} \\
$\sum_i^N (b_i^2 + c_i^2)$        & 0,13\epm{1}{0.44}{0.10} & 0,12\epm{1}{0.33}{0.16} & 0,90\epm{.8}{1.17}{0.57} & 1,18\epm{1}{1.25}{0.75} \\
$\sum_i^N d_i^2$\Bstrut           & 0,54\epm{1}{1.40}{0.38} & 0,51\epm{1}{1.42}{0.33} & 0,41\epm{1}{1.25}{0.33}  & 0,77\epm{1}{1.60}{0.72} \\
     \hline
    $|V_{cb}|_{\text{c}} $\Tstrut &                         &                         &         38,40(78)        &         38,36(78)       \\
    $R(D^\ast)$\Bstrut            &       0,265(13)         &       0,274(10)         &        0,2484(13)        &        0,2492(12)       \\
     \hline
     \hline
    \end{tabular*}
\end{table*}

\subsubsection{The \ensuremath{z} expansion with an improved CLN parametrization}
\label{CLNCompSec}

For the sake of completeness, we offer an alternative analysis, replacing the BGL parametrization with CLN.
In the CLN parametrization, the form factor $h_{A_1}$ is expressed as a polynomial in $z$,
and the other form factors appear as ratios with respect to $h_{A_1}$:
\begin{align}
R_0 =& \frac{1}{1+r}\left(w+1 + w\frac{rh_{A_2}-h_{A_3}}{h_{A_1}} - \frac{h_{A_2}-rh_{A_3}}{h_{A_1}}\right), \\
R_1 =& \frac{h_V}{h_{A_1}}, \\
R_2 =& \frac{rh_{A_2}+h_{A_3}}{h_{A_1}}. 
\end{align}

We include a few improvements to address the weak points of CLN: first we extract the full covariance matrix relating the parameters
$\rho_{A_1}$ and $c_{A_1}$ from the original article~\cite{Caprini:1997mu} using the data given for the one-sigma ellipsoids.  With
the full covariance matrix, we can account for the strong correlations between $\rho^2_{A_1}$, $c_{A_1}$ and $d_{A_1}$, which allows
for small variations of the fixed relations often used in CLN fits.  Second, we use updated results for the expansions in $w-1$ of the
ratios $R_j$. The form factors to be fit are
\begin{align}
    h_{A_1}(z) &= h_{A_1}(1)\Big[1 - 8\rho_{A_1}^2z + \left(64c_{A_1} - 16\rho_{A_1}^2\right)z^2 \nonumber \\
               &\phantom{=} + \left(512d_{A_1} + 256c_{A_1} - 24\rho_{A_1}^2\right)z^3\Big],
    \label{hA1CLN}\\
    R_0(w)     &=  1.25(35) - 0.183(77)(w-1) + 0.063(23)(w-1)^2, \\
    R_1(w)     &=  1.28(36) - 0.101(51)(w-1) + 0.066(24)(w-1)^2, \\
    R_2(w)     &= 0.740(44) + 0.128(38)(w-1) - 0.079(19)(w-1)^2,
    \label{R2CLN}
\end{align}
where to fit $R_0$ to experimental data requires measurements of the $\tau$ final state.  The full correlation matrix of
$\rho^2_{A_1}$ and $c_{A_1}$ is given in Table~\ref{CorrMat}, and we follow Ref.~\cite{Caprini:1997mu} to calculate
$d_{A_1}$. In the following, we refer to the CLN parametrization as ``improved'' when using
Eqs.~\eqref{hA1CLN}-\eqref{R2CLN} and ``base'' when using the coefficients of the original paper~\cite{Caprini:1997mu}.
\begin{table}
    \caption{Correlation matrix relating $\rho^2_{A_1}$ and $c_{A_1}$ in the CLN parametrization.}
    \label{CorrMat}
    \begin{tabular}{cSS}
        \hline\hline
        \Tstrut\Bstrut        &{\Hs{0.5cm}$\rho^2_{A_1}$}&{\Hs{0.5cm}$c_{A_1}$} \\
        \hline
        $\rho^2_{A_1}$\Tstrut &          1.0000          &         0.9834       \\
        $c_{A_1}$     \Bstrut &          0.9834          &         1.0000       \\
        \hline\hline
    \end{tabular}
\end{table}

Fitting our lattice data with the improved CLN parametrization yields a result for $R(D^\ast)$ that is compatible with
Eq.~\eqref{RDstLatVal}, but the $\chi^2$/\text{dof} of the fit increases spectacularly to $\chi^2/\text{dof}=25.7/1$.  A base
CLN fit is similarly bad, with $\chi^2/\text{dof}=31.9/7$.  The combined fit using lattice and Belle data again yields a
compatible $|V_{cb}|$ and is also very bad, with $\chi^2/\text{dof}=133.5/80$.  Here too the base and the improved versions of
CLN are equally incompatible with the lattice-QCD form factors and Belle data.
The CLN fits involving only lattice-QCD data violate the kinematic constraint in Eq.~\eqref{KinF2} by
$2.7\sigma$.  The combined CLN fits satisfy it at around $1\sigma$.  In light of these issues, the BGL parametrization provides a
much superior $z$~expansion, so we have chosen it in our main analysis.\footnote{The fits discussed include the Coulomb factor.
Its removal does not change the $\chi^2/\text{dof}$ significantly.}

\begin{figure}
    \centering
    \includegraphics[width=\linewidth]{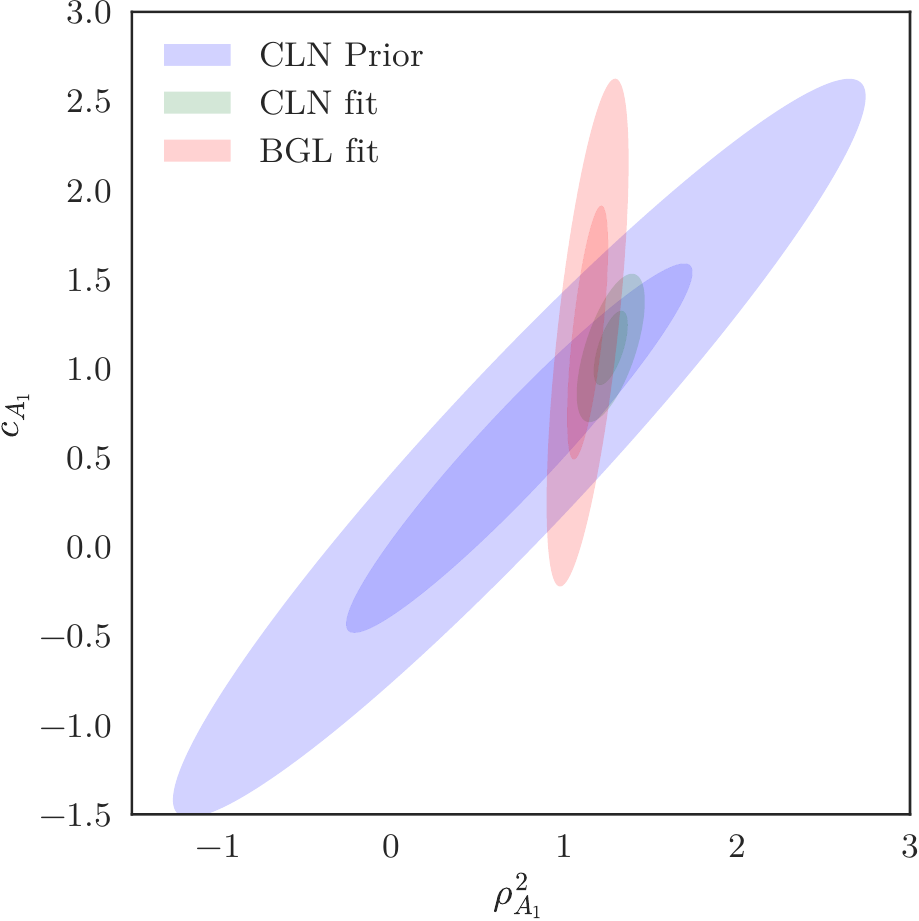}
    \caption{Contour plot of the slope ($\rho^2_{A_1}$) versus curvature ($c_{A_1}$) of the $h_{A_1}$ form factor in the CLN and 
        BGL cases. We show the ellipsoids corresponding to the CLN priors extracted from~\cite{Caprini:1997mu} and the results of 
        the improved CLN and BGL fits at one and two sigmas. There is good agreement in all cases, and a numerical test shows that 
        the differences in $\rho^2_{A_1}$, $c_{A_1}$ and $d_{A_1}$ between the fit results of the improved CLN parametrization and 
        the BGL parametrization is close to $1\sigma$.}
   \label{ClnVsBgl-Prm}
\end{figure}

One can wonder why the improved CLN fit performs so poorly.  One possible point of tension between our data and the improved CLN
ansatz is the relationship between the slope, the curvature and the cubic coefficient in $h_{A_1}$, which is much more constrained
than in the BGL parametrization.  We compare our fit results to only lattice data for both parametrizations by calculating a Taylor
expansion around $z=0$ up to cubic order of our BGL result for $h_{A_1}\propto f$, and we present in Fig.~\ref{ClnVsBgl-Prm} contour
plots of the CLN priors, the improved CLN fit result, and the BGL fit result.  Our BGL results for $\rho_{A_1}^2$, $c_{A_1}$ and
$d_{A_1}$ are compatible with our improved CLN results within one sigma, suggesting that the tensions with the improved CLN
parametrization come from the other form factors.

\begin{figure}
    \centering
    \includegraphics[width=\linewidth]{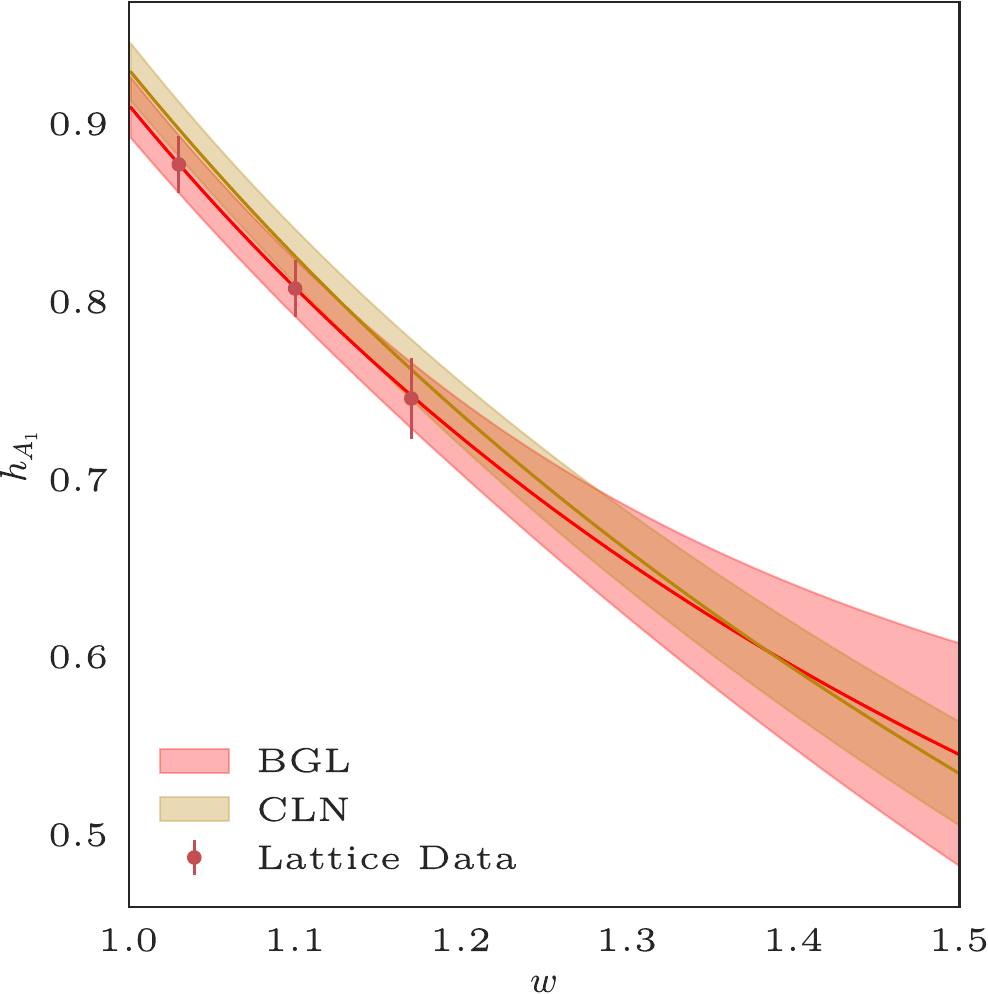} \hfill
    \includegraphics[width=\linewidth]{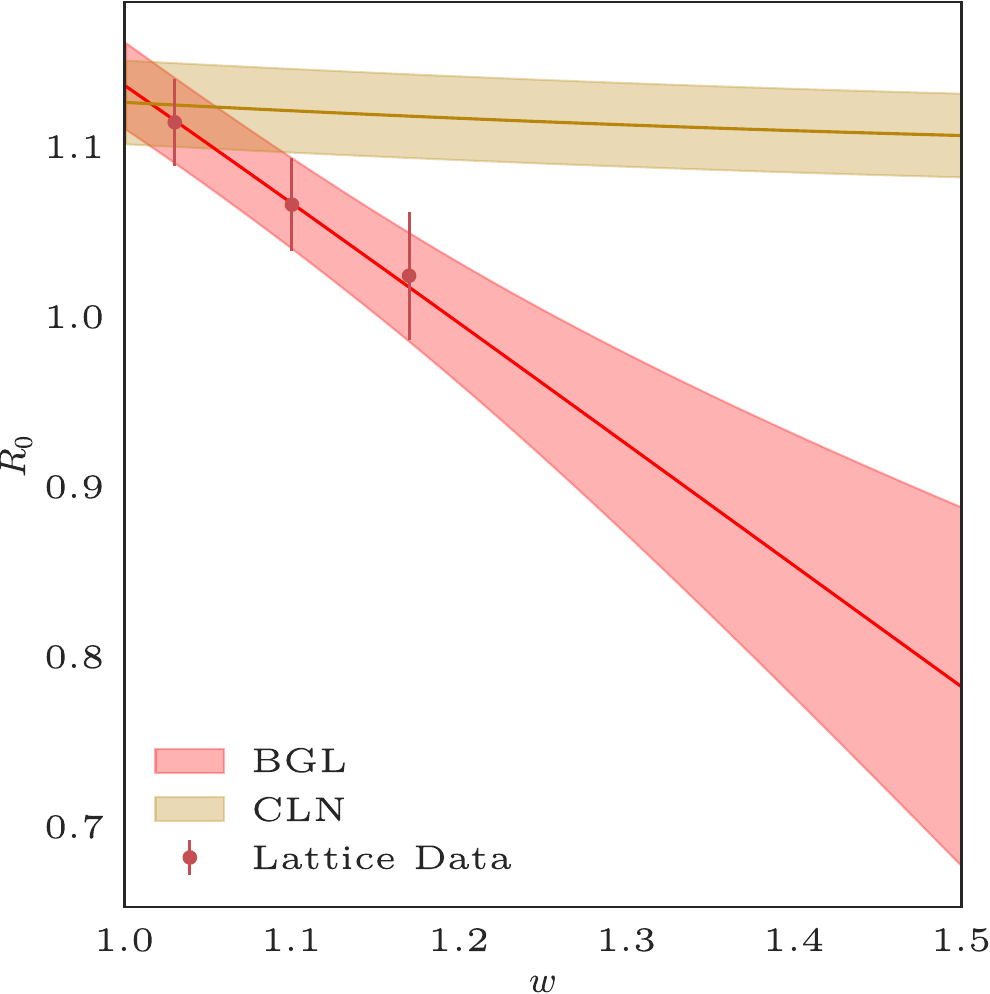}
    \caption{Comparison between our fits to lattice data using BGL and the improved CLN parametrization. We show $h_{A_1}$ (top)
    and $R_0$ (bottom). The differences between BGL and CLN are mild for $h_{A_1}$, at least in the region where 
    there are data. In contrast, the improved CLN fit to $R_0$ shows large tensions with the lattice data.}
    \label{ClnVsBgl-I}
\end{figure}

\begin{figure}
    \centering
    \includegraphics[width=\linewidth]{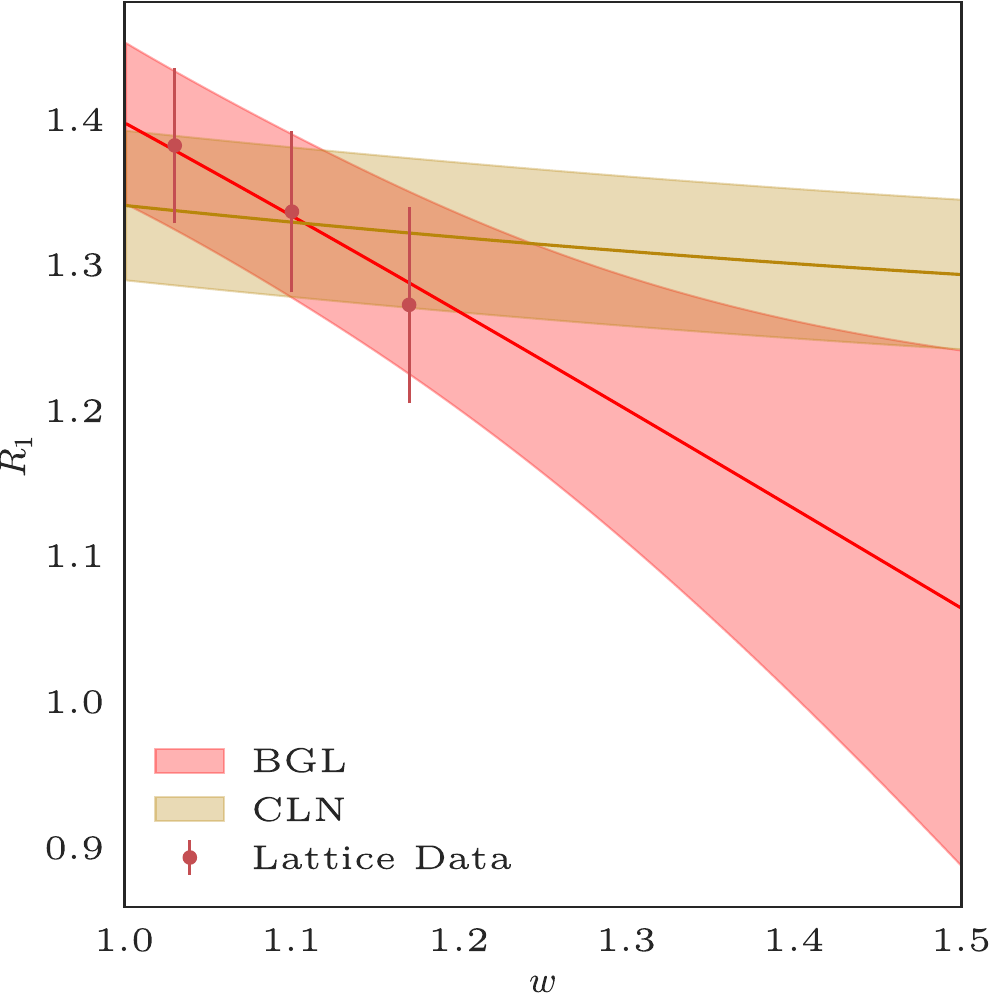} \hfill
    \includegraphics[width=\linewidth]{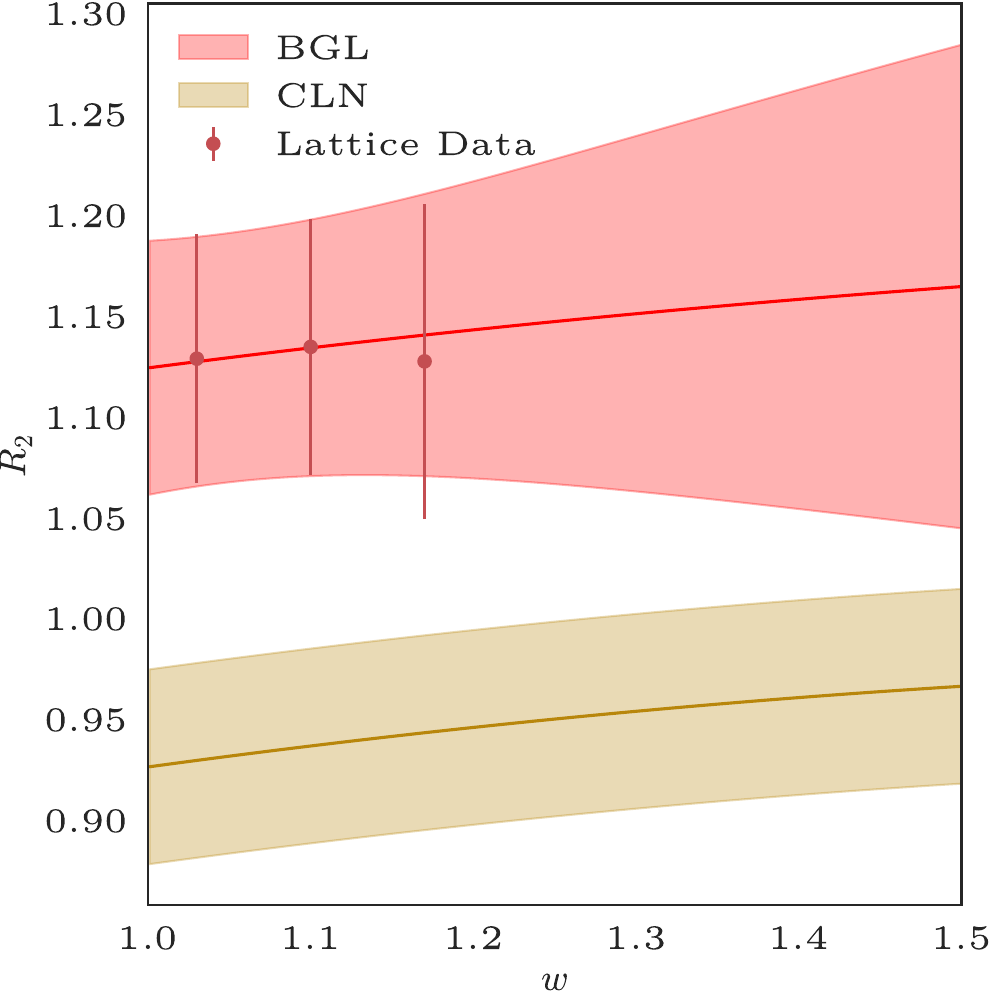}
    \caption{Comparison between our fits to lattice data using BGL and the improved CLN parametrization. Here we show $R_1$ (top) 
        and $R_2$ (bottom). In this case, the fit to the $R_1$ data looks acceptable in both cases, but the CLN fit to
        $R_2$ shows large differences with the data.}
    \label{ClnVsBgl-II}
\end{figure}

Figures~\ref{ClnVsBgl-I} and \ref{ClnVsBgl-II} show the results for $h_{A_1}$ and $R_{0,1,2}$ in the BGL and the improved CLN cases,
along with the lattice data used in the fit.  In general, the BGL fit shows a much better agreement with our lattice data, and the
ratios $R_{0,2}$ are poorly fitted with the improved CLN ansatz.  Also, the improved CLN fit violates the constraint given by
Eq.~\eqref{KinF2}.  Imposing the constraint does not improve the fit quality, or reduce the tensions in the $R_{0,2}$ ratios.

Following Ref.~\cite{Gambino:2020jvv}, we advocate either a revision or a deprecation of the CLN parametrization in favor of a truly
model-independent parametrization, such as BGL. Even if we had found a good quality of fit with CLN, we would still prefer the BGL
results on theoretical grounds. The main reason not to use the CLN parametrization in a first principles analysis like this one is to
test the theoretical assumptions that makes CLN different from BGL. CLN imposes constraints\linebreak through crossing symmetry
from the physical region of the cross channel and constraints from heavy-quark effective theory.  The BGL parametrization does not.
With the latter we are able to verify {\it ex post facto} the reliability of these constraints.

Other limitations of the base CLN, such as an update of its inputs, and a more careful treatment of the errors and correlations
of the CLN coefficients, have been addressed in a few recent papers~\cite{Bernlochner:2017jka,Bordone:2019vic,Bordone:2019guc}.  The
HQET parametrization discussed in those works also include corrections of order $1/m_c^2$.  We refer the reader to those papers for
further discussion.

\subsubsection{Comparison with LCSR}

We can also test the validity of the light cone sum rules (LCSR), often employed to constrain the form factors at maximum recoil.
To this end, we take the latest results from Ref.~\cite{Gubernari:2018wyi}.  They present results for the form factors in a different
notation.  For the reader's convenience, we provide the conversion formulae:
\begin{align}
V\,(w_{\textrm{Max}}) = 0.69(13), & \quad V\,(w) = h_V(w)\frac{1+r}{2\sqrt{r}},                                \label{VLCSR} \\
A_1(w_{\textrm{Max}}) = 0.60( 9), & \quad A_1(w) = h_{A_1}(w) \frac{(1+w)\sqrt{r}}{1+r},                       \label{A1LCSR} \\
A_2(w_{\textrm{Max}}) = 0.51( 9), & \nonumber \\
                                  & \hspace{-1cm} A_2(w) = \left(rh_{A_2}(w) + h_{A_3}(w)\right)\frac{1+r}{2\sqrt{r}}. \label{A2LCSR}
\end{align}
Our lattice-QCD only results for the aforementioned form factors are:
\begin{align}
V\,(w_{\textrm{Max}}) =& 0.65(10),  \\
A_1(w_{\textrm{Max}}) =& 0.608(71), \\
A_2(w_{\textrm{Max}}) =& 0.71(11).
\end{align}
The agreement is excellent for $A_1$ and $V$, and the $A_2$ form factor also agrees within $1.4\sigma$.

\section{Discussion and outlook}
\label{Conclusions}

Using the first unquenched lattice-QCD calculation of the form factors describing the decay $B\to D^\ast\ell\nu$ at nonzero recoil,
together with 2018 Belle~\cite{Waheed:2018djm} and 2019 BaBar~\cite{Dey:2019bgc} measurements, we obtain the following results for
the CKM matrix element $|V_{cb}|$:
\begin{equation}
    |V_{cb}| = (38.40 \pm 0.78) \times 10^{-3},
    \label{Vcbfinal}
\end{equation}
which includes the Coulomb correction for neutral $B$-meson decays. Omitting this correction leads to
$|V_{cb}|=(38.74 \pm 0.78) \times 10^{-3}$, which is the result that should be compared with  inclusive decays.
The quoted uncertainty is from experiment and theory together. As discussed in Sec.~\ref{sec:Vcb}, this result comes from a fit
exhibiting tension among the datasets, $\chi^2/\text{dof}=126/84$. The tension is motivation for better experimental measurements and
lattice-QCD calculations, as a $\sim2\sigma$ effect it is inconclusive. In order to disentangle the contributions from experiment and theory,
we run a BGL fit as described in Sec.~\ref{zExpSec}, but assuming very small errors on the synthetic lattice-QCD data, and then we run
analogous fits assuming very small experimental errors. By looking at how the final error changes, we estimate each contribution, finding
$0.34\times 10^{-3}$ from experiment, $0.67\times 10^{-3}$ from lattice QCD, $0.10\times 10^{-3}$ from the truncation of the BGL expansion,
and $0.18\times 10^{-13}$ due to EW+EM effects.
These partial values are an approximation to guide how much improvement we can expect from future calculations and experiments.
The final error in Eq.~\eqref{Vcbfinal} is an accurate estimate of the full uncertainty.

Belle II is expected to deliver experimental data for this decay, although until there is a better understanding of its detector
performance and the systematics of the experiment, it is not clear how much improvement on the error for $|V_{cb}|$ could come from
these high statistics data~\cite{Kou:2018nap}.  An improved result can be obtained from this or any other forthcoming data, in
combination with the synthetic data points and covariance matrix that fully describe the output of our chiral-continuum
extrapolation, given in the ancillary files as described in \ref{ApFullRes}.

The main result of this article is the behavior of the form factors parametrizing semileptonic $B\to D^\ast\ell\nu$ decays at small,
but nonzero recoil.  It allows us to perform a much more robust analysis of $|V_{cb}|$ than when using just the value of the decay
amplitude at zero recoil.  We find excellent agreement between our current and previous results, both from $B\to D^*\ell\nu$ at zero
recoil~\cite{Bailey:2014tva} and $B\to D\ell\nu$ at nonzero recoil~\cite{Lattice:2015rga}.  Our result also agrees with other recent
$B\to D^{(\ast)}\ell\nu$ exclusive determinations~\cite{Harrison:2017fmw,Na:2015kha}.  It is also compatible with the recent
extraction based on $B_s\to D_s^{(*)}\ell\nu$ decays measured by LHCb~\cite{Aaij:2020hsi}, and with form factors recently calculated
by HPQCD~\cite{McLean:2019sds,McLean:2019qcx}, but with much smaller errors.

Our result for $|V_{cb}|$ does not change the current status of the inclusive-exclusive puzzle. The inclusive determination was recently updated
in~\cite{Bordone:2021oof}, slightly reducing the uncertainty, and preliminary results from Belle based on a data-driven approach~\cite{Fael:2018vsp},
and compatible with the current inclusive world average~\cite{Amhis:2019ckw}, have been presented~\cite{Bernlochner:2022ucr}.
On the other hand, it has been argued that it is very challenging for BSM physics to accommodate such a tension~\cite{Crivellin:2014zpa,Jung:2018lfu}.
A further reduction in errors is necessary to extract conclusions.
Inclusive calculations of $|V_{cb}|$ might also benefit in the longer term from recent ideas in lattice QCD~\cite{Hashimoto:2017wqo,Gambino:2020crt},
which might lead to calculations with very different systematics.

During the implementation of the joint experimental-lattice fits we found issues in both experimental datasets used: the BaBar
collaboration does not provide unfolded data, and their $z$~expansion uses only five coefficients to describe three form factors, as
opposed to the nine coefficients we use for the same form factors in our lattice data only fit.  As a result, we are concerned that
the truncation errors in the $z$~expansion could be underestimated in the synthetic data generated from BaBar fits.
However, the $|V_{cb}|$ and $R(D^\ast)$ fits are dominated by the Belle and lattice-QCD data, and the effect of including BaBar data
is just a small reduction in the total error in Eq.~\eqref{Vcbfinal}.
The Belle collaboration provides unfolded data, but as pointed out in Ref.~\cite{Bobeth:2021lya}, the statistical correlation
matrices given in Ref.~\cite{Waheed:2018djm} seem inconsistent.  We also checked that this potential problem has no significant
impact on our results for either $|V_{cb}|$ or $R(D^*)$ (with form factors from the $|V_{cb}|$ fit).
Thus, an improvement in the presentation of the data from both collaborations would be very welcome.

Another benefit of the knowledge of all form factors at nonzero recoil is the possibility of calculating $R(D^\ast)$ from first
principles.  Our result
\begin{equation}
    R(D^\ast)_{\text{Lat}} = 0.265 \pm 0.013
    \label{FinalRDst}
\end{equation}
reaches a similar precision to that of the $B \to D\ell\nu$ analysis for $R(D)$.  Even though our calculation of $R(D^\ast)$
involves integrals of extrapolated quantities with large errors, the combined error is relatively small due to the large
correlations between $\mathcal{B}(B\to D^\ast\tau\nu)$ and $\mathcal{B}(B\to D^\ast\ell\nu)$.  In this case, the form factor that
enter only in the determination of the decay to a $\tau$ was computed with lattice input only.  We have also calculated a more
precise value using form factors from the $|V_{cb}|$ fit, $R(D^\ast)_{\text{Lat+Exp}} = 0.2484(13)$.  Our preferred SM value is the
one given in Eq.~\eqref{FinalRDst}, which comes exclusively from lattice QCD and avoids any experimental decay-rate input.

The result in Eq.~(\ref{FinalRDst}) confirms previous theoretical estimates of $R(D^\ast)$, as well as the current tension between
the SM and experiment in the $R(D)$-$R(D^\ast)$ plane.  Recent experimental determinations of $R(D^\ast)$ tend to reduce the
tension, however.  In fact, before Belle published results from their untagged dataset~\cite{Belle:2019rba}, the tension was as
large as $4\sigma$, but the newest analysis has reduced it to $3\sigma$, and remaining tensions come mainly because of the influence
of the BaBar $R(D)$ result~\cite{Lees:2012xj}.  An updated measurement of $R(D)$ could cast some light on the current tensions.
Also, future high-precision experimental measurements from Belle~II and LHCb are bound to become critical to determine whether these
quantities will agree with the SM in the end.

Together with more precise experimental measurements, lattice-QCD form factors with a smaller uncertainty will also be crucial for
shedding light to this theory-experiment tension, as well as to the exclusive-inclusive tension in the determination of $|V_{cb}|$.
We expect to reduce the uncertainties in the form factors at nonzero recoil in future work, the first of which is already in
progress at the time of this publication.  The main sources of uncertainty in this work come from statistics and the quark
discretization errors.  An improvement in this area will require a modification in both the light and the heavy-quark actions to
allow for smaller systematics. Chiral-continuum-extrapolation errors can be reduced by using a better discretization for light quarks and
physical pion masses.
Another area where we can reduce errors is the renormalization, by using a nonperturbative calculation of the
renormalization factors.  Further validation of our lattice-QCD result will come with independent analyses currently in progress by
other lattice-QCD collaborations~\cite{Kaneko:2019vkx}.  Similar improvements, with an expected reduction of errors, can be applied
to our calculation of $B\to D\ell\nu$ form factors at nonzero recoil~\cite{Lattice:2015rga}.  A correlated analysis of both decays
will allow a correlated determination of $R(D)$ and $R(D^*)$ that could provide tighter theoretical constraints.

In this work, we also assess the impact of using an improved CLN parametrization to describe the shape of the form factors.
Instead of using the fixed coefficients published in Ref.~\cite{Caprini:1997mu}, we employ the full covariance matrix that relates
the slope and curvature coefficients of the reference form factor using data from the original CLN paper~\cite{Caprini:1997mu}, and
pass this information as priors to a CLN fit.  We also compute the errors in the cubic coefficient, and update the values of the
ratios with respect to the reference form factor with values coming from one of the latest HQET calculations~\cite{Bigi:2017jbd},
assuming a 20\% error on each coefficient.  Our updated CLN parametrization gives a very similar central value and error bar,
compared with that of the BGL parametrization, but the quality of fit decreases greatly when the lattice-QCD data are included.  CLN
is very restrictive with the shape of certain form factors, and because the lattice-QCD data have relatively small errors, they
introduce serious constraints in the parametrization.  Our findings reinforce the current consensus of the
community~\cite{Gambino:2020jvv} to abandon CLN in favor of the more flexible and rigorous BGL parametrization.  The impact of using
improved HQE parametrizations, such as the one in Refs.~\cite{Bernlochner:2017jka,Bordone:2019vic,Bordone:2019guc}, should be
nevertheless investigated.

\section*{Acknowledgments}
We thank Claude Bernard for early contributions to this project, in particular for laying the foundations for the chiral-continuum
extrapolations employed in this analysis.
We thank
Biplab Dey,
Danny van~Dyk, % lower-case ``van'', alphabetizes as Dyk
Paolo Gambino,
Martin Jung,
Laurent Lellouch,
William Marciano,
Christoph Schwanda,
Phillip Urquijo, 
and Eiasha Waheed
for useful discussions.
We thank Biplab Dey especially for providing additional information on the BaBar $B\to D^\ast\ell\nu$ analysis~\cite{Dey:2019bgc},
and Martin Jung for his careful reading of the manuscript.
We thank Jon Bailey and Chris Bouchard for generating some of the correlator data.
Computations for this work were carried out with resources provided by the USQCD Collaboration, the National Energy Research
Scientific Computing Center, and the Argonne Leadership Computing Facility, which are funded by the Office of Science of the U.S.\
Department of Energy; and with resources provided by the National Institute for Computational Science and the Texas Advanced
Computing Center, which are funded through the National Science Foundation’s Teragrid/XSEDE Program.
This work was supported in part by the U.S.\ Department of Energy under Awards No.\ DE-FG02-13ER41976 (D.T.), No.\ DE-SC0009998 (J.L.),
%No.\ DE-SC0010005 (E.T.N.),
No.\ DE-SC0010120 (S.G.), and No.\ DE-SC0015655 (A.X.K.,Z.G.);
by the U.S.\ National Science Foundation under Grants No.\ PHY17-19626 (C.D., A.V.), and No.\ PHY14-17805 (J.L.);
by SRA (Spain) under Grant No.\ PID2019-\-106087GB-\-C21 / 10.13039 / 501100011033 (E.G.);
by the Junta de Andalucía (Spain) under Grants No.\ FQM- 101, A-FQM-467-UGR18, and P18-FR-4314 (FEDER) (E.G.);
by the Fermilab Distinguished Scholars Program (A.X.K.).
This document was prepared by the Fermilab Lattice and MILC Collaborations using the resources of the Fermi National Accelerator
Laboratory (Fermilab), a U.S.\ Department of Energy, Office of Science, HEP User Facility.  Fermilab is managed by Fermi Research
Alliance, LLC (FRA), acting under Contract No.\ DE-AC02-07CH11359.

\appendix
\section{The chiral logs in our chiral-continuum extrapolation}
\label{ApChiral}

In this Appendix, we give further details on the chiral-continuum extrapolation. In particular, we discuss the staggered version of the chiral logs corresponding to the $B\to D^\ast\ell\nu$ case at nonzero recoil.

The chiral logs employed in Eq.~\eqref{NLOChCt} are derived following Refs.~\cite{Chow:1993hr,Laiho:2005ue}. Since in this analysis we do not use partially quenched ensembles, the expressions employed are those of
full QCD with 2+1 flavors. Here we reproduce the expression of the logs for the different form factors for the sake of completeness,
\begin{align}
{\rm logs^Y_{SU(3)}} = & \frac{1}{16}\sum_{\Theta}\left(2\bar{F}_{\pi_\Theta}^Y + \bar{F}_{K_\Theta}^Y\right) - \frac{1}{2}\bar{F}^Y_{\pi_I} + \frac{1}{6}\bar{F}^Y_{\eta_I} \nonumber \\
                       & \hspace{-8mm} + \sum_{\Xi=V,A} a^2\delta'_\Xi\Bigg(\frac{m^2_{S_\Xi} - m^2_{\pi_\Xi}}{(m^2_{\eta_\Xi} - m^2_{\pi_\Xi})(m^2_{\pi_\Xi} - m^2_{\eta'_\Xi})}\bar{F}^Y_{\pi_\Xi} \nonumber \\
                       & \hspace{-8mm} + \frac{m^2_{\eta_\Xi} - m^2_{S_\Xi}}{(m^2_{\eta_\Xi} - m^2_{\eta'_\Xi})(m^2_{\eta_\Xi} - m^2_{\pi_\Xi})}\bar{F}^Y_{\eta_\Xi} \nonumber \\
                       & \hspace{-8mm} + \frac{m^2_{S_\Xi} - m^2_{\eta'_\Xi}}{(m^2_{\eta_\Xi} - m^2_{\eta'_\Xi})(m^2_{\eta'_\Xi} - m^2_{\pi_\Xi})}\bar{F}^Y_{\eta'_\xi}\Bigg). \label{ChLogDef}
\end{align}
In the expression above $Y=A_{1,2,3}$, and for the vector form factor ${\rm logs^V_{SU(3)}}={\rm logs^{A_1}_{SU(3)}}$.
The index $\Theta$ goes over all pseudoscalar meson fields of the effective theory, whereas $\Xi$ labels vector and axial counterparts.
Explicit expressions for the mass of a pseudoscalar meson $P$ with taste $\Xi$, $m_{P_\Xi}$, in terms of the parameters of the rooted staggered theory can be found in Ref.~\cite{Aubin:2003mg}.
The hairpin parameters, which come from $\chi$PT disconnected diagrams, are marked as $\delta'_{V,A}$.
Both vector and axial hairpin parameters are defined for the lightest coarse ensemble $a\approx 0.12$fm to be $\delta'_A = -0.28(6)$ and $\delta'_V=0.00(7)$, and their value is obtained for other ensembles by rescaling this number,
assuming the hairpin parameter scales as the root mean square of the taste splittings.
The taste splittings along with the tree-level LEC $B_0$ depend solely on the lattice spacing and are given in Table~\ref{ChPTinput}.
The values for the hairpin parameters, as well as those quoted in Table~\ref{ChPTinput}, were determined from fits to light-quark quantities by the MILC collaboration in Ref.~\cite{Bazavov:2009bb}.
Our results are insensitive to the errors in the parameters quoted in the table, as well as the errors in the hairpin parameters, since the chiral logs are a very small contribution to our fit function in the
region where we have data.
%the following table:
\begin{table}
  \scriptsize
  \setlength\tabcolsep{0pt} 
  \caption{Taste splittings and $B_0$ LEC used to compute the meson masses at tree level in $\chi$PT.\label{ChPTinput}}
  \begin{tabular}{SSSSSS}
  \hline
  \hline
  {\Hs{3mm}$a$(fm)}\Tstrut&{\Hs{5mm}$r_1^2a^2\Delta_I$} &{\Hs{5mm}$r_1^2a^2\Delta_V$}&{\Hs{5mm}$r_1^2a^2\Delta_T$}&{\Hs{5mm}$r_1^2a^2\Delta_A$}&{\Hs{5mm}$r_1B_0$}\\
  \hline
    0,15                  &           0,9851            &           0,7962           &           0,6178           &           0,3915           &       6,761      \\
    0,12                  &           0,6008            &           0,4803           &           0,3662           &           0,2270           &       6,832      \\
    0,09                  &           0,2207            &           0,1593           &           0,1238           &           0,0747           &       6,639      \\
    0,06                  &           0,0704            &           0,0574           &           0,0430           &           0,0263           &       6,487      \\
    0,045                 &           0,0278            &           0,0227           &           0,0170           &           0,0104           &       6,417      \\
  {Continuum}             &         {\Hs{5mm}---}       &         {\Hs{5mm}---}      &         {\Hs{5mm}---}      &       {\Hs{5mm}---}        &       6,015      \\
  \hline
  \hline
  \end{tabular}
\end{table}

The functions appearing in Eq.~(\ref{ChLogDef}) are given by
\begin{equation}
\bar{F}^Y_j \equiv F^Y(w,m_j,-\Delta^{(c)}/m_j),
\end{equation}
where $\Delta^{(c)}$ is the $D$-$D^*$ mass splitting that we take from the PDG~\cite{Zyla:2020zbs}, assuming a charged meson.
The error from the PDG uncertainty, even if enlarged to cover a neutral-meson mass difference, also has a negligible impact in the final fit.
The $F^Y$ functions are defined as
\begin{align}
F^{A_1}(w,m,x) &= -2\Big[I_1(w,m,x) - \frac{1}{2}I_3(w,m,x) \nonumber \\
               &\hspace{-0.5cm} + (w+1)I_1(w,m,0) + (w^2-1)I_2(w,m,0) \nonumber \\
               &\hspace{-0.5cm} - \frac{5}{2}I_3(w,m,0)\Big], \\
F^{A_2}(w,m,x) &= -2\bigg[I_1(w,m,x) + (w+1)I_2(w,m,x) \nonumber \\
               &\hspace{-0.5cm} - I_1(w,m,0) - (w+1)I_2(w,m,0)\bigg], \\
F^{A_3}(w,m,x) &= -2\Big[-(w+1)I_2(w,m,x) \nonumber \\
               &\hspace{-0.5cm} - \frac{1}{2}I_3(w,m,x) + (w+2) I_1(w,m,0) \nonumber \\
               &\hspace{-0.5cm} + w(w+1)I_2(w,m,0) - \frac{5}{2}I_3(w,m,0)\Big],
\end{align}
with
\begin{align}
I_j(w,m,x) &= - \Big[m^2 x E_j(w) + m^2x^2 {\rm ln}\left(\frac{m^2}{\Lambda_\chi^2}\right)G_j(w) \nonumber \\
           &\hspace*{2.5cm} + m^2x^2F_j(w,x)\Big].
\end{align}
The functions $E_j(w)$ and $G_j(w)$ are
\begin{align}
E_1(w) =& \frac{\pi}{w+1},      & G_1(w) =& \frac{r(w)-w}{2(w^2-1)}, \\ 
E_2(w) =& \frac{-\pi}{(w+1)^2}, & G_2(w) =& \frac{w^2 + 2 - 3wr(w)}{2(w^2-1)^2}, \\ 
E_3(w) =& \pi,                  & G_3(w) =& -1.
\end{align}
Here $r(w)$ is
\begin{equation}
r(w) = \frac{1}{\sqrt{w^2-1}}\ln(w + \sqrt{w^2-1}).
\end{equation}
The $F_j$ functions in general cannot be expressed in closed form, and are defined as integrals,
\begin{strip}
\begin{align}
F_1(w,x) =& \frac{1}{x^2}\int^{\pi/2}_0 d\theta\frac{a}{ 1+w\sin 2\theta   }\left\{\pi\left(\sqrt{1-a^2}-1\right)-2\left[\frac{1}{2}\sqrt{a^2-1}\ln\left(1-2a(a+\sqrt{a^2-1})\right) - a\right]\right\}, \\
F_2(w,x) =& \frac{1}{x^2}\int^{\pi/2}_0 d\theta\frac{a\sin 2\theta}{(1+w\sin 2\theta)^2}\left\{-\frac{3\pi}{2}\left(\sqrt{1-a^2}-1\right) + \frac{\pi a^2}{2\sqrt{1-a^2}}\right. \nonumber \\
          & + \left.\left[\left(\frac{3-4a^2}{\sqrt{a^2-1}}\right)\left(-\frac{1}{2}\ln\left(1-2a(a+\sqrt{a^2-1})\right)\right) - 3a\right]\right\}, \\
F_3(w,x) =& \frac{1}{x}\left\{\pi\left(\sqrt{1-x^2}-1\right) - 2\left[\frac{1}{2}\sqrt{x^2-1}\ln\left(1-2x(x+\sqrt{x^2-1})\right) - x\right]\right\},
\end{align}
\end{strip}
where
\begin{equation}
a = \frac{x\cos\theta}{\sqrt{1+w\sin 2\theta}}\,.
\end{equation}
%Pag 35 B to pi universal functions

\section{Estimation of the matching factors and their errors}
\label{ApMatch}

In this Appendix, we derive the renormalization factors for the vector and axial-vector currents using heavy-quark effective 
theory (HQET) as an intermediary between lattice gauge theory and continuum QCD~\cite{Kronfeld:2000ck,Harada:2001fj}.
We have calculated the renormalization factors at one loop in perturbation theory, sometimes using an expedient approximation 
described below.
These one-loop results are shown in Table~\ref{TableMatchingFactors}.
Here, we use the notation $p_B=p$, $v_B=v$, $p_{D^\ast}=p'$, and $v_{D^\ast}=v'$.

\subsection{HQET Matching}
\label{sec:matching}

The HQET description of lattice-QCD currents is~\cite{Harada:2001fj}
\begin{align}
    V^\mu &\doteq \bar{C}_{V_\parallel}^\text{LGT}(w)v^\mu\bar{c}_{v'}b_v +
        \bar{C}_{V_\perp}^\text{LGT}(w)\bar{c}_{v'}i\gamma^\mu_\perp b_v \nonumber \\
        &\phantom{=} + \bar{C}_{V_{v'}}^\text{LGT}(w){v'_\perp\!}^\mu\bar{c}_{v'}b_v , \label{eq:VLGT} \\
    A^\mu &\doteq \bar{C}_{V_\perp}^\text{LGT}(w)\bar{c}_{v'}i\gamma^\mu_\perp\gamma^5b_v -
        \bar{C}_{V_\parallel}(w)^\text{LGT}v^\mu\bar{c}_{v'}\gamma^5b_v \nonumber \\
        &\phantom{=} - \bar{C}_{V_{v'}}^\text{LGT}(w){v'_\perp\!}^\mu\bar{c}_{v'}\gamma^5b_v . \label{eq:ALGT} 
\end{align}
where $\doteq$ means ``has the same matrix elements as''. The currents on the left-hand side are lattice operators, while the
bilinears on the right-hand side are HQET operators (e.g., built from fields satisfying $\vslash b_v=ib_v$).
Similarly, HQET can be used to describe continuum-QCD currents.
\begin{align}
    \mathscr{V}^\mu &\doteq \bar{C}_{V_\parallel}(w)v^\mu\bar{c}_{v'}b_v +
        \bar{C}_{V_\perp}(w)\bar{c}_{v'}i\gamma^\mu_\perp b_v \nonumber \\
        &\phantom{=} + \bar{C}_{V_{v'}}(w){v'_\perp\!}^\mu\bar{c}_{v'}b_v , \label{eq:VQCD} \\
    \mathscr{A}^\mu &\doteq \bar{C}_{V_\perp}(w)\bar{c}_{v'}i\gamma^\mu_\perp\gamma^5b_v -
        \bar{C}_{V_\parallel}(w)v^\mu\bar{c}_{v'}\gamma^5b_v \nonumber \\
        &\phantom{=} - \bar{C}_{V_{v'}}(w){v'_\perp\!}^\mu\bar{c}_{v'}\gamma^5b_v , \label{eq:AQCD} 
\end{align}
The only difference between the lattice and the continuum currents lies in the short-distance coefficients;
for Wilson-like fermions, as used here,
\begin{equation}
    \lim_{a\to0} \bar{C}_J^\text{LGT}(w) = \bar{C}_J(w).
\end{equation}
Eqs.~(\ref{eq:VLGT})--(\ref{eq:AQCD}) are valid at zeroth order in $1/m_Q$, $Q=c,b$, which suffices here.%
\footnote{Matching at the next order in $1/m_Q$ is possible in principle but cumbersome beyond the tree level.}

To continue with the analysis, it is advantageous to write explicitly the velocity and polarization vectors of our mesons.
Assuming the $B$~meson to be at rest, and the $D^\ast$ to have momentum $p'$ along the $z$ axis:
\begin{align}
    v  &= (0,0,0,i) = p/M_B , \\
    v' &= (0,0,\sqrt{w^2-1},iw) = p'/M_{D^\ast}, \\
    v'_\perp &= (0,0,\sqrt{w^2-1},0) , \\
    \hat{v}'_\perp &\equiv v'_\perp/(v'_\perp\cdot v'_\perp)^{1/2} = (0,0,1,0) , \label{eq:unit-p} \\
    \epsilon^\pm &= (\bm{\epsilon}^\pm,0,0) , \\
    \epsilon^3   &= (0,0,w,i\sqrt{w^2-1}) , \\
    \epsilon^s   &= v',
\end{align}
where $\bm{\epsilon}^\pm$ are two-component unit vectors in the $xy$ plane.
For momenta in other directions, such as $\bm{v}'\propto(1,1,0)$ or $(1,-1,1)$, one can rotate the spatial components accordingly.
Note that we pick $\hat{v}'_\perp$, and then we deduce $\epsilon^\pm$.

In this notation, $v\cdot v=-1$, $v'\cdot v'=-1$, and $v\cdot v'=-w$.%
\footnote{These conventions also hold in Minkowski space when using the metric $g^{\mu\nu}=\diag(-1,1,1,1)$ \cite{Kronfeld:2000ck}.
NB: in Minkowski space, $\mu\in\{0,1,2,3\}$; in Euclidean space, $g^{\mu\nu}=\delta^{\mu\nu}$ and $\mu\in\{1,2,3,4\}$; $x^4=ix^0$.}
The polarization vectors satisfy $\bar{\epsilon}^m\cdot\epsilon^n=g^{mn}$ for $(m,n)\in\{+,-,3,s\}$ with
$g^{mn}=\diag(1,1,1,-1)$.
The bar on a polarization vector means to complex-conjugate the spatial components, which arises if (as usual) $\epsilon^\pm$
corresponds to $D^\ast$ helicity~$\pm1$.
(We use linear polarizations with real $\bm{\epsilon}$ but keep the $\pm$ notation.)
The polarization vectors satisfy $v'\cdot\epsilon^{\pm,3}=0$, and also $v\cdot\epsilon^{\pm}=0$.

These vectors can be used to isolate parts of the currents with different matching factors:
\begin{align}
    -v            \cdot V &\doteq \bar{C}_{V_\parallel}^\text{LGT}(w)\bar{c}_{v'}b_v , \\
    \hat{v}'_\perp\cdot V &\doteq \bar{C}_{V_\perp}^\text{LGT}(w)\bar{c}_{v'}i\vhslash'_\perp   b_v +
        \sqrt{w^2-1}\bar{C}_{V_{v'}}^\text{LGT}(w)\bar{c}_{v'}b_v, \\
    \epsilon^\pm  \cdot V &\doteq \bar{C}_{V_\perp}^\text{LGT}(w)\bar{c}_{v'}i\eslash^\pm_\perp b_v ,
\end{align}
and similarly for $\mathscr{V}$ (without the superscript ``LGT'') and for $A$ and $\mathscr{A}$ with
$\bar{c}_{v'}\to\bar{c}_{v'}\gamma^5$.

In HQET, the matrix elements can be worked out with the ``trace formalism'' \cite{Falk:1992wt}:
\begin{equation}
    \langle D^\ast|\bar{c}_{v'}\Gamma b_v|B\rangle = \sqrt{M_{D^\ast}M_B}\,
        \tr\left[\bar{\mathscr{M}}_{D^\ast}\Gamma\mathscr{M}_B\right] \xi(w) ,
\end{equation}
where $\xi(w)$ is a form factor known as the ``Isgur-Wise function'', and
\begin{equation}
    \mathscr{M}_B = -\half(1-i\vslash)\gamma^5, \qquad
    \bar{\mathscr{M}}_{D^\ast} = \ebslash\half(1-i\vslash'),
\end{equation}
and similarly for $\bar{\mathscr{M}}_D$ and $\mathscr{M}_{D^\ast}$.
Then
\begin{subequations}
    \label{eq:traces}
\begin{align}
    \tr\left[\bar{\mathscr{M}}_{D^\ast}\mathscr{M}_B\right] &= 0 , \\
    \tr\left[\bar{\mathscr{M}}_{D^\ast}i\gamma^\mu_\perp        \mathscr{M}_B\right] &=
        \varepsilon^{\mu\nu\alpha\beta} \bar{\epsilon}_\nu v'_\alpha v_\beta , \\
    \tr\left[\bar{\mathscr{M}}_{D^\ast}(-\gamma^5)\mathscr{M}_B\right] &= i\bar{\epsilon}\cdot v , \\
    \tr\left[\bar{\mathscr{M}}_{D^\ast}i\gamma^\mu_\perp\gamma^5\mathscr{M}_B\right] &=
        -i\left[(w+1)\bar{\epsilon}^\mu_\perp + \bar{\epsilon}\cdot v v'_\perp\right] , \\[0.25em]
    \tr\left[\bar{\mathscr{M}}_D\mathscr{M}_B\right] &= w+1 , \\
    \tr\left[\bar{\mathscr{M}}_Di\gamma^\mu_\perp        \mathscr{M}_B\right] &= {v'_\perp}^\mu , \\
    \tr\left[\bar{\mathscr{M}}_D(-\gamma^5)\mathscr{M}_B\right] &= 0 , \\
    \tr\left[\bar{\mathscr{M}}_Di\gamma^\mu_\perp\gamma^5\mathscr{M}_B\right] &= 0 , \\[0.25em]
    \tr\left[\bar{\mathscr{M}}_{D^\ast}\mathscr{M}_{B^*}\right] &= (w+1)\bar{\epsilon}'\cdot\epsilon +
        \bar{\epsilon}'\cdot v\,\epsilon\cdot v' , \\
    \tr\left[\bar{\mathscr{M}}_{D^\ast}i\gamma^\mu_\perp        \mathscr{M}_{B^*}\right] &=
        v_\perp^{\prime\,\mu} \bar{\epsilon}'\cdot\epsilon
        - \epsilon_\perp^\mu \bar{\epsilon}'\cdot v - \bar{\epsilon}^{\prime\,\mu}_\perp \epsilon\cdot v', \\
%     v'_{\perp\,\mu}\tr\left[\bar{\mathscr{M}}_{D^\ast}i\gamma^\mu_\perp        \mathscr{M}_{B^*}\right] &= (w-1)\left[
%         (w+1)\bar{\epsilon}'\cdot\epsilon + \bar{\epsilon}'\cdot v\,\epsilon\cdot v' \right], \\
    \tr\left[\bar{\mathscr{M}}_{D^\ast}(-\gamma^5)\mathscr{M}_{B^*}\right] &= 
        \varepsilon^{\nu\rho\alpha\beta}\bar{\epsilon}'_\nu\epsilon_\rho v'_\alpha v_\beta , \\
    \tr\left[\bar{\mathscr{M}}_{D^\ast}i\gamma^\mu_\perp\gamma^5\mathscr{M}_{B^*}\right] &= (\delta^\mu_{\,\tau}+v^\mu v_\tau)
        \varepsilon^{\tau\nu\rho\alpha}\bar{\epsilon}'_\nu(v+v')_\alpha\epsilon_\rho.
\end{align}
\end{subequations}
With these results, physical combinations of the decompositions in Eqs.~(\ref{eq:VLGT})--(\ref{eq:AQCD}) can be more easily
tracked.
Note that the sign convention for $\varepsilon$ cancels in ratios introduced below.

These expressions can be used to relate matrix elements of continuum and LGT currents to each other, via HQET and
Eqs.~(\ref{eq:VLGT})--(\ref{eq:AQCD}).
Thus, $Z_VV$ and $Z_AA$ have the same matrix elements as $\mathscr{V}$ and $\mathscr{A}$ if one chooses matching factors such that
\begin{subequations}
\begin{align}
    \bar{Z}_{V_\parallel}(w)    v         \cdot V &\doteq      v        \cdot \mathscr{V} , \\
    \bar{Z}_{V_{v'}}     (w)\hat{v}'_\perp\cdot V &\doteq \hat{v}'_\perp\cdot \mathscr{V} , \\
    \bar{Z}_{V_\perp}    (w)\epsilon^\pm  \cdot V &\doteq \epsilon^\pm  \cdot \mathscr{V} , \\
    \bar{Z}_{A_\parallel}(w)    v         \cdot A &\doteq      v        \cdot \mathscr{A} , \\
    \bar{Z}_{A_{v'}}     (w)\hat{v}'_\perp\cdot A &\doteq \hat{v}'_\perp\cdot \mathscr{A} , \\
    \bar{Z}_{A_\perp}    (w)\epsilon^\pm  \cdot A &\doteq \epsilon^\pm  \cdot \mathscr{A} . 
\end{align}
\end{subequations}
From these requirements one finds
\begin{subequations}
\begin{align}
    \bar{Z}_{V_\parallel} (w) &= \frac{\bar{C}_{V_\parallel}(w)}{\bar{C}^\text{LGT}_{V_\parallel}(w)} , \\
    \bar{Z}_{V_{v'}}      (w) &= \frac{\bar{C}_{V_\perp}(w)+(w+1)\bar{C}_{V_{v'}}}
        {\bar{C}^\text{LGT}_{V_\perp}(w)+(w+1)\bar{C}^\text{LGT}_{V_{v'}}} ,  \\
    \bar{Z}_{V_\perp}     (w) &= \frac{\bar{C}_{V_\perp}    (w)}{\bar{C}^\text{LGT}_{V_\perp}    (w)} , \\
    \bar{Z}_{A_\parallel} (w) &= \frac{\bar{C}_{A_\parallel}(w)}{\bar{C}^\text{LGT}_{A_\parallel}(w)} , \\
    \bar{Z}_{A_{v'}}      (w) &= \frac{\bar{C}_{A_\perp}(w)+(w-1)\bar{C}_{A_{v'}}}
        {\bar{C}^\text{LGT}_{A_\perp}(w)+(w-1)\bar{C}^\text{LGT}_{A_{v'}}} ,  \\
    \bar{Z}_{A_\perp}     (w) &= \frac{\bar{C}_{A_\perp}    (w)}{\bar{C}^\text{LGT}_{A_\perp}    (w)} ,
\end{align}
\end{subequations}
In anticipation of one-loop perturbative calculations, it is convenient to define
\begin{equation}
    \bar{\rho}_J(w) = \frac{\bar{Z}_J(w)}{\bar{Z}_{V_\parallel,\bar{b}b}^{1/2}(1) \bar{Z}_{V_\parallel,\bar{c}c}^{1/2}(1)}
\end{equation}
to cancel conventional field-normalization factors as well as potentially large tadpole diagrams.
The denominators can be computed nonperturbatively.
With a quantitative method to compute matrix elements, one can obtain the matching factors.
For example, one can use quark states and expand them in perturbative QCD.

\subsection{Useful Ratios}
\label{sec:ratios}

We start with the original double ratio
\begin{equation}
    |R_{A_1}(1)|^2 = \frac{
        \langle D^\ast(\bm{0},\epsilon)|\epsilon\!\cdot\!A|B(\bm{0})\rangle\,
        \langle B  (\bm{0})|\bar{\epsilon}\!\cdot\!A|D^\ast(\bm{0},\epsilon)\rangle}{
        \langle D^\ast(\bm{0},\epsilon)|v\cdot V|D^\ast(\bm{0},\epsilon)\rangle\hfill
        \langle B  (\bm{0})|v\cdot V|B  (\bm{0})\rangle} ,
    \label{eq:RA11}    
\end{equation}
which requires the matching factor
\begin{equation}
    \bar{\rho}^2_{A_\perp}(1) = \frac{\bar{Z}^2_{A_\perp}(1)}{\bar{Z}_{V_\parallel,\bar{b}b}(1)\bar{Z}_{V_\parallel,\bar{c}c}(1)}.
    \label{eq:ZRA11}
\end{equation}

To obtain the $w$ dependence of all four form factors, we define several further ratios:
\begin{subequations}
\label{eq:ratios}
\begin{align}
    Q_{A_1}(w) &= \frac{\langle D^\ast(\bm{p},\epsilon)|\epsilon^\pm\!\cdot\!A|B(\bm{0})\rangle}{
                        \langle D^\ast(\bm{0},\epsilon)|\epsilon^\pm\!\cdot\!A|B(\bm{0})\rangle}, \label{eq:QA1w} \\
    X_{0}  (w) &= \frac{\langle D^\ast(\bm{p},\epsilon)|(-v\!\cdot\!A)|B(\bm{0})\rangle}{
                        \langle D^\ast(\bm{p},\epsilon)|\epsilon^\pm\!\cdot\!A|B(\bm{0})\rangle}, \label{eq:R0w} \\
    X_{1}  (w) &= \frac{\langle D^\ast(\bm{p},\epsilon)|\hat{v}'_\perp\!\cdot\!A|B(\bm{0})\rangle}{
                        \langle D^\ast(\bm{p},\epsilon)|\epsilon^\pm\!\cdot\!A|B(\bm{0})\rangle}, \label{eq:R1w} \\
    X_{V}  (w) &= \frac{\langle D^\ast(\bm{p},\epsilon)|\epsilon^\pm\!\cdot\!V|B(\bm{0})\rangle}{
                        \langle D^\ast(\bm{p},\epsilon)|\epsilon^\pm\!\cdot\!A|B(\bm{0})\rangle}, \label{eq:XVw}
\end{align}
\end{subequations}
which require, respectively, the matching factors
\begin{subequations}
\label{eq:Zbar}
\begin{align}
    \frac{\bar{Z}_{A_\perp}    (w)}{\bar{Z}_{A_\perp}(1)} &= \frac{\bar{\rho}_{A_\perp}    (w)}{\bar{\rho}_{A_\perp}(1)} ,
    \label{eq:ZbarQA1w} \\
    \frac{\bar{Z}_{A_\parallel}(w)}{\bar{Z}_{A_\perp}(w)} &= \frac{\bar{\rho}_{A_\parallel}(w)}{\bar{\rho}_{A_\perp}(w)} ,
    \label{eq:ZbarR0w} \\
    \frac{\bar{Z}_{A_{v'}}     (w)}{\bar{Z}_{A_\perp}(w)} &= \frac{\bar{\rho}_{A_{v'}}     (w)}{\bar{\rho}_{A_\perp}(w)} ,
    \label{eq:ZbarR1w} \\
    \frac{\bar{Z}_{V_\perp}    (w)}{\bar{Z}_{A_\perp}(w)} &= \frac{\bar{\rho}_{V_\perp}    (w)}{\bar{\rho}_{A_\perp}(w)} ,
    \label{eq:ZbarXVw}
\end{align}
\end{subequations}
Note that the form factor $A_1(q^2)\propto h_{A_1}(w)$, defined in Eq.~\eqref{A1LCSR}, comes directly from\linebreak $R_{A_1}Q_{A_1}(w)$,
and $V(q^2)\propto h_V(w)$, defined in Eq.~\eqref{VLCSR}, comes directly from $R_{A_1}Q_{A_1}(w)X_V(w)$, while the helicity amplitudes
$H_0$ and $H_s$ are linear combinations of $R_{A_1}Q_{A_1}(w)X_0(w)$ and $R_{A_1}Q_{A_1}(w)X_1(w)$.
The helicity amplitudes $H_\pm$ come from $A_1$ and $V$, but it is presumably more convenient (or just as convenient) to keep the
axial and vector parts separate.

For the dynamic velocity consider the vector-current matrix element
\begin{align} 
    \langle D^\ast(\bm{p},\epsilon')|\mathscr{V}^\mu|D^\ast(\bm{0},\epsilon)\rangle =&
        \quad\bar{\epsilon}'\cdot\epsilon (p'+p)^\mu f_1(w) \nonumber \\
        & \hspace{-1cm} + \text{terms proportional to } v'\cdot\epsilon, v\cdot\epsilon' .
\end{align}
If we take the same transverse polarization, $\epsilon'=\epsilon=\epsilon^\pm$, for the initial and final $D^\ast$s, then
\begin{align}
    \langle D^\ast(\bm{p})|(-v           \cdot\mathscr{V})|D^\ast(\bm{0})\rangle &= M_{D^\ast} (w + 1)      f_1(w) , \\
    \langle D^\ast(\bm{p})|\hat{v}'_\perp\cdot\mathscr{V} |D^\ast(\bm{0})\rangle &= M_{D^\ast} \sqrt{w^2-1} f_1(w) ,
\end{align}
where $\hat{v}'_\perp$, defined in Eq.~(\ref{eq:unit-p}), is a unit vector in the direction of~$\bm{p}$, such as $(0,0,1)$ or 
$(1,1,0)/\sqrt{2}$.
On the right-hand side of the second equation, $\sqrt{w^2-1}$ is nothing but $|v'_\perp|$, the magnitude of $v'_\perp$; that is, 
$v'_\perp=\sqrt{w^2-1}\hat{v}'_\perp$.
Because $\mathscr{V}$ is properly normalized, it measures the flavor charge, so the form factor satisfies\linebreak $f_1(1)=1$.

Using the trace formalism, it is easy to show
\begin{align}
    \langle D^\ast(\bm{p})|(-v           \cdot V)|D^\ast(\bm{0})\rangle &= M_{D^\ast} (w + 1)
        \bar{C}_{V_\parallel}^\text{LGT}(w)\xi(w) , \\
    \langle D^\ast(\bm{p})|\hat{v}'_\perp\cdot V |D^\ast(\bm{0})\rangle &= M_{D^\ast} \sqrt{w^2-1} \nonumber \\
        &\hspace{-1cm} \left[\bar{C}_{V_\perp}^\text{LGT}(w) + (w+1)\bar{C}_{V_{v'}}^\text{LGT}(w)\right]\xi(w) .
\end{align}
Taking the ratio, the form factor drops out:
\begin{align}
    \frac{\langle D^\ast(\bm{p})|\hat{v}'_\perp\cdot V|D^\ast(\bm{0})\rangle}
         {\langle D^\ast(\bm{p})|(-v\cdot V)|D^\ast(\bm{0})\rangle} = & \nonumber \\
         & \hspace*{-2cm} \frac{|v'_\perp|}{w+1}
           \frac{\bar{C}_{V_\perp}^\text{LGT}(w) + (w+1)\bar{C}_{V_{v'}}^\text{LGT}(w)}{\bar{C}_{V_\parallel}^\text{LGT}(w)},
\end{align}
but a matching factor remains.%
\footnote{We overlooked this factor in previous papers~\cite{Bailey:2012rr,Bailey:2012jg,Lattice:2015rga}.}
The analogous equations for $\mathscr{V}$ shows that $f_1(w)=\bar{C}_{V_\parallel}(w)\xi(w)$ and
$\bar{C}_{V_\parallel}(w)=\bar{C}_{V_\perp}(w) + (w+1)\bar{C}_{V_{v'}}(w)$.

\subsection{Expedient Approximation}
\label{sec:approx}

Calculating the full $w$ dependence of the $\bar{Z}_J$ at the one-loop level is, in general, very cumbersome.
From Ref.~\cite{Harada:2001fj}, however, we have
\begin{align}
    \lim_{m_ca\to0} \bar{Z}_{J_\parallel}(w) &= Z_{J_\parallel} , \\
    \lim_{m_ca\to0} \bar{Z}_{J_\perp}    (w) &= Z_{J_\perp}     , \\
    \lim_{m_ca\to0} \bar{Z}_{J_\perp} \bar{C}^\text{LGT}_{J_{v'}}(w) &= Z_{J_\perp} \bar{C}_{J_{v'}}(w) \nonumber \\
    &\hspace*{1mm}\Rightarrow \lim_{m_ca\to0} \bar{Z}_{J_{v'}}     (w)  = Z_{J_\perp} .
\end{align}
Then we can approximate,
\begin{subequations}
\label{eq:Zratios}
\begin{align}
    \frac{\bar{\rho}_{A_\perp}    (w)}{\bar{\rho}_{A_\perp}(1)} &\approx 1 \pm
        \alpha_V(q^*) \rho^{[1]}_\text{max}(w-1)m_{2c}a ,
    \label{eq:ZQA1w} \\
    \frac{\bar{\rho}_{A_\parallel}(w)}{\bar{\rho}_{A_\perp}(w)} &\approx \frac{\rho_{A_\parallel}}{\rho_{A_\perp}} \pm
        \alpha_V(q^*) \rho^{[1]}_\text{max}m_{2c}a ,
    \label{eq:ZR0w} \\
    \frac{\bar{\rho}_{A_{v'}}     (w)}{\bar{\rho}_{A_\perp}(w)} &\approx 1 \pm
        \alpha_V(q^*) \rho^{[1]}_\text{max}m_{2c}a ,
    \label{eq:ZR1w} \\
    \frac{\bar{\rho}_{V_\perp}    (w)}{\bar{\rho}_{A_\perp}(w)} &\approx \frac{\rho_{V_\perp}    }{\rho_{A_\perp}} \pm
        \alpha_V(q^*) \rho^{[1]}_\text{max}m_{2c}a ,
    \label{eq:ZXVw}
\end{align}
\end{subequations}
where $\rho^{[1]}_\text{max} = 0.352$ is the largest one-loop coefficient that we find among the computable one-loop coefficients.
For lack of a better choice, we set $q^*=2/a$, as in other papers.

In the limit $m_ca\to0$, the matching factor in the velocity tends to~$1$.
We could use an uncertainty like that in Eq.~(\ref{eq:ZR1w}), but that seems to be an unecessary complication.
A little algebra shows that the mismatch in $w$ is very small:
\begin{equation}
    w \approx w^\text{LGT} \pm \alpha_V(q^*) \rho^{[1]}_\text{max}(w^2-1)m_{2c}a,
\end{equation}
while the mismatch in $z$ is
\begin{equation}
    z \approx z^\text{LGT} \left[1 \pm \alpha_V(q^*) 2\rho^{[1]}_\text{max}\frac{1+z}{1-z}m_{2c}a\right].
\end{equation}

\section{Heavy Quark mistuning corrections procedure}
\label{ApHQCorr}

We tune the heavy-quark masses ($\kappa_b$ and $\kappa_c$) for clover quarks in the Fermilab interpretation using the procedure described in Ref.~\cite{Bailey:2014tva}.
In brief, in a mass-\-in\-de\-pend\-ent scheme for the lattice scale, set by $r_1 = 0.3117(22)$ fm~\cite{Bazavov:2011aa}, we tune the heavy-quark masses so that the kinetic masses $M_2$ 
of the $B_s$ and $D_s$ mesons agree with their experimental values~\cite{Bernard:2010fr,Bazavov:2011aa}.
Historically, the tuning was done in two stages.
A preliminary, lower-statistics study set the heavy-quark masses used in the simulation to generate the two- and three-point correlators in the present study.
A subsequent, higher-statistics study was then possible, and it gave slightly different kappa values, as shown in Table~\ref{kappaValues}.
We denote the simulation values with $\kappa_c^\prime$ and $\kappa_b^\prime$ and the refined values with $\kappa_c$ and $\kappa_b$. 
As in Ref.~\cite{Bailey:2014tva}, we proceed to correct our results for the slight mistuning and, in the process, estimate the systematic error associated with the correction.

The correction procedure is based on a single $a\approx 0.12$ fm ensemble.
A full set of two- and three-point correlation functions were calculated at slightly shifted values of $\kappa_b^\prime$ and $\kappa_c^\prime$, as shown in Table~\ref{kCorrEns}.
The shifted values were chosen close to both the corrected and simulation kappa values. With these data, we then
estimate the derivatives of the form factors and the recoil parameter with respect to the heavy quark masses. By expressing these results in dimensionless terms, we can
 extrapolate them to other ensembles with different light-quark masses and lattice spacings. As a departure from Ref.~\cite{Lattice:2015rga}, we perform the correction
after the renormalization factors have been applied.

\begin{table}
  \scriptsize
  \setlength\tabcolsep{2pt}
  \centering
  \caption{Simulation values $\kappa'_c$ and $\kappa'_b$ employed in the calculation of the form factors compared with their
    more precisely tuned values $\kappa_c$ and $\kappa_b$ \cite{Bailey:2014tva}.
    The first error in $\kappa_b$ and $\kappa_c$ includes statistical and fitting contributions; the second one comes from scale setting.}
  \label{kappaValues}
  \begin{tabular}{cccccc}
    \hline
    \hline
            $a$ (fm)          &   $am_l/am_s$  & $\kappa'_b$ &  $\kappa_b$   & $\kappa'_c$ &   $\kappa_c$    \\
    \hline
         $\approx 0.15$       &  0.0097/0.0484 &    0.0781   & 0.0775(16)(3) &    0.1218   & 0.12237(26)(20) \\
    \hline
\MulRow{4}{*}{$\approx 0.12$} &    0.02/0.05   &    0.0918   & 0.0868(9)(3)  &    0.1259   & 0.12423(15)(16) \\
                              &    0.01/0.05   &    0.0901   & 0.0868(9)(3)  &    0.1254   & 0.12423(15)(16) \\
                              &   0.007/0.05   &    0.0901   & 0.0868(9)(3)  &    0.1254   & 0.12423(15)(16) \\
                              &   0.005/0.05   &    0.0901   & 0.0879(9)(3)  &    0.1254   & 0.12452(15)(16) \\
    \hline
\MulRow{5}{*}{$\approx 0.09$} &  0.0124/0.031  &    0.0982   & 0.0964(7)(3)  &    0.1277   & 0.12710(9)(14) \\
                              &  0.0062/0.031  &    0.0979   & 0.0965(7)(3)  &    0.1276   & 0.12714(9)(14) \\
                              & 0.00465/0.031  &    0.0977   & 0.0966(7)(3)  &    0.1275   & 0.12718(9)(14) \\
                              &  0.0031/0.031  &    0.0976   & 0.0967(7)(3)  &    0.1275   & 0.12722(9)(14) \\
                              & 0.00155/0.031  &    0.0976   & 0.0972(7)(3)  &    0.1275   & 0.12737(9)(14) \\
    \hline
\MulRow{4}{*}{$\approx 0.06$} &  0.0072/0.018  &    0.1048   & 0.1050(5)(2)  &    0.1295   & 0.12955(4)(11) \\
                              &  0.0036/0.018  &    0.1052   & 0.1051(5)(2)  &    0.1296   & 0.12957(4)(11) \\
                              &  0.0025/0.018  &    0.1052   & 0.1052(5)(2)  &    0.1296   & 0.12960(4)(11) \\
                              &  0.0018/0.018  &    0.1052   & 0.1054(5)(2)  &    0.1296   & 0.12964(4)(11) \\
    \hline
         $\approx 0.045$      &  0.0028/0.014  &    0.1143   & 0.1116(3)(2)  &    0.1310   & 0.130921(16)(7) \\
    \hline
    \hline
  \end{tabular}
\end{table}

\begin{table}
  \scriptsize
  \setlength\tabcolsep{2pt}
  \centering
  \caption{Ensemble and kappa parameters used to calculate the mistuning correction~\cite{Bailey:2014tva}.
  The values in bold are the uncorrected simulation values for this ensemble.
  Not all combinations are available, since we change only one quark mass at a time while keeping the other quark mass fixed to its uncorrected value.}
  \label{kCorrEns}
  \begin{tabular}{cccccc}
    \hline
    \hline
            $a$ (fm)          & $am_l/am_s$&     Available $\kappa'_b$    &      Available $\kappa'_c$    \\
    \hline
         $\approx 0.12$       &  0.01/0.05 & {\bf 0.0901}, 0.860, 0.820   &  {\bf 0.1254}, 0.1280, 0.1230 \\
    \hline
  \end{tabular}
\end{table}

The first step is to construct the combinations of ratios we need to correct. Clever combinations can remove a dependence on the recoil parameter or vanish at zero recoil.  We can use this information to improve the precision of the corrections. In this case, we define the following quantities:
\begin{align}
A   =& \,(1-x_f^2) R_{A_1},      \label{HQA}  \\
V   =& \,\frac{X_v}{x_f},        \label{HQV}  \\
B_0 =& \,\frac{X_0}{x_f},        \label{HQB0} \\
B_1 =& \,\frac{X_1 - 1}{2x_f^2}, \label{HQB1} \\
C_1 =& \,\frac{X_1 + 1}{2},      \label{HQC1}  
\end{align}
At zero recoil we expect $C_1\to 1$, since $X_1\to 1$ in that limit. Although $x_f$ vanishes when $w\to 1$, the combinations are designed to give a finite value at zero recoil. In fact, it is easy to reconstruct the form factors using these building blocks,
\begin{align}
h_{A_1} =& \,A,  \label{HQ-hA1} \\
h_V     =& \,AV, \label{HQ-hV}  \\ 
h_{A_2} =& \,A(C_1 + B_1 - B_0), \label{HQ-hA2} \\
h_{A_3} =& \,A(C_1 - B_1).       \label{HQ-hA3}
\end{align}
The last quantity corrected is the recoil parameter, defined as in Eq.~\eqref{wDef0}, with a trivial dependence on $\kappa_b$.

We estimate the derivative by taking finite differences between the observables in Eqs.~(\ref{HQA})-(\ref{HQC1}) calculated in the standard run and one of the correction runs listed in Table~\ref{kCorrEns}. The derivative is taken with respect to $\xi_\alpha = 1/am_{2\alpha}$ with $\alpha=b,c$, where the kinetic mass $am_{2\alpha}$ is defined as
\begin{equation}
\frac{1}{am_{2\alpha}} = \frac{2}{am_{0\alpha}(2+am_{0\alpha})} + \frac{1}{1+am_{0\alpha}}, \label{HQ-qKin}
\end{equation}
and the bare quark mass is calculated using the tadpole-improved, tree-level formula
\begin{equation}
am_{0\alpha} = \frac{1}{u_0}\left(\frac{1}{2\kappa_\alpha} - \frac{1}{2\kappa_{cr}}\right). \label{HQ-qBMs}
\end{equation}
Here $u_0$ is the tadpole parameter, and $\kappa_{cr}$ is the value of $\kappa$ at which the clover-quark pion becomes massless. We calculate the derivatives of each observable for each available value of the recoil parameter. When computing the correction for the $b$ quark, since the $\kappa'_b=0.860$
simulation value is so close to the tuned value of $\kappa_b$, we do not use the data coming from $\kappa'_b=0.820$.

In general, the calculated derivative is small for all recoil values, and the errors are large enough that we can consider a linear dependence on $w-1$ for all derivatives. Nonetheless, we can exploit some good properties of our observables in order to reduce the number of free parameters. For instance,
$dC_1/d\xi_\alpha(w=1) = 0$, since $C_1=1$ in that limit, so we can set the constant term to zero. Also, the derivatives of $V,B_{0,1}$ change very little with $w-1$ compared with their errors. Thus, we can safely assume these derivatives behave as constants in the small recoil range we are considering. In fact,
these quantities are derived from ratios of form factors $h_X/h_{A_1}$, which should depend only weakly on $w$. These simplifications are not only welcome, they are necessary since for $V,B_{0,1},C_1$ we have only two values of the recoil parameter available, so our fits can have one degree of freedom. The
observable $A$ is not a ratio {\it per se}, but it is calculated at zero recoil as well, and a clear dependence with $w$ arises. Finally, it is obvious that $dw/d\xi_c(w=1) = 0$, and we can apply the same treatment as for $C_1$. Summarizing, we fit our data to the expressions,

%\vspace{-0.1cm}

  \begin{align}
    \frac{dw}  {d\xi_c} =& \,a_{w,  c} (w-1), \\
    \frac{dA}  {d\xi_c} =& \,a_{A,  c} (w-1) + b_{A,c}, \\
    \frac{dV}  {d\xi_c} =& \,b_{V,  c}, \\
    \frac{dB_0}{d\xi_c} =& \,b_{B_0,c}, \\
    \frac{dB_1}{d\xi_c} =& \,b_{B_1,c}, \\
    \frac{dC_1}{d\xi_c} =& \,a_{C_1,c} (w-1), \\
    \frac{dw}  {d\xi_b} =& \,0, \\
    \frac{dA}  {d\xi_b} =& \,a_{A,  b} (w-1) + b_{A,b}, \\
    \frac{dV}  {d\xi_b} =& \,b_{V,  b}, \\
    \frac{dB_0}{d\xi_b} =& \,b_{B_0,b}, \\
    \frac{dB_1}{d\xi_b} =& \,b_{B_1,b},
  \end{align}
  \begin{align}
    \frac{dC_1}{d\xi_b} =& \,a_{C_1,b} (w-1).
  \end{align}

All fits are unconstrained and result in a $p$~value exceeding 0.5. Our final corrections are gathered in Table~\ref{HQcResults}. These values are used to correct the form factors for all ensembles. As an example, we show in Table~\ref{HQcExample} how this tuning modifies the calculated form factors
for the ensemble used in the correction.

\begin{table*}[h]%[H]
  \centering
  \caption{Results of the $\kappa$ correction fits.\label{HQcResults}}
  \centering
  \begin{tabular*}{\textwidth}{@{\extracolsep{\fill}}cccccccc}
    \hline
    \hline
    Heavy quark  &   $a_w$   &   $a_A$   &   $b_A$   &   $b_V$  & $b_{B_0}$ & $b_{B_1}$ & $a_{C_1}$ \\
    \hline 
    Bottom       &     -     &  1.6(1.5) & -0.23(13) & 0.37(14) & -0.65(15) & -0.35(37) & -0.17(17) \\
    Charm        &  1.14(11) & -0.79(55) & -0.010(31)& 0.186(84)& 0.297(78) & -0.47(30) & -0.38(14) \\
    \hline
  \end{tabular*}
\end{table*}

\begin{table*}[h]%[H]
  \caption{Effect of the mistuning adjustments in the form factors and the recoil parameter of the $a\approx 0.12$ fm ensemble used to compute the correction.
           Only the final error is shown in the table. The error for $w$ seems to decrease with the correction, but the relevant quantity is $w-1$, for which
           the error either stays the same or increases. The correction (and its error) is highly correlated with $w-1$, as it is directly proportional to it.}
  \label{HQcExample}
  \centering
  \begin{tabular*}{\textwidth}{@{\extracolsep{\fill}}ccccccc}
    \hline
    \hline
    \MulRow{2}{*}{Momentum} & \MulCol{2}{c}{$(0,0,0)$} & \MulCol{2}{c}{$(1,0,0)$} & \MulCol{2}{c}{$(2,0,0)$} \\
                            &     Raw    &    Tuned    &     Raw     &   Tuned    &     Raw     &   Tuned    \\
    \hline
               $w$          &   1.0(0)   &    1.0(0)   & 1.0402(40)  & 1.0374(37) &  1.134(21)  &  1.125(20) \\
            $h_{A_1}$       & 0.901(13)  &  0.906(13)  &  0.858(21)  &  0.863(21) &  0.816(30)  &  0.824(30) \\
            $h_{A_2}$       &     -      &      -      &  -0.43(15)  &  -0.40(16) &  -0.49(17)  &  -0.46(17) \\
            $h_{A_3}$       &     -      &      -      &   0.98(12)  &   0.96(12) &   1.02(12)  &   1.01(12) \\
            $h_{ V }$       &     -      &      -      &  1.106(85)  &  1.098(85) &   1.05(11)  &   1.04(11) \\
    \hline
    \hline
  \end{tabular*}
\end{table*}

\section{Full covariance matrix of the chiral-continuum extrapolation results}
\label{ApFullRes}

Future work with the lattice-QCD form factors computed here can start with the synthetic data generated for the $z$~expansion.  We
provide these data for the form factors in the BGL and HQET notation, along with their correlation matrix, in the ancillary
files included in this paper.  We also include the values and correlation matrix of the BGL $z$~expansion
coefficients resulting from our preferred fits from Secs.~\ref{sec:synth} and ~\ref{sec:Vcb}, using only lattice-QCD data (the
quadratic fit in Table~\ref{SynthResults}) and lattice QCD plus experimental data (the last column in Table~\ref{CompResults}).  Our
BGL lattice-QCD-only fit can be reproduced by fitting the included synthetic data in the BGL notation to the BGL expansion described
in Sec.~\ref{zExpSec}, and using the constraint at zero recoil given in Eq.~\eqref{KinF1} to remove $c_0$.

\subsection{Reading the data}

The data is provided in Python format using the {\tt gvar} package~\cite{peter_lepage_2021_4695132}.  The synthetic data points can
be read from the file {\tt SynthData.PyDat} using the following code:
\begin{center}
\begin{minipage}{\linewidth}
% \centering
\begin{lstlisting}[language=python, frame=single]
import gvar

data = gvar.load("SynthData.PyDat")
corr = gvar.evalcorr(data)
\end{lstlisting}
\end{minipage}
\end{center}
After execution, {\tt data} holds a dictionary whose keys are in the format {\tt F(w)}, where {\tt F $=$ g, f, F1 or F2} for the $g$, $f$,
$\mathcal{F}_1$ and $\mathcal{F}_2$ form factors in the BGL notation, or {\tt F $=$ hA1, hA2, hA3 or V} for the $h_{A_i}$ and $h_V$
form factors in the HQET notation, and {\tt w} is the recoil parameter, $w = 1.03, 1.10, 1.17$.  The $h_X$ form factors are also
available at zero recoil, but the extra data point at $w = 1.00$ is not independent and should be used only if one of the other
available points is removed.  The last line stores the correlations between the different synthetic data points in a dictionary
called {\tt corr}.  One can access any correlation by invoking the pair {\tt corr[F(w1),G(w2)]}, where {\tt F} and {\tt G} are the
two relevant form factors, and {\tt w1} and {\tt w2} are the recoil parameters at which the correlation should be evaluated.

The $z$-fit results are stored in the file\linebreak {\tt FitResults.PyDat} and can be read following the same procedure.  The dictionary keys in
this case are in the format {\tt FIT\_xj}, where {\tt FIT} can be either {\tt LQCD} for the lattice-QCD-only fit, or {\tt LQCDEXP}
for the fit including lattice-QCD and experimental data, and\linebreak {\tt xj = a0, a1}\ldots are the different coefficients of the
$z$~expansion $a_0$, $a_1$\ldots Our final result for $|\eta_{\text{EW}}|^2|V_{cb}|^2$ is stored under the key {\tt LQCDEXP\_eVcb},
and the results for $R(D^\ast)$ are stored as {\tt LQCD\_RDst} and {\tt LQCDEXP\_RDst}.

\bibliographystyle{spphys}
\bibliography{Bibliography}

\begin{thebibliography}{100}
\providecommand{\url}[1]{{#1}}
\providecommand{\urlprefix}{URL }
\expandafter\ifx\csname urlstyle\endcsname\relax
  \providecommand{\doi}[1]{DOI \discretionary{}{}{}#1}\else
  \providecommand{\doi}{DOI \discretionary{}{}{}\begingroup
  \urlstyle{rm}\Url}\fi

\bibitem{Amhis:2019ckw}
Y.S. Amhis, et~al., Eur. Phys. J. C \textbf{81}(3), 226 (2021).
\newblock \doi{10.1140/epjc/s10052-020-8156-7}

\bibitem{Zyla:2020zbs}
P.A. Zyla, et~al., PTEP \textbf{2020}(8), 083C01 (2020).
\newblock \doi{10.1093/ptep/ptaa104}

\bibitem{Crivellin:2014zpa}
A.~Crivellin, S.~Pokorski, Phys. Rev. Lett. \textbf{114}(1), 011802 (2015).
\newblock \doi{10.1103/PhysRevLett.114.011802}

\bibitem{Jung:2018lfu}
M.~Jung, D.M. Straub, JHEP \textbf{01}, 009 (2019).
\newblock \doi{10.1007/JHEP01(2019)009}

\bibitem{Boyd:1995sq}
C.G. Boyd, B.~Grinstein, R.F. Lebed, Nucl. Phys. \textbf{B461}, 493 (1996).
\newblock \doi{10.1016/0550-3213(95)00653-2}

\bibitem{Boyd:1995cf}
C.G. Boyd, B.~Grinstein, R.F. Lebed, Phys. Lett. B \textbf{353}, 306 (1995).
\newblock \doi{10.1016/0370-2693(95)00480-9}

\bibitem{Boyd:1997kz}
C.G. Boyd, B.~Grinstein, R.F. Lebed, Phys. Rev. D \textbf{56}, 6895 (1997).
\newblock \doi{10.1103/PhysRevD.56.6895}

\bibitem{Caprini:1997mu}
I.~Caprini, L.~Lellouch, M.~Neubert, Nucl. Phys. \textbf{B530}, 153 (1998).
\newblock \doi{10.1016/S0550-3213(98)00350-2}

\bibitem{Aubert:2007qs}
B.~Aubert, et~al., Phys. Rev. Lett. \textbf{100}, 231803 (2008).
\newblock \doi{10.1103/PhysRevLett.100.231803}

\bibitem{Adam:2002uw}
N.E. Adam, et~al., Phys. Rev. D \textbf{67}, 032001 (2003).
\newblock \doi{10.1103/PhysRevD.67.032001}

\bibitem{Aubert:2008yv}
B.~Aubert, et~al., Phys. Rev. D \textbf{79}, 012002 (2009).
\newblock \doi{10.1103/PhysRevD.79.012002}

\bibitem{Amhis:2012bh}
Y.~Amhis, et~al.,   (2012).
\newblock ArXiv:1207.1158

\bibitem{Abdesselam:2017kjf}
A.~Abdesselam, et~al.,   (2017).
\newblock ArXiv:1702.01521

\bibitem{Bigi:2017njr}
D.~Bigi, P.~Gambino, S.~Schacht, Phys. Lett. \textbf{B769}, 441 (2017).
\newblock \doi{10.1016/j.physletb.2017.04.022}

\bibitem{Bigi:2017jbd}
D.~Bigi, P.~Gambino, S.~Schacht, JHEP \textbf{11}, 061 (2017).
\newblock \doi{10.1007/JHEP11(2017)061}

\bibitem{Grinstein:2017nlq}
B.~Grinstein, A.~Kobach, Phys. Lett. \textbf{B771}, 359 (2017).
\newblock \doi{10.1016/j.physletb.2017.05.078}

\bibitem{Jaiswal:2017rve}
S.~Jaiswal, S.~Nandi, S.K. Patra, JHEP \textbf{12}, 060 (2017).
\newblock \doi{10.1007/JHEP12(2017)060}

\bibitem{Waheed:2018djm}
E.~Waheed, et~al., Phys. Rev. D \textbf{100}(5), 052007 (2019).
\newblock \doi{10.1103/PhysRevD.100.052007}

\bibitem{Dey:2019bgc}
J.~Lees, et~al., Phys. Rev. Lett. \textbf{123}(9), 091801 (2019).
\newblock \doi{10.1103/PhysRevLett.123.091801}

\bibitem{Gambino:2019sif}
P.~Gambino, M.~Jung, S.~Schacht, Phys. Lett. B \textbf{795}, 386 (2019).
\newblock \doi{10.1016/j.physletb.2019.06.039}

\bibitem{Jaiswal:2020wer}
S.~Jaiswal, S.~Nandi, S.K. Patra, JHEP \textbf{06}, 165 (2020).
\newblock \doi{10.1007/JHEP06(2020)165}

\bibitem{Bailey:2014tva}
J.A. Bailey, et~al., Phys. Rev. \textbf{D89}(11), 114504 (2014).
\newblock \doi{10.1103/PhysRevD.89.114504}

\bibitem{McLean:2019sds}
E.~McLean, C.T.H. Davies, A.T. Lytle, J.~Koponen, Phys. Rev. D \textbf{99}(11),
  114512 (2019).
\newblock \doi{10.1103/PhysRevD.99.114512}

\bibitem{Bifani:2018zmi}
S.~Bifani, S.~Descotes-Genon, A.~Romero~Vidal, M.H. Schune, J. Phys. G
  \textbf{46}(2), 023001 (2019).
\newblock \doi{10.1088/1361-6471/aaf5de}

\bibitem{Hirose:2016wfn}
S.~Hirose, et~al., Phys. Rev. Lett. \textbf{118}(21), 211801 (2017).
\newblock \doi{10.1103/PhysRevLett.118.211801}

\bibitem{Hirose:2017dxl}
S.~Hirose, et~al., Phys. Rev. \textbf{D97}(1), 012004 (2018).
\newblock \doi{10.1103/PhysRevD.97.012004}

\bibitem{Aaij:2017uff}
R.~Aaij, et~al., Phys. Rev. Lett. \textbf{120}(17), 171802 (2018).
\newblock \doi{10.1103/PhysRevLett.120.171802}

\bibitem{Aaij:2017deq}
R.~Aaij, et~al., Phys. Rev. \textbf{D97}(7), 072013 (2018).
\newblock \doi{10.1103/PhysRevD.97.072013}

\bibitem{Belle:2019rba}
G.~Caria, et~al., Phys. Rev. Lett. \textbf{124}(16), 161803 (2020).
\newblock \doi{10.1103/PhysRevLett.124.161803}

\bibitem{Hashimoto:2001nb}
S.~Hashimoto, A.S. Kronfeld, P.B. Mackenzie, S.M. Ryan, J.N. Simone, Phys. Rev.
  \textbf{D66}, 014503 (2002).
\newblock \doi{10.1103/PhysRevD.66.014503}

\bibitem{Bernard:2008dn}
C.~Bernard, et~al., Phys. Rev. \textbf{D79}, 014506 (2009).
\newblock \doi{10.1103/PhysRevD.79.014506}

\bibitem{Harrison:2017fmw}
J.~Harrison, C.~Davies, M.~Wingate, Phys. Rev. D \textbf{97}(5), 054502 (2018).
\newblock \doi{10.1103/PhysRevD.97.054502}

\bibitem{deDivitiis:2008df}
G.M. de~Divitiis, R.~Petronzio, N.~Tantalo, Nucl. Phys. B \textbf{807}, 373
  (2009).
\newblock \doi{10.1016/j.nuclphysb.2008.09.013}

\bibitem{Qiu:2013ofa}
S.W. Qiu, C.~DeTar, A.X. El-Khadra, A.S. Kronfeld, J.~Laiho, R.S. Van~de Water,
  PoS \textbf{LATTICE2013}, 385 (2014).
\newblock \doi{10.22323/1.187.0385}

\bibitem{Aviles-Casco:2017nge}
A.~Vaquero Avil\'es-Casco, C.~DeTar, D.~Du, A.~El-Khadra, A.S. Kronfeld,
  J.~Laiho, R.S. Van~de Water, EPJ Web Conf. \textbf{175}, 13003 (2018).
\newblock \doi{10.1051/epjconf/201817513003}

\bibitem{Aviles-Casco:2019vin}
A.V. Avil\'es-Casco, C.~DeTar, A.X. El-Khadra, A.S. Kronfeld, J.~Laiho, R.S.
  Van~de Water, PoS \textbf{LATTICE2018}, 282 (2019).
\newblock \doi{10.22323/1.334.0282}

\bibitem{Vaquero:2019ary}
A.~Vaquero, C.~DeTar, A.X. El-Khadra, A.S. Kronfeld, J.~Laiho, R.S. Van~de
  Water, in \emph{{17th Conference on Flavor Physics and CP Violation}} (2019).
\newblock ArXiv:1906.01019

\bibitem{Aviles-Casco:2019zop}
A.V. Avil\'es-Casco, C.~DeTar, A.X. El-Khadra, A.S. Kronfeld, J.~Laiho, R.S.
  Van~de Water, PoS \textbf{LATTICE2019}, 049 (2019).
\newblock \doi{10.22323/1.363.0049}

\bibitem{Lattice:2015tia}
J.A. Bailey, et~al., Phys. Rev. D \textbf{92}(1), 014024 (2015).
\newblock \doi{10.1103/PhysRevD.92.014024}

\bibitem{Bailey:2015nbd}
J.A. Bailey, et~al., Phys. Rev. Lett. \textbf{115}(15), 152002 (2015).
\newblock \doi{10.1103/PhysRevLett.115.152002}

\bibitem{Bailey:2015dka}
J.A. Bailey, et~al., Phys. Rev. D \textbf{93}(2), 025026 (2016).
\newblock \doi{10.1103/PhysRevD.93.025026}

\bibitem{Bazavov:2019aom}
A.~Bazavov, et~al., Phys. Rev. D \textbf{100}(3), 034501 (2019).
\newblock \doi{10.1103/PhysRevD.100.034501}

\bibitem{Falk:1992wt}
A.F. Falk, M.~Neubert, Phys. Rev. D \textbf{47}, 2965 (1993).
\newblock \doi{10.1103/PhysRevD.47.2965}

\bibitem{Korner:1989qb}
J.G. Körner, G.A. Schuler, Z. Phys. C \textbf{46}, 93 (1990).
\newblock \doi{10.1007/BF02440838}

\bibitem{Sirlin:1981ie}
A.~Sirlin, Nucl. Phys. \textbf{B196}, 83 (1982).
\newblock \doi{10.1016/0550-3213(82)90303-0}

\bibitem{Ginsberg:1968pz}
E.S. Ginsberg, Phys. Rev. \textbf{171}, 1675 (1968).
\newblock \doi{10.1103/PhysRev.171.1675, 10.1103/PhysRev.174.2169.3}.
\newblock [Erratum: Phys. Rev.174,2169(1968)]

\bibitem{Ginsberg:1969jh}
E.S. Ginsberg, Phys. Rev. \textbf{162}, 1570 (1967).
\newblock \doi{10.1103/physrev.187.2280.2, 10.1103/PhysRev.162.1570}.
\newblock [Erratum: Phys. Rev.187,2280(1969)]

\bibitem{Atwood:1989em}
D.~Atwood, W.J. Marciano, Phys. Rev. \textbf{D41}, 1736 (1990).
\newblock \doi{10.1103/PhysRevD.41.1736}

\bibitem{ElKhadra:1996mp}
A.X. El-Khadra, A.S. Kronfeld, P.B. Mackenzie, Phys. Rev. \textbf{D55}, 3933
  (1997).
\newblock \doi{10.1103/PhysRevD.55.3933}

\bibitem{Harada:2001fj}
J.~Harada, S.~Hashimoto, A.S. Kronfeld, T.~Onogi, Phys. Rev. \textbf{D65},
  094514 (2002).
\newblock \doi{10.1103/PhysRevD.65.094514}

\bibitem{ElKhadra:2001rv}
A.X. El-Khadra, A.S. Kronfeld, P.B. Mackenzie, S.M. Ryan, J.N. Simone, Phys.
  Rev. \textbf{D64}, 014502 (2001).
\newblock \doi{10.1103/PhysRevD.64.014502}

\bibitem{Bailey:2012rr}
J.A. Bailey, et~al., Phys. Rev. D \textbf{85}, 114502 (2012).
\newblock \doi{10.1103/PhysRevD.85.114502}.
\newblock [Erratum: Phys.Rev.D 86, 039904 (2012)]

\bibitem{Lattice:2015rga}
J.A. Bailey, et~al., Phys. Rev. \textbf{D92}(3), 034506 (2015).
\newblock \doi{10.1103/PhysRevD.92.034506}

\bibitem{Bazavov:2009bb}
A.~Bazavov, et~al., Rev. Mod. Phys. \textbf{82}, 1349 (2010).
\newblock \doi{10.1103/RevModPhys.82.1349}

\bibitem{Aubin:2004wf}
C.~Aubin, C.~Bernard, C.~DeTar, J.~Osborn, S.~Gottlieb, E.B. Gregory,
  D.~Toussaint, U.M. Heller, J.E. Hetrick, R.~Sugar, Phys. Rev. D \textbf{70},
  094505 (2004).
\newblock \doi{10.1103/PhysRevD.70.094505}

\bibitem{Bernard:2001av}
C.W. Bernard, T.~Burch, K.~Orginos, D.~Toussaint, T.A. DeGrand, C.E. Detar,
  S.~Datta, S.A. Gottlieb, U.M. Heller, R.~Sugar, Phys. Rev. D \textbf{64},
  054506 (2001).
\newblock \doi{10.1103/PhysRevD.64.054506}

\bibitem{Wingate:2002fh}
M.~Wingate, J.~Shigemitsu, C.T.H. Davies, G.P. Lepage, H.D. Trottier, Phys.
  Rev. \textbf{D67}, 054505 (2003).
\newblock \doi{10.1103/PhysRevD.67.054505}

\bibitem{Bazavov:2016nty}
A.~Bazavov, et~al., Phys. Rev. D \textbf{93}(11), 113016 (2016).
\newblock \doi{10.1103/PhysRevD.93.113016}

\bibitem{Bazavov:2017weg}
A.~Bazavov, et~al., Phys. Rev. D \textbf{97}(3), 034513 (2018).
\newblock \doi{10.1103/PhysRevD.97.034513}

\bibitem{MILC:2010iid}
A.~Bazavov, et~al., Phys. Rev. D \textbf{81}, 114501 (2010).
\newblock \doi{10.1103/PhysRevD.81.114501}

\bibitem{FermilabLattice:2019ycs}
R.~Li, et~al., PoS \textbf{LATTICE2018}, 269 (2019).
\newblock \doi{10.22323/1.334.0269}

\bibitem{Bazavov:2017lyh}
A.~Bazavov, et~al., Phys. Rev. D \textbf{98}(7), 074512 (2018).
\newblock \doi{10.1103/PhysRevD.98.074512}

\bibitem{Richardson:1978bt}
J.L. Richardson, Phys. Lett. B \textbf{82}, 272 (1979).
\newblock \doi{10.1016/0370-2693(79)90753-6}

\bibitem{Bazavov:2011aa}
A.~Bazavov, et~al., Phys. Rev. \textbf{D85}, 114506 (2012).
\newblock \doi{10.1103/PhysRevD.85.114506}

\bibitem{Lepage:2001ym}
G.P. Lepage, B.~Clark, C.T.H. Davies, K.~Hornbostel, P.B. Mackenzie,
  C.~Morningstar, H.~Trottier, Nucl. Phys. B Proc. Suppl. \textbf{106}, 12
  (2002).
\newblock \doi{10.1016/S0920-5632(01)01638-3}

\bibitem{Toussaint:2008ke}
D.~Toussaint, W.~Freeman,   (2008).
\newblock ArXiv:0808.2211

\bibitem{Oktay:2008ex}
M.B. Oktay, A.S. Kronfeld, Phys. Rev. D \textbf{78}, 014504 (2008).
\newblock \doi{10.1103/PhysRevD.78.014504}

\bibitem{Mertens:1997wx}
B.P.G. Mertens, A.S. Kronfeld, A.X. El-Khadra, Phys. Rev. D \textbf{58}, 034505
  (1998).
\newblock \doi{10.1103/PhysRevD.58.034505}

\bibitem{Kronfeld:1996uy}
A.S. Kronfeld, Nucl. Phys. B Proc. Suppl. \textbf{53}, 401 (1997).
\newblock \doi{10.1016/S0920-5632(96)00671-8}

\bibitem{Bernard:2010fr}
C.~Bernard, et~al., Phys. Rev. D \textbf{83}, 034503 (2011).
\newblock \doi{10.1103/PhysRevD.83.034503}

\bibitem{Brodsky:1982gc}
S.J. Brodsky, G.P. Lepage, P.B. Mackenzie, Phys. Rev. D \textbf{28}, 228
  (1983).
\newblock \doi{10.1103/PhysRevD.28.228}

\bibitem{Lepage:1992xa}
G.P. Lepage, P.B. Mackenzie, Phys. Rev. D \textbf{48}, 2250 (1993).
\newblock \doi{10.1103/PhysRevD.48.2250}

\bibitem{Bailey:2008wp}
J.A. Bailey, et~al., Phys. Rev. \textbf{D79}, 054507 (2009).
\newblock \doi{10.1103/PhysRevD.79.054507}

\bibitem{Harada:2001fi}
J.~Harada, S.~Hashimoto, K.I. Ishikawa, A.S. Kronfeld, T.~Onogi, N.~Yamada,
  Phys. Rev. \textbf{D65}, 094513 (2002).
\newblock \doi{10.1103/PhysRevD.71.019903, 10.1103/PhysRevD.65.094513}.
\newblock [Erratum: Phys. Rev.D71,019903(2005)]

\bibitem{Aubin:2005aq}
C.~Aubin, C.~Bernard, Phys. Rev. \textbf{D73}, 014515 (2006).
\newblock \doi{10.1103/PhysRevD.73.014515}

\bibitem{Prelovsek:2005rf}
S.~Prelovsek, Phys. Rev. D \textbf{73}, 014506 (2006).
\newblock \doi{10.1103/PhysRevD.73.014506}

\bibitem{Bernard:2006gj}
C.W. Bernard, C.E. DeTar, Z.~Fu, S.~Prelovsek, PoS \textbf{LAT2006}, 173
  (2006).
\newblock \doi{10.22323/1.032.0173}

\bibitem{Bernard:2006zw}
C.~Bernard, Phys. Rev. D \textbf{73}, 114503 (2006).
\newblock \doi{10.1103/PhysRevD.73.114503}

\bibitem{Bernard:2007qf}
C.~Bernard, C.E. DeTar, Z.~Fu, S.~Prelovsek, Phys. Rev. D \textbf{76}, 094504
  (2007).
\newblock \doi{10.1103/PhysRevD.76.094504}

\bibitem{Aubin:2008wk}
C.~Aubin, J.~Laiho, R.S. Van~de Water, Phys. Rev. D \textbf{77}, 114501 (2008).
\newblock \doi{10.1103/PhysRevD.77.114501}

\bibitem{Bernard:2006ee}
C.~Bernard, M.~Golterman, Y.~Shamir, Phys. Rev. D \textbf{73}, 114511 (2006).
\newblock \doi{10.1103/PhysRevD.73.114511}

\bibitem{Shamir:2006nj}
Y.~Shamir, Phys. Rev. D \textbf{75}, 054503 (2007).
\newblock \doi{10.1103/PhysRevD.75.054503}

\bibitem{Shamir:2004zc}
Y.~Shamir, Phys. Rev. D \textbf{71}, 034509 (2005).
\newblock \doi{10.1103/PhysRevD.71.034509}

\bibitem{Durr:2005ax}
S.~Dürr, PoS \textbf{LAT2005}, 021 (2006).
\newblock \doi{10.22323/1.020.0021}

\bibitem{Sharpe:2006re}
S.R. Sharpe, PoS \textbf{LAT2006}, 022 (2006).
\newblock \doi{10.22323/1.032.0022}

\bibitem{Kronfeld:2007ek}
A.S. Kronfeld, PoS \textbf{LATTICE2007}, 016 (2007).
\newblock \doi{10.22323/1.042.0016}

\bibitem{Laiho:2005ue}
J.~Laiho, R.S. Van~de Water, Phys. Rev. \textbf{D73}, 054501 (2006).
\newblock \doi{10.1103/PhysRevD.73.054501}

\bibitem{Anastassov:2001cw}
A.~Anastassov, et~al., Phys. Rev. \textbf{D65}, 032003 (2002).
\newblock \doi{10.1103/PhysRevD.65.032003}

\bibitem{Lees:2013uxa}
J.P. Lees, et~al., Phys. Rev. \textbf{D88}(5), 052003 (2013).
\newblock \doi{10.1103/PhysRevD.88.079902, 10.1103/PhysRevD.88.052003}.
\newblock [Erratum: Phys. Rev.D88,no.7,079902(2013)]

\bibitem{Lees:2013zna}
J.P. Lees, et~al., Phys. Rev. Lett. \textbf{111}(11), 111801 (2013).
\newblock \doi{10.1103/PhysRevLett.111.111801, 10.1103/PhysRevLett.111.169902}

\bibitem{Detmold:2011bp}
W.~Detmold, C.J.D. Lin, S.~Meinel, Phys. Rev. Lett. \textbf{108}, 172003
  (2012).
\newblock \doi{10.1103/PhysRevLett.108.172003}

\bibitem{Can:2012tx}
K.U. Can, G.~Erkol, M.~Oka, A.~Ozpineci, T.T. Takahashi, Phys. Lett.
  \textbf{B719}, 103 (2013).
\newblock \doi{10.1016/j.physletb.2012.12.050}

\bibitem{Becirevic:2012pf}
D.~Becirevic, F.~Sanfilippo, Phys. Lett. \textbf{B721}, 94 (2013).
\newblock \doi{10.1016/j.physletb.2013.03.004}

\bibitem{Flynn:2015xna}
J.M. Flynn, P.~Fritzsch, T.~Kawanai, C.~Lehner, B.~Samways, C.T. Sachrajda,
  R.S. Van~de Water, O.~Witzel, Phys. Rev. \textbf{D93}(1), 014510 (2016).
\newblock \doi{10.1103/PhysRevD.93.014510}

\bibitem{Bernardoni:2014kla}
F.~Bernardoni, J.~Bulava, M.~Donnellan, R.~Sommer, Phys. Lett. \textbf{B740},
  278 (2015).
\newblock \doi{10.1016/j.physletb.2014.11.051}

\bibitem{Detmold:2012ge}
W.~Detmold, C.J.D. Lin, S.~Meinel, Phys. Rev. \textbf{D85}, 114508 (2012).
\newblock \doi{10.1103/PhysRevD.85.114508}

\bibitem{Aoki:2019cca}
S.~Aoki, et~al., Eur. Phys. J. C \textbf{80}(2), 113 (2020).
\newblock \doi{10.1140/epjc/s10052-019-7354-7}

\bibitem{Luke:1990eg}
M.E. Luke, Phys. Lett. B \textbf{252}, 447 (1990).
\newblock \doi{10.1016/0370-2693(90)90568-Q}

\bibitem{Ledoit2003}
O.~Ledoit, M.~Wolf, J. Empirical Finance \textbf{10}(5), 603 (2003).
\newblock \doi{10.1016/S0927-5398(03)00007-0}

\bibitem{Schaefer2005}
J.~Schaefer, K.~Strimmer, Statistical Applications in Genetics and Molecular
  Biology \textbf{4}, 32 (2005).
\newblock \doi{10.2202/1544-6115.1175}.
\newblock
  \urlprefix\url{https://www.degruyter.com/view/journals/sagmb/4/1/article-sagmb.2005.4.1.1175.xml.xml}

\bibitem{Ledoit2017}
O.~Ledoit, M.~Wolf,  (264) (2017).
\newblock \doi{10.5167/uzh-139880}.
\newblock \urlprefix\url{http://hdl.handle.net/10419/173422}

\bibitem{Simone2017}
J.N. Simone.
\newblock
  \href{https://cafpe.ugr.es/lattice2017/indico/session/102/contribution/386.html}{Improved
  data covariance estimation techniques in lattice {QCD}} (2017).
\newblock
  \urlprefix\url{https://cafpe.ugr.es/lattice2017/indico/session/102/contribution/386.html}.
\newblock {talk at \href{https://cafpe.ugr.es/lattice2017/}{35th International
  Symposium on Lattice Field Theory}}

\bibitem{Kronfeld:2000ck}
A.S. Kronfeld, Phys. Rev. \textbf{D62}, 014505 (2000).
\newblock \doi{10.1103/PhysRevD.62.014505}

\bibitem{Bazavov:2010hj}
A.~Bazavov, et~al., PoS \textbf{LATTICE2010}, 074 (2010).
\newblock \doi{10.22323/1.105.0074}

\bibitem{Sommer:1993ce}
R.~Sommer, Nucl. Phys. B \textbf{411}, 839 (1994).
\newblock \doi{10.1016/0550-3213(94)90473-1}

\bibitem{Bernard:2000gd}
C.W. Bernard, T.~Burch, K.~Orginos, D.~Toussaint, T.A. DeGrand, C.E. DeTar,
  S.A. Gottlieb, U.M. Heller, J.E. Hetrick, B.~Sugar, Phys. Rev. D \textbf{62},
  034503 (2000).
\newblock \doi{10.1103/PhysRevD.62.034503}

\bibitem{Arndt:2004bg}
D.~Arndt, C.J.D. Lin, Phys. Rev. \textbf{D70}, 014503 (2004).
\newblock \doi{10.1103/PhysRevD.70.014503}

\bibitem{Gambino:2020jvv}
P.~Gambino, et~al., Eur. Phys. J. C \textbf{80}(10), 966 (2020).
\newblock \doi{10.1140/epjc/s10052-020-08490-x}

\bibitem{Bourrely:2008za}
C.~Bourrely, I.~Caprini, L.~Lellouch, Phys. Rev. D \textbf{79}, 013008 (2009).
\newblock \doi{10.1103/PhysRevD.82.099902}.
\newblock [Erratum: Phys.Rev.D 82, 099902 (2010)]

\bibitem{Ferlewicz:2020lxm}
D.~Ferlewicz, P.~Urquijo, E.~Waheed, Phys. Rev. D \textbf{103}(7), 073005
  (2021).
\newblock \doi{10.1103/PhysRevD.103.073005}

\bibitem{Bigi:2016mdz}
D.~Bigi, P.~Gambino, Phys. Rev. D \textbf{94}(9), 094008 (2016).
\newblock \doi{10.1103/PhysRevD.94.094008}

\bibitem{Dungel:2010uk}
W.~Dungel, et~al., Phys. Rev. D \textbf{82}, 112007 (2010).
\newblock \doi{10.1103/PhysRevD.82.112007}

\bibitem{deBoer:2018ipi}
S.~de~Boer, T.~Kitahara, I.~Nisandzic, Phys. Rev. Lett. \textbf{120}(26),
  261804 (2018).
\newblock \doi{10.1103/PhysRevLett.120.261804}

\bibitem{Cali:2019nwp}
S.~Cal\'\i{}, S.~Klaver, M.~Rotondo, B.~Sciascia, Eur. Phys. J. C
  \textbf{79}(9), 744 (2019).
\newblock \doi{10.1140/epjc/s10052-019-7254-x}

\bibitem{Amhis:2016xyh}
Y.~Amhis, et~al., Eur. Phys. J. \textbf{C77}(12), 895 (2017).
\newblock \doi{10.1140/epjc/s10052-017-5058-4}

\bibitem{Bobeth:2021lya}
C.~Bobeth, M.~Bordone, N.~Gubernari, M.~Jung, D.~van Dyk, Eur. Phys. J. C
  \textbf{81}(11), 984 (2021).
\newblock \doi{10.1140/epjc/s10052-021-09724-2}

\bibitem{BaBar:2015zkb}
J.P. Lees, et~al., Phys. Rev. Lett. \textbf{116}(4), 041801 (2016).
\newblock \doi{10.1103/PhysRevLett.116.041801}

\bibitem{Fajfer:2012vx}
S.~Fajfer, J.F. Kamenik, I.~Nisandzic, Phys. Rev. D \textbf{85}, 094025 (2012).
\newblock \doi{10.1103/PhysRevD.85.094025}

\bibitem{Bernlochner:2017jka}
F.U. Bernlochner, Z.~Ligeti, M.~Papucci, D.J. Robinson, Phys. Rev. D
  \textbf{95}(11), 115008 (2017).
\newblock \doi{10.1103/PhysRevD.95.115008}.
\newblock [Erratum: Phys.Rev.D 97, 059902 (2018)]

\bibitem{Bordone:2019vic}
M.~Bordone, M.~Jung, D.~van Dyk, Eur. Phys. J. C \textbf{80}(2), 74 (2020).
\newblock \doi{10.1140/epjc/s10052-020-7616-4}

\bibitem{Bordone:2019guc}
M.~Bordone, N.~Gubernari, D.~van Dyk, M.~Jung, Eur. Phys. J. C \textbf{80}(4),
  347 (2020).
\newblock \doi{10.1140/epjc/s10052-020-7850-9}

\bibitem{Gubernari:2018wyi}
N.~Gubernari, A.~Kokulu, D.~van Dyk, JHEP \textbf{01}, 150 (2019).
\newblock \doi{10.1007/JHEP01(2019)150}

\bibitem{Kou:2018nap}
W.~Altmannshofer, et~al., PTEP \textbf{2019}(12), 123C01 (2019).
\newblock \doi{10.1093/ptep/ptz106}.
\newblock [Erratum: PTEP 2020, 029201 (2020)]

\bibitem{Na:2015kha}
H.~Na, C.M. Bouchard, G.P. Lepage, C.~Monahan, J.~Shigemitsu, Phys. Rev. D
  \textbf{92}(5), 054510 (2015).
\newblock \doi{10.1103/PhysRevD.93.119906}.
\newblock [Erratum: Phys.Rev.D 93, 119906 (2016)]

\bibitem{Aaij:2020hsi}
R.~Aaij, et~al., Phys. Rev. D \textbf{101}(7), 072004 (2020).
\newblock \doi{10.1103/PhysRevD.101.072004}

\bibitem{McLean:2019qcx}
E.~McLean, C.T.H. Davies, J.~Koponen, A.T. Lytle, Phys. Rev. D \textbf{101}(7),
  074513 (2020).
\newblock \doi{10.1103/PhysRevD.101.074513}

\bibitem{Bordone:2021oof}
M.~Bordone, B.~Capdevila, P.~Gambino, Phys. Lett. B \textbf{822}, 136679
  (2021).
\newblock \doi{10.1016/j.physletb.2021.136679}

\bibitem{Fael:2018vsp}
M.~Fael, T.~Mannel, K.K. Vos, JHEP \textbf{02}, 177 (2019).
\newblock \doi{10.1007/JHEP02(2019)177}

\bibitem{Bernlochner:2022ucr}
F.~Bernlochner, M.~Fael, K.~Olschewsky, E.~Persson, R.~van Tonder, K.K. Vos,
  M.~Welsch, JHEP \textbf{10}, 068 (2022).
\newblock \doi{10.1007/JHEP10(2022)068}

\bibitem{Hashimoto:2017wqo}
S.~Hashimoto, PTEP \textbf{2017}(5), 053B03 (2017).
\newblock \doi{10.1093/ptep/ptx052}

\bibitem{Gambino:2020crt}
P.~Gambino, S.~Hashimoto, Phys. Rev. Lett. \textbf{125}(3), 032001 (2020).
\newblock \doi{10.1103/PhysRevLett.125.032001}

\bibitem{Lees:2012xj}
J.~Lees, et~al., Phys. Rev. Lett. \textbf{109}, 101802 (2012).
\newblock \doi{10.1103/PhysRevLett.109.101802}

\bibitem{Kaneko:2019vkx}
T.~Kaneko, Y.~Aoki, G.~Bailas, B.~Colquhoun, H.~Fukaya, S.~Hashimoto,
  J.~Koponen, PoS \textbf{LATTICE2019}, 139 (2019).
\newblock \doi{10.22323/1.363.0139}

\bibitem{Chow:1993hr}
C.K. Chow, M.B. Wise, Phys. Rev. D \textbf{48}, 5202 (1993).
\newblock \doi{10.1103/PhysRevD.48.5202}

\bibitem{Aubin:2003mg}
C.~Aubin, C.~Bernard, Phys. Rev. D \textbf{68}, 034014 (2003).
\newblock \doi{10.1103/PhysRevD.68.034014}

\bibitem{Bailey:2012jg}
J.A. Bailey, et~al., Phys. Rev. Lett. \textbf{109}, 071802 (2012).
\newblock \doi{10.1103/PhysRevLett.109.071802}

\bibitem{peter_lepage_2021_4695132}
P.~Lepage, C.~Gohlke, D.~Hackett.
\newblock gplepage/gvar: gvar version 11.9.2 (2021).
\newblock \doi{10.5281/zenodo.4695132}.
\newblock \urlprefix\url{https://doi.org/10.5281/zenodo.4695132}

\end{thebibliography}

\end{document}